\input amstexl


\catcode`\@=11
\ifx\amstexloaded@\relax\else
 \errmessage{AmS-TeX must be loaded before LamS-TeX}\fi
\ifx\laxread@\undefined\else\catcode`\@=\active \fi
\def\err@#1{\errmessage{LamS-TeX error: #1}}
\def^^L{\par}
\let\+\tabalign
\def\newcount{\alloc@0\count\countdef\insc@unt}
\def\newdimen{\alloc@1\dimen\dimendef\insc@unt}
\def\newskip{\alloc@2\skip\skipdef\insc@unt}
\def\newmuskip{\alloc@3\muskip\muskipdef\@cclvi}
\def\newbox{\alloc@4\box\chardef\insc@unt}
\let\newtoks\relax
\def\newhelp#1#2{\newtoks#1#1\expandafter{\csname#2\endcsname}}
\def\newtoks{\alloc@5\toks\toksdef\@cclvi}
\def\newread{\alloc@6\read\chardef\sixt@@n}
\def\newwrite{\alloc@7\write\chardef\sixt@@n}
\def\newfam{\alloc@8\fam\chardef\sixt@@n}
\def\newlanguage{\alloc@9\language\chardef\@cclvi}
\def\newinsert#1{\global\advance\insc@unt by\m@ne
  \ch@ck0\insc@unt\count
  \ch@ck1\insc@unt\dimen
  \ch@ck2\insc@unt\skip
  \ch@ck4\insc@unt\box
  \allocationnumber=\insc@unt
  \global\chardef#1=\allocationnumber
  \wlog{\string#1=\string\insert\the\allocationnumber}}
\def\newif#1{\count@\escapechar \escapechar\m@ne
  \expandafter\expandafter\expandafter
   \edef\@if#1{true}{\let\noexpand#1=\noexpand\iftrue}%
  \expandafter\expandafter\expandafter
   \edef\@if#1{false}{\let\noexpand#1=\noexpand\iffalse}%
  \@if#1{false}\escapechar\count@}

\def\Err@#1{\errhelp\defaulthelp@\err@{#1}}
{\catcode`\@=\active
 \edef\next{\gdef\noexpand@{\futurelet\noexpand\next
  \csname at\string@\endcsname}}
 \next
}
\def\at@{\ifcat\noexpand\next a\let\next@\at@@\else
 \ifcat\noexpand\next0\let\next@\at@@\else
 \ifcat\noexpand\next\relax\let\next@\at@@\else
 \let\next@\at@@@\fi\fi\fi\next@}
\def\at@@@{\errhelp\athelp@\err@{Invalid use of @}}
\def\at@@#1{\expandafter
 \ifx\csname\string#1@at\endcsname\relax\let\next@\at@@@\else
 \DN@{\csname\string#1@at\endcsname}\fi\next@}
\def\atdef@#1{\expandafter\def\csname\string#1@at\endcsname}
\newif\iftest@
\def\tagin@#1{\tagin@false
 \DN@##1\tag##2##3\next@{\test@true\ifx\tagin@##2\test@false\fi}%
 \next@#1\tag\tagin@\next@\tagin@false\iftest@\tagin@true\fi}
\let\lkerns@\relax
\def\nolinebreak{\RIfM@\mathmodeerr@\nolinebreak\else
 \ifhmode\saveskip@\lastskip\unskip
 \nobreak\ifdim\saveskip@>\z@\hskip\saveskip@\fi\lkerns@
 \else\vmodeerr@\nolinebreak\fi\fi}
\def\allowlinebreak{\RIfM@\mathmodeerr@\allowlinebreak\else
 \ifhmode\saveskip@\lastskip\unskip
 \allowbreak\ifdim\saveskip@>\z@\hskip\saveskip@\fi\lkerns@
 \else\vmodeerr@\allowlinebreak\fi\fi}
\def\linebreak{\RIfM@\mathmodeerr@\linebreak\else
 \ifhmode\unskip\unkern\break\lkerns@
 \else\vmodeerr@\linebreak\fi\fi}
\let\nkerns@\relax
\def\newline{\RIfM@\mathmodeerr@\newline\else
 \ifhmode\unskip\unkern\null\hfill\break\nkerns@
 \else\vmodeerr@\newline\fi\fi}%
\def\newbox@{\alloc@@4\box\chardef\insc@unt}
\def\newcount@{\alloc@@0\count\countdef\insc@unt}
\def\accentedsymbol#1#2{\expandafter\newbox@\csname\exstring@#1@box\endcsname
 \setbox\csname\exstring@#1@box\endcsname\hbox{$\m@th#2$}%
 \define#1{\copy\csname\exstring@#1@box\endcsname{}}}
\def\rightadd@#1\to#2{\toks@{\\#1}\toks@@\expandafter{#2}\xdef#2{\the\toks@@
 \the\toks@}\toks@{}\toks@@{}}
\def\fontlist@{\\\tenrm\\\sevenrm\\\fiverm\\\teni\\\seveni\\\fivei
 \\\tensy\\\sevensy\\\fivesy\\\tenex\\\tenbf\\\sevenbf\\\fivebf
 \\\tensl\\\tenit}
\def\font@#1=#2 {\rightadd@#1\to\fontlist@\font#1=#2 }
\def\ismember@#1#2{\global\let\Next@ F\let\next@= #2%
 {\def\\##1{\let\nextii@##1\ifx\nextii@\next@\global\let\Next@ T\fi}#1}%
 \test@false\ifx\Next@ T\test@true\fi\let\next@\relax}
\def\FNSS@#1{\let\FNSS@@#1\FN@\FNSS@@@}
\def\FNSS@@@{\ifx\next\space@\def\FNSS@@@@. {\FN@\FNSS@@@}\else
 \def\FNSS@@@@.{\FNSS@@}\fi\FNSS@@@@.}
\atdef@"{\unskip
 \DN@{\ifx\next`\DN@`{\FN@\nextii@}%
  \else\ifx\next\lq\DN@\lq{\FN@\nextii@}%
  \else\DN@####1{\FN@\nextiii@}\fi\fi
  \next@}%
 \DNii@{\ifx\next`\DN@`{\sldl@``}%
  \else\ifx\next\lq\DN@\lq{\sldl@``}%
  \else\DN@{\dlsl@`}\fi\fi\next@}%
 \def\nextiii@{\ifx\next'\DN@'{\srdr@''}%
  \else\ifx\next\rq\DN@\rq{\srdr@''}%
  \else\DN@{\drsr@'}\fi\fi\next@}%
 \FNSS@\next@}
\def\root{%
  \DN@{\ifx\next\uproot\let\next@\nextii@\else
   \ifx\next\leftroot\let\next@\nextiii@\else
   \let\next@\plainroot@\fi\fi\next@}%
  \DNii@\uproot##1{\uproot@##1\relax\FNSS@\nextiv@}%
  \def\nextiv@{\ifx\next\leftroot\let\next@\nextv@\else
   \let\next@\plainroot@\fi\next@}%
  \def\nextv@\leftroot##1{\leftroot@##1\relax\plainroot@}%
  \def\nextiii@\leftroot##1{\leftroot@##1\relax\FNSS@\nextvi@}%
  \def\nextvi@{\ifx\next\uproot\let\next@\nextvii@\else
   \let\next@\plainroot@\fi\next@}%
  \def\nextvii@\uproot##1{\uproot@##1\relax\plainroot@}%
  \bgroup\uproot@\z@\leftroot@\z@
 \FNSS@\next@}
\def\loop#1\repeat{\def\iterate{#1\relax\expandafter\iterate\fi}%
 \iterate\let\iterate\relax}
\def\gloop@#1\repeat{\gdef\iterate@{#1\relax\expandafter\iterate@\fi}%
 \iterate@\global\let\iterate@\relax}
\def\printoptions{\W@{Do you want S(yntax check),
  G(alleys) or P(ages)?^^JType S, G or P, follow by <return>: }\loop
 \read\m@ne to\ans@
 \edef\next@{\def\noexpand\Ans@{\ans@}}\uppercase\expandafter{\next@}%
 \ifx\Ans@\S@\test@true\syntax\else
 \ifx\Ans@\G@\test@true\galleys\else
 \ifx\Ans@\P@\test@true\else
 \test@false\fi\fi\fi
 \iftest@\else\W@{Type S, G or P, follow by <return>: }%
 \repeat}
\expandafter\let\csname A@;\endcsname;
\expandafter\let\csname A@:\endcsname:
\expandafter\let\csname A@?\endcsname?
\expandafter\let\csname A@!\endcsname!
\def\APdef#1{\def\next@{\expandafter\let\csname A@\string#1\endcsname#1}%
 \afterassignment\next@\def#1}
\let\fextra@\,
\def\tdots@{\unskip
 \DN@{$\m@th\mathinner{\ldotp\ldotp\ldotp}\,
   \ifx\next,\,$\else\ifx\next.\,$\else
   \ifx\next;\,$\else
   \expandafter\ifx\csname A@\string;\endcsname\next\fextra@$\else
   \ifx\next:\,$\else
   \expandafter\ifx\csname A@\string:\endcsname\next\fextra@$\else
   \ifx\next?\,$\else
   \expandafter\ifx\csname A@\string?\endcsname\next\fextra@$\else
   \ifx\next!\,$\else
   \expandafter\ifx\csname A@\string!\endcsname\next\fextra@$\else
   $ \fi\fi\fi\fi\fi\fi\fi\fi\fi\fi}%
 \ \FN@\next@}
\def\extrap@#1{%
 \ifx\next,\DN@{#1\,}\else
 \ifx\next;\DN@{#1\,}\else
 \expandafter\ifx\csname A@\string;\endcsname\next\DN@{#1\fextra@}\else
 \ifx\next.\DN@{#1\,}\else\extra@
 \ifextra@\DN@{#1\,}\else
 \let\next@#1\fi\fi\fi\fi\fi\next@}
\def\dotsc{\DN@{\ifx\next;\plainldots@\,\else
 \expandafter\ifx\csname A@\string;\endcsname\next\plainldots@\fextra@\else
 \ifx\next.\plainldots@\,\else\extra@\plainldots@
 \ifextra@\,\fi\fi\fi\fi}%
 \FN@\next@}
\def\keybin@{\keybin@true
 \ifx\next+\else\ifx\next=\else\ifx\next<\else\ifx\next>\else\ifx\next-\else
 \ifx\next*\else\ifx\next:\else
 \expandafter\ifx\csname A@\string;\endcsname\next\else
 \keybin@false\fi\fi\fi\fi\fi\fi\fi\fi}
\def\boldkey#1{\ifcat\noexpand#1A%
  \ifcmmibloaded@{\fam\cmmibfam#1}\else
   \Err@{First bold symbol font not loaded}\fi
 \else
 \let\next=#1%
 \ifx#1!\mathchar"5\bffam@21 \else
 \expandafter\ifx\csname A@\string!\endcsname\next\mathchar"5\bffam@21 \else
 \ifx#1(\mathchar"4\bffam@28 \else\ifx#1)\mathchar"5\bffam@29 \else
 \ifx#1+\mathchar"2\bffam@2B \else\ifx#1:\mathchar"3\bffam@3A \else
 \expandafter\ifx\csname A@\string:\endcsname\next\mathchar"3\bffam@3A \else
 \ifx#1;\mathchar"6\bffam@3B \else
 \expandafter\ifx\csname A@\string;\endcsname\next\mathchar"6\bffam@3B \else
 \ifx#1=\mathchar"3\bffam@3D \else
 \ifx#1?\mathchar"5\bffam@3F \else
 \expandafter\ifx\csname A@\string?\endcsname\next\mathchar"5\bffam@3F \else
 \ifx#1[\mathchar"4\bffam@5B \else
 \ifx#1]\mathchar"5\bffam@5D \else
 \ifx#1,\mathchari@63B \else
 \ifx#1-\mathcharii@200 \else
 \ifx#1.\mathchari@03A \else
 \ifx#1/\mathchari@03D \else
 \ifx#1<\mathchari@33C \else
 \ifx#1>\mathchari@33E \else
 \ifx#1*\mathcharii@203 \else
 \ifx#1|\mathcharii@06A \else
 \ifx#10\bold0\else\ifx#11\bold1\else\ifx#12\bold2\else\ifx#13\bold3\else
 \ifx#14\bold4\else\ifx#15\bold5\else\ifx#16\bold6\else\ifx#17\bold7\else
 \ifx#18\bold8\else\ifx#19\bold9\else
  \Err@{\noexpand\boldkey can't be used with #1}%
 \fi\fi\fi\fi\fi\fi\fi\fi\fi\fi\fi\fi\fi\fi\fi
 \fi\fi\fi\fi\fi\fi\fi\fi\fi\fi\fi\fi\fi\fi\fi\fi\fi\fi}
\def\arabic#1{#1}
\def\alph#1{\count@#1\relax\advance\count@96 \ifnum\count@>122
 \Err@{\noexpand\alph invalid for numbers > 26}\else\char\count@\fi}
\def\Alph#1{\count@#1\relax\advance\count@64 \ifnum\count@>90
 \Err@{\noexpand\Alph invalid for numbers > 26}\else\char\count@\fi}

\def\Roman#1{\uppercase\expandafter{\romannumeral#1}}
\def\fnsymbol#1{\count@#1\relax
 \count@@\count@
 \advance\count@\m@ne\divide\count@7
 \count@@@\count@\advance\count@@@\@ne
 \multiply\count@7 \advance\count@@-\count@
 \count@\count@@@
 {\loop
  \ifcase\count@@\or*\or\dag\or\ddag\or\P\or\S\or\text{$\|$}\or\#\fi
  \advance\count@\m@ne\ifnum\count@>\z@\repeat}}
\def\cardnine@#1{\ifcase#1\or one\or two\or three\or four\or five\or
 six\or seven\or eight\or nine\fi}
\let\alloc@\alloc@@
\newcount\ten@
\ten@10
\def\cardinal#1{\count@#1\relax
 \ifnum\count@>99 \number\count@
 \else
  \ifnum\count@=\z@ zero%
  \else
   \ifnum\count@<\ten@\cardnine@\count@
   \else
    \ifnum\count@<20
     \advance\count@-\ten@
     \ifcase\count@ ten\or eleven\or twelve\or thirteen\or fourteen\or
      fifteen\or sixteen\or seventeen\or eighteen\or nineteen\fi
    \else
     \count@@\count@\count@@@\count@@
     \divide\count@\ten@\multiply\count@\ten@
     \advance\count@@@-\count@\divide\count@\ten@
     \ifcase\count@\or\or twenty\or thirty\or forty\or fifty\or sixty\or
      seventy\or eighty\or ninety\fi
     \ifnum\count@@@=\z@\else-\cardnine@\count@@@\fi
    \fi
   \fi
  \fi
 \fi}
\def\ordnine@#1{\ifcase#1\or first\or second\or third\or fourth\or fifth\or
 sixth\or seventh\or eighth\or ninth\fi}
\newcount\count@@@@
\def\ordsuffix@{\count@@@@\count@
 \divide\count@\ten@
 \count@@@\count@\count@@\count@
 \divide\count@@\ten@\multiply\count@@\ten@
 \advance\count@@@-\count@@
 \ifnum\count@@@=\@ne th%
 \else
  \count@@@\count@@@@
  \count@@\count@@@@
  \divide\count@@\ten@\multiply\count@@\ten@
  \advance\count@@@-\count@@
  \ifcase\count@@@ th\or st\or nd\or rd\else th\fi
 \fi}
\def\nordinal#1{\count@#1\relax\number\count@\ordsuffix@}
\def\spordinal#1{\count@#1\relax\number\count@$^{\text{\ordsuffix@}}$}
\def\ordinal#1{\count@#1\relax
 \ifnum\count@>99 \number\count@\ordsuffix@
 \else
   \ifnum\count@=\z@ zeroth%
  \else
    \ifnum\count@<\ten@\ordnine@\count@
    \else
     \ifnum\count@<20 \advance\count@-\ten@
      \ifcase\count@ tenth\or eleventh\or twelfth\or thirteenth\or
       fourteenth\or fifteenth\or sixteenth\or seventeenth\or eighteenth\or
       nineteenth\fi
     \else
      \count@@\count@
      \divide\count@\ten@\multiply\count@\ten@
      \count@@@\count@@\advance\count@@@-\count@
      \divide\count@\ten@
      \ifcase\count@\or\or twent\or thirt\or fort\or fift\or sixt\or sevent\or
       eight\or ninet\fi
      \ifnum\count@@@=\z@ ieth\else y-\ordnine@\count@@@\fi
     \fi
    \fi
  \fi
 \fi}
\font@\tensmc=cmcsc10
\textonlyfont@\smc\tensmc
\newtoks\noexpandtoks@
\noexpandtoks@{\let\arabic\relax\let\alph\relax\let\Alph\relax
 \let\Roman\relax\let\fnsymbol\relax\let\rm\relax
 \let\it\relax\let\bf\relax\let\sl\relax\let\smc\relax
 \let\/\relax\let\null\relax}
\def\noexpands@{\the\noexpandtoks@}
\def\Nonexpanding#1{\global\noexpandtoks@
 \expandafter{\the\noexpandtoks@\let#1\relax}}
\def\prevanish@{\saveskip@\z@\ifhmode\saveskip@\lastskip\unskip\fi}
\def\postvanish@{\ifdim\saveskip@>\z@\hskip\saveskip@\fi\FN@\postvanish@@}
\def\postvanish@@{\DN@.{}%
 \ifx\next\space@\ifdim\saveskip@>\z@\DN@. {}\fi\fi\next@.}
\def\invisible#1{\prevanish@\ignorespaces#1\unskip\postvanish@}
\def\vanishlist@{\\\invisible}
\let\noindent@\noindent
\def\noindent{\par\noindent@\FN@\pretendspace@}
\def\pretendspace@{\ismember@\vanishlist@\next
 \iftest@\nobreak\hskip-\p@\hskip\p@\fi}
\let\flushpar\noindent
\newtoks\everypartoks@
\def\noindent@@{\par\everypartoks@\expandafter{\the\everypar}\everypar{}%
 \noindent@\everypar\expandafter{\the\everypartoks@}}
\def\page{\Err@{\noexpand\page has no meaning by itself}}
\let\page@C\pageno
\let\page@P\empty
\let\page@Q\empty
\def\page@S#1{#1\/}
\def\page@F{\rm}
\def\page@N{\arabic}   
\newif\ifindexing@
\def\indexfile{\ifindexing@\else
 \alloc@@7\write\chardef\sixt@@n\ndx@
 \immediate\openout\ndx@=\jobname.ndx
 \global\indexing@true\fi}
\global\advance\insc@unt\m@ne
\ch@ck0\insc@unt\count
\ch@ck1\insc@unt\dimen
\ch@ck2\insc@unt\skip
\ch@ck4\insc@unt\box
\allocationnumber\insc@unt
\global\chardef\margin@\allocationnumber
\dimen\margin@\maxdimen
\count\margin@\z@
\skip\margin@\z@
\newif\ifindexproofing@
\def\indexproofing{\indexproofing@true}
\def\noindexproofing{\indexproofing@false}
\def\unmacro@#1:#2->#3\unmacro@{\def\macpar@{#2}\def\macdef@{#3}}
\def\starparts@#1{\def\stari@{#1}\def\starii@{#1}\let\stariii@\empty
 \test@false
 \DN@##1*##2##3\next@{\ifx\starparts@##2\test@false\else\test@true\fi}%
 \next@#1*\starparts@\next@
 \iftest@\DN@{\starparts@@#1\starparts@@}\else\let\next@\relax\fi\next@}
\def\starparts@@#1*#2\starparts@@{\def\starii@{#1}\def\stariii@{*#2}}
\def\windex@{\ifindexing@
 \expandafter\unmacro@\meaning\stari@\unmacro@
 \edef\macdef@{\string"\macdef@\string"}%
 \edef\next@{\write\ndx@{\macdef@}}\next@
 \write\ndx@{{\number\pageno}{\page@N}{\page@P}{\page@Q}}%
 \fi
 \ifindexproofing@
  \ifx\stariii@\empty\else
   \expandafter\unmacro@\meaning\stariii@\unmacro@\fi
  \insert\margin@{\hbox{\rm\vrule\height9\p@\depth2\p@\width\z@\starii@
  \ifx\stariii@\empty\else\tt\macdef@\fi}}\fi}
\catcode`\"=\active
\def"{\FN@\quote@}
\def\quote@{\ifx\next"\expandafter\quote@@\else\expandafter\quote@@@\fi}
\def\quote@@@#1"{\starparts@{#1}\starii@\windex@}
\def\quote@@"#1"{\prevanish@\starparts@{#1}\windex@\FN@\quote@@@@}
\def\quote@@@@{\ifx\next"\DN@"{\postvanish@}\else
 \let\next@\postvanish@\fi\next@}
\rightadd@"\to\vanishlist@
\def\idefine#1{\DN@{#1}\DNii@{\noexpand#1}%
 \afterassignment\idefine@\def\nextiii@}
\def\idefine@{\ifindexing@
 \expandafter\let\next@\nextiii@
 \expandafter\unmacro@\meaning\nextiii@\unmacro@
 \immediate\write\ndx@{\noexpand\define\nextii@\macpar@{\macdef@}}\fi}
\def\iabbrev*#1#2{\ifindexing@\toks@{#2}%
 \immediate\write\ndx@{\noexpand\abbrev*\noexpand#1{\the\toks@}}\fi}
\newread\laxread@
\newwrite\laxwrite@
\let\fnpages@\empty
\def\Finit@#1#2\Finit@{\let\nextii@#1\def\nextiii@{#2}}
\catcode`\~=11
\def\getparts@ @#1~#2~#3~#4~#5~#6{\def\nextiv@{#1}%
 \def\nextiii@{#2~#3~#4~#5~}\count@#6\relax}
\newif\ifdocument@
\def\document{\ifdocument@\else\global\document@true
 \let\fontlist@\empty
 \immediate\openin\laxread@=\jobname.lax\relax
 {\endlinechar\m@ne\noexpands@\catcode`\@=11 \catcode`\~=11
  \loop\ifeof\laxread@\else
   \read\laxread@ to\next@
   \ifx\next@\empty
   \else
    \expandafter\Finit@\next@\Finit@
    \if\nextii@ F%
     \expandafter\rightadd@\nextiii@\to\fnpages@
    \else
     \expandafter\getparts@\next@
     \edef\next@{\gdef\csname\nextiv@ @L\endcsname{\nextiii@\number\count@}}%
     \next@
    \fi
   \fi
  \repeat}%
 \immediate\closein\laxread@
 \immediate\openout\laxwrite@=\jobname.lax\relax\fi}
\let\thelabel@\relax
\def\thelabels@{\thelabel@ ~\thelabel@@ ~\thelabel@@@ ~\thelabel@@@@ ~}
\def\label#1{\prevanish@
 \ifx\thelabel@\relax
  \Err@{There's nothing here to be labelled}%
 \else
  {\noexpands@
  \expandafter\ifx\csname#1@L\endcsname\relax
   \expandafter\xdef\csname#1@L\endcsname{\thelabels@0}%
   \immediate\write\laxwrite@{@#1~\thelabels@1}%
  \else
   \edef\next@{@~\csname#1@L\endcsname}%
    \expandafter\getparts@\next@
    \ifodd\count@
    \expandafter\xdef\csname#1@L\endcsname{\thelabels@0}%
    \immediate\write\laxwrite@{@#1~\thelabels@1}%
   \else
    \Err@{Label #1 already used}%
   \fi
  \fi
  }%
 \fi
 \postvanish@}
\rightadd@\label\to\vanishlist@
\def\thepages@{\page@N{\number\page@C}~%
 \page@S{\page@P\page@N{\number\page@C}\page@Q}~%
 \number\page@C ~\page@P\page@N{\number\page@C}\page@Q ~}
\def\pagelabel#1{\prevanish@
 \expandafter\ifx\csname#1@L\endcsname\relax
  {\noexpands@
  \expandafter\xdef\csname#1@L\endcsname{\thepages@2}}%
  \write\laxwrite@{@#1~\thepages@3}%
 \else
  {\noexpands@
  \edef\next@{@~\csname#1@L\endcsname}%
  \expandafter\getparts@\next@
  \ifodd\count@
   \ifnum\count@=\@ne
    \expandafter\xdef\csname#1@L\endcsname{\thelabels@2}%
   \fi
   \write\laxwrite@{@#1~\thepages@3}%
  \else
   \Err@{Label #1 already used}%
  \fi
  }%
 \fi
 \postvanish@}
\rightadd@\pagelabel\to\vanishlist@
\newif\ifreferr@
\referr@true
\def\RefErrors{\global\referr@true}
\def\RefWarnings{\global\referr@false}
\setbox\z@\hbox{\global\count@=`^^30}
\ifnum\count@=48 \let\versionthree@\relax\fi
\def\nolabel@#1#2#3{\expandafter\ifx\csname#2@L\endcsname\relax
 \ifreferr@\Err@{No \noexpand\label found for #2}\else
 \W@{Warning: No \noexpand\label found for #2.}%
 \ifx\versionthree@\relax\W@{l.\number\inputlineno\space ... \string#1{#2}}\fi
 \fi#3\else}
\def\csL@#1{{\noexpands@\xdef\Next@{\csname#1@L\endcsname}}}
\def\ref#1{\nolabel@\ref{#1}\relax
 \DNii@##1~##2\nextii@{##1}%
 \csL@{#1}\expandafter\nextii@\Next@\nextii@\fi}
\def\Ref#1{\nolabel@\Ref{#1}\relax
 \DNii@##1~##2~##3\nextii@{##2}%
 \csL@{#1}\expandafter\nextii@\Next@\nextii@\fi}
\def\nref#1{\nolabel@\nref{#1}\relax
 \DNii@##1~##2~##3~##4\nextii@{##3}%
 \csL@{#1}\expandafter\nextii@\Next@\nextii@\fi}
\def\pref#1{\nolabel@\pref{#1}\relax
 \DNii@##1~##2~##3~##4~##5\nextii@{##4}%
 \csL@{#1}\expandafter\nextii@\Next@\nextii@\fi}
\let\pref@\pref
\def\Evaluatenref#1{\nolabel@\Evaluatenref{#1}{\gdef\Nref{-10000 }}%
 \DNii@##1~##2~##3~##4\nextii@{\DNii@{##3}}%
 \csL@{#1}\expandafter\nextii@\Next@\nextii@
 \xdef\Nref{\nextii@}\fi}
\def\Evaluatepref#1{\nolabel@\Evaluatepref{#1}{\global\let\Pref\empty}%
 \DNii@##1~##2~##3~##4~##5\nextii@{\DNii@{##4}}%
 \csL@{#1}\expandafter\nextii@\Next@\nextii@
 \xdef\Pref{\nextii@}\fi}
\def\readlax#1{\immediate\openin\laxread@=#1.lax\relax
 \ifeof\laxread@\W@{}\W@{File #1.lax not found.}\W@{}\fi
 {\endlinechar\m@ne\noexpands@\catcode`\@=11 \catcode`\~=11
  \loop\ifeof\laxread@\else
   \read\laxread@ to\nextv@
   \ifx\nextv@\empty
   \else
    \expandafter\Finit@\nextv@\Finit@
    \ifx\nextii@ F%
    \else
     \expandafter\getparts@\nextv@
     \expandafter\ifx\csname\nextiv@ @L\endcsname\relax
      \edef\next@{\gdef\csname\nextiv@ @L\endcsname
       {\nextiii@\ifnum\count@=\@ne0\else2\fi}}%
      \next@
     \else
      \Err@{Label \nextiv@\space in #1.lax already used}%
     \fi
    \fi
   \fi
  \repeat}%
 \immediate\closein\laxread@}
\catcode`\~=\active

\def\FNSSP@{\FNSS@\pretendspace@}
\everydisplay{\csname displaymath \endcsname}
\expandafter\def\csname displaymath \endcsname#1$${#1$$\FNSSP@}
\def\locallabel@{\let\thelabel@\Thelabel@\let\thelabel@@\Thelabel@@
 \let\thelabel@@@\Thelabel@@@\let\thelabel@@@@\Thelabel@@@@}
\newcount\tag@C
\tag@C\z@
\let\tag@P\empty
\let\tag@Q\empty
\def\tag@S#1{{\rm(}{#1\/}{\rm)}}
\let\tag@N\arabic
\def\tag@F{\rm}
\def\maketag@{\FN@\maketag@@}
\def\maketag@@{\ifx\next\relax\DN@\relax{\FN@\maketag@@}\else
 \ifx\next"\let\next@\maketag@@@\else
 \let\next@\maketag@@@@\fi\fi\next@}
\def\xdefThelabel@#1{\xdef\Thelabel@{#1{\Thelabel@@@}}}
\def\xdefThelabel@@#1{\xdef\Thelabel@@{#1{\Thelabel@@@@}}}
\def\maketag@@@@#1\maketag@{\global\advance\tag@C\@ne
 {\noexpands@
  \xdef\Thelabel@@@{\number\tag@C}%
  \xdefThelabel@\tag@N
  \xdef\Thelabel@@@@{\ifmathtags@$\tag@P\Thelabel@\tag@Q$\else
   \tag@P\Thelabel@\tag@Q\fi}%
  \xdefThelabel@@\tag@S
  }%
 \locallabel@
 \hbox{\tag@F\thelabel@@}%
 #1}
\def\Qlabel@#1{{\noexpands@\xdef\Thelabel@@{#1}%
 \let\style\empty\xdef\Thelabel@@@@{#1}%
 \let\pre\empty\let\post\empty\xdef\Thelabel@{#1}%
 \let\numstyle\empty\xdef\Thelabel@@@{#1}}}
\def\maketag@@@"#1"#2\maketag@{%
 {\let\pre\tag@P\let\post\tag@Q\let\style\tag@S\let\numstyle\tag@N
  \hbox{\tag@F#1}%
  \noexpands@
  \Qlabel@{#1}%
  }%
 \locallabel@
 #2}
\def\align@{\inalign@true\inany@true
 \vspace@\allowdisplaybreak@\displaybreak@\intertext@
 \def\tag{\global\tag@true\ifnum\and@=\z@
  \DN@{&\omit\global\rwidth@\z@&\relax}\else
  \DN@{&\relax}\fi\next@}%
 \iftagsleft@\DN@{\csname align \endcsname}\else
  \DN@{\csname align \space\endcsname}\fi\next@}
\def\noset@{\def\Offset##1##2{\prevanish@\postvanish@}%
 \def\Reset##1##2{\prevanish@\postvanish@}}
\def\measure@#1\endalign{\global\lwidth@\z@\global\rwidth@\z@
 \global\maxlwidth@\z@\global\maxrwidth@\z@
 \global\and@\z@
 \setbox\z@\vbox
  {\noset@\everycr{\noalign{\global\tag@false\global\and@\z@}}\Let@
  \halign{\setboxz@h{$\m@th\displaystyle{\@lign##}$}%
   \global\lwidth@\wdz@
   \ifdim\lwidth@>\maxlwidth@\global\maxlwidth@\lwidth@\fi
   \global\advance\and@\@ne
   &\setboxz@h{$\m@th\displaystyle{{}\@lign##}$}\global\rwidth@\wdz@
   \ifdim\rwidth@>\maxrwidth@\global\maxrwidth@\rwidth@\fi
   \global\advance\and@\@ne
   &\Tag@\eat@{##}\crcr#1\crcr}}%
 \totwidth@\maxlwidth@\advance\totwidth@\maxrwidth@}
\def\prepost@{\global\let\tag@P@\tag@P\global\let\tag@Q@\tag@Q}
\def\reprepost@{\let\tag@P\tag@P@\let\tag@Q\tag@Q@}
\expandafter\def\csname align \space\endcsname#1\endalign
 {\measure@#1\endalign\global\and@\z@
 \ifingather@\everycr{\noalign{\global\and@\z@}}\else\displ@y@\fi
 \Let@\tabskip\centering@
 \halign to\displaywidth
  {\hfil\strut@\setboxz@h{$\m@th\displaystyle{\@lign##\prepost@}$}%
  \boxz@\global\advance\and@\@ne
  \tabskip\z@skip
  &\setboxz@h{$\m@th\displaystyle{{}\@lign##\prepost@}$}%
  \global\rwidth@\wdz@\boxz@\hfil\global\advance\and@\@ne
  \tabskip\centering@
  &\setboxz@h{\@lign\strut@\reprepost@\maketag@##\maketag@}%
  \dimen@\displaywidth\advance\dimen@-\totwidth@
  \divide\dimen@\tw@\advance\dimen@\maxrwidth@\advance\dimen@-\rwidth@
  \ifdim\dimen@<\tw@\wdz@\llap{\vtop{\normalbaselines\null\boxz@}}%
  \else\llap{\boxz@}\fi
  \tabskip\z@skip
  \crcr#1\crcr
  \black@\totwidth@}}
\expandafter\def\csname align \endcsname#1\endalign{\measure@#1\endalign
 \global\and@\z@
 \ifdim\totwidth@>\displaywidth\let\displaywidth@\totwidth@\else
  \let\displaywidth@\displaywidth\fi
 \ifingather@\everycr{\noalign{\global\and@\z@}}\else\displ@y@\fi
 \Let@\tabskip\centering@\halign to\displaywidth
  {\hfil\strut@\setboxz@h{$\m@th\displaystyle{\@lign##\prepost@}$}%
  \global\lwidth@\wdz@\global\lineht@\ht\z@
  \boxz@\global\advance\and@\@ne
  \tabskip\z@skip&\setboxz@h{$\m@th\displaystyle{{}\@lign##\prepost@}$}%
  \ifdim\ht\z@>\lineht@\global\lineht@\ht\z@\fi
  \boxz@\hfil\global\advance\and@\@ne
  \tabskip\centering@&\kern-\displaywidth@
  \setboxz@h{\@lign\strut@\reprepost@\maketag@##\maketag@}%
  \dimen@\displaywidth\advance\dimen@-\totwidth@
  \divide\dimen@\tw@\advance\dimen@\maxlwidth@\advance\dimen@-\lwidth@
  \ifdim\dimen@<\tw@\wdz@
   \rlap{\vbox{\normalbaselines\boxz@\vbox to\lineht@{}}}\else
   \rlap{\boxz@}\fi
  \tabskip\displaywidth@\crcr#1\crcr\black@\totwidth@}}
\def\attag@#1{\let\Maketag@\maketag@\let\TAG@\Tag@
 \let\Prepost@\prepost@\let\Reprepost@\reprepost@
 \let\Tag@\relax\let\maketag@\relax
 \let\prepost@\relax\let\reprepost@\relax
 \ifmeasuring@
  \def\llap@##1{\setboxz@h{##1}\hbox to\tw@\wdz@{}}%
  \def\rlap@##1{\setboxz@h{##1}\hbox to\tw@\wdz@{}}%
 \else\let\llap@\llap\let\rlap@\rlap\fi
 \toks@{\hfil\strut@
  $\m@th\displaystyle{\@lign\the\hashtoks@\prepost@}$%
  \tabskip\z@skip\global\advance\and@\@ne&
  $\m@th\displaystyle{{}\@lign\the\hashtoks@\prepost@}$\hfil
  \ifxat@\tabskip\centering@\fi\global\advance\and@\@ne}%
 \iftagsleft@
  \toks@@{\tabskip\centering@&\Tag@\kern-\displaywidth
   \rlap@{\@lign\reprepost@\maketag@\the\hashtoks@\maketag@}%
   \global\advance\and@\@ne\tabskip\displaywidth}\else
  \toks@@{\tabskip\centering@&\Tag@\llap@{\@lign\reprepost@\maketag@
   \the\hashtoks@\maketag@}\global\advance\and@\@ne\tabskip\z@skip}\fi
 \atcount@#1\relax\advance\atcount@\m@ne
 \loop\ifnum\atcount@>\z@
  \toks@\expandafter{\the\toks@&\hfil$\m@th\displaystyle{\@lign
  \the\hashtoks@\prepost@}$\global\advance\and@\@ne
  \tabskip\z@skip
  &$\m@th\displaystyle{{}\@lign\the\hashtoks@\prepost@}$\hfil\ifxat@
  \tabskip\centering@\fi\global\advance\and@\@ne}\advance\atcount@\m@ne
 \repeat
 \edef\preamble@{\the\toks@\the\toks@@}%
 \edef\preamble@@{\preamble@}%
 \let\maketag@\Maketag@\let\Tag@\TAG@
 \let\prepost@\Prepost@\let\reprepost@\Reprepost@}
\def\unlabel@{\def\label##1{\prevanish@\postvanish@}%
 \def\pagelabel##1{\prevanish@\postvanish@}}
\newcount\tag@CC
\expandafter\def\csname alignat \endcsname#1#2\endalignat
 {\inany@true\xat@false
 \def\tag{\global\tag@true
  \count@#1\relax\multiply\count@\tw@\advance\count@\m@ne
  \gdef\tag@{&}%
  \loop\ifnum\count@>\and@\xdef\tag@{&\omit\tag@}%
  \advance\count@\m@ne\repeat
  \tag@\relax}%
 \vspace@\allowdisplaybreak@\displaybreak@\intertext@
 \displ@y@\measuring@true\tag@CC\tag@C
 \setbox\savealignat@\hbox{\noset@\unlabel@$\m@th\displaystyle\Let@
  \attag@{#1}\vbox{\halign{\span\preamble@@\crcr#2\crcr}}$}%
 \measuring@false
 \Let@\attag@{#1}\tag@C\tag@CC
 \tabskip\centering@\halign to\displaywidth
  {\span\preamble@@\crcr#2\crcr\black@{\wd\savealignat@}}}
\expandafter\def\csname xalignat \endcsname#1#2\endxalignat
 {\inany@true\xat@true
 \def\tag{\global\tag@true
  \count@#1\relax\multiply\count@\tw@\advance\count@\m@ne
  \gdef\tag@{&}%
  \loop\ifnum\count@>\and@\xdef\tag@{&\omit\tag@}%
  \advance\count@\m@ne\repeat
  \tag@\relax}%
 \vspace@\allowdisplaybreak@\displaybreak@\intertext@
 \displ@y@\measuring@true\tag@CC\tag@C
 \setbox\savealignat@\hbox{\noset@\unlabel@$\m@th\displaystyle\Let@
  \attag@{#1}\vbox{\halign{\span\preamble@@\crcr#2\crcr}}$}%
 \measuring@false\Let@\attag@{#1}\tag@C\tag@CC
 \tabskip\centering@\halign to\displaywidth
 {\span\preamble@@\crcr#2\crcr\black@{\wd\savealignat@}}}
\def\gather{\RIfMIfI@\DN@{\onlydmatherr@\gather}\else
 \ingather@true\inany@true\def\tag{&\relax}%
 \vspace@\allowdisplaybreak@\displaybreak@\intertext@
 \displ@y\Let@
 \iftagsleft@\DN@{\csname gather \endcsname}\else
  \DN@{\csname gather \space\endcsname}\fi\fi
 \else\DN@{\onlydmatherr@\gather}\fi\next@}
\def\exstring@{\expandafter\eat@\string}
\def\newcounter#1{\define#1{}%
 \edef\next@{\def\noexpand#1{\futurelet\noexpand\next
  \csname\exstring@#1@Z\endcsname}}\next@
 \edef\next@{\def\csname\exstring@#1@Z\endcsname
  {\global\advance\csname\exstring@#1@C\endcsname\@ne
  {\csname\exstring@#1@F\endcsname\csname\exstring@#1@S\endcsname
   {\csname\exstring@#1@P\endcsname\csname\exstring@#1@N\endcsname
   {\noexpand\number\csname\exstring@#1@C\endcsname}%
   \csname\exstring@#1@Q\endcsname}}%
  \noexpand\ifx\noexpand\next\noexpand\label
   \def\noexpand\next@\noexpand\label########1{{\noexpand\noexpands@
    \xdef\noexpand\Thelabel@{\csname\exstring@#1@N\endcsname
     {\noexpand\number\csname\exstring@#1@C\endcsname}}%
    \xdef\noexpand\Thelabel@@@{\noexpand\number
     \csname\exstring@#1@C\endcsname}%
    \xdef\noexpand\Thelabel@@{\csname\exstring@#1@S\endcsname
     {\csname\exstring@#1@P\endcsname
     \csname\exstring@#1@N\endcsname
     {\noexpand\number\csname\exstring@#1@C\endcsname}%
     \csname\exstring@#1@Q\endcsname}}%
    \xdef\noexpand\Thelabel@@@@{\csname\exstring@#1@P\endcsname
     \csname\exstring@#1@N\endcsname
     {\noexpand\number\csname\exstring@#1@C\endcsname}%
     \csname\exstring@#1@Q\endcsname}}%
    {\noexpand\locallabel@\noexpand\label{########1}}}%
   \noexpand\else\let\noexpand\next@\relax\noexpand\fi\noexpand\next@}}\next@
 \expandafter\newcount@\csname\exstring@#1@C\endcsname
 \expandafter\let\csname\exstring@#1@N\endcsname\arabic
 \expandafter\def\csname\exstring@#1@S\endcsname##1{##1\/}%
 \expandafter\let\csname\exstring@#1@P\endcsname\empty
 \expandafter\let\csname\exstring@#1@Q\endcsname\empty
 \expandafter\def\csname\exstring@#1@F\endcsname{\rm}%
 }
\def\HASH@#1#2{\ifnum#2=\z@\else
 \edef\next@{\toks@{\the\toks@\the\hashtoks@#2}%
 \toks@@{\the\toks@@{\the\hashtoks@#2}}}\next@\expandafter\HASH@\fi}
\def\HASH@@{\toks@{}\toks@@{}\expandafter\HASH@\macpar@00}
\def\usecounter#1#2{\expandafter\ifx\csname\exstring@#1@Z\endcsname
 \relax\Err@{\noexpand#1not created with \string\newcounter}\fi
 \expandafter\let\csname\exstring@#1@@Z\endcsname\relax
 \expandafter\let\csname\exstring@#1@@Z@\endcsname\relax
 \expandafter\let\csname\exstring@#1@@Z@@\endcsname\relax
 \edef\next@{\def\noexpand#2{\futurelet\noexpand\next
  \csname\exstring@#1@@Z\endcsname}}\next@
 \edef\next@{\def\csname\exstring@#1@@Z\endcsname{\noexpand\ifx
  \noexpand\next\noexpand\label\def\noexpand\next@\noexpand\label
   ########1{\csname\exstring@#1@@Z@\endcsname
   {\noexpand#1\noexpand\label{########1}}}%
   \noexpand\else\noexpand\ifx\noexpand\next
   \noexpand"\def\noexpand\next@\noexpand"########1\noexpand"%
   {\csname\exstring@#1@@Z@\endcsname{{\expandafter\noexpand
   \csname\exstring@#1@F\endcsname
   \let\noexpand\pre\expandafter\noexpand\csname\exstring@#1@P\endcsname
   \let\noexpand\post\expandafter\noexpand\csname\exstring@#1@Q\endcsname
   \let\noexpand\style\expandafter\noexpand\csname\exstring@#1@S\endcsname
   \let\noexpand\numstyle\expandafter\noexpand\csname\exstring@#1@N\endcsname
   ########1}}}\noexpand\else
   \def\noexpand\next@{\csname\exstring@#1@@Z@\endcsname{\noexpand#1}}%
   \noexpand\fi\noexpand\fi\noexpand\next@}}\next@
 \def\next@{\expandafter\expandafter\expandafter\unmacro@\expandafter
  \meaning\csname\exstring@#1@@Z@@\endcsname\unmacro@
  \HASH@@
  \edef\next@{\def\csname\exstring@#1@@Z@\endcsname\the\toks@{%
   \expandafter\noexpand\csname\exstring@#1@@Z@@\endcsname\the\toks@@
   \noexpand\FNSSP@}}\next@}%
 \afterassignment\next@
 \expandafter\def\csname\exstring@#1@@Z@@\endcsname}
\def\listbi@{\penalty50 \medskip}
\def\listbii@{\penalty100 \smallskip}
\let\listbiii@\relax
\let\listbiv@\relax
\let\listbv@\relax
\def\listmi@{\advance\leftskip30\p@\relax}
\let\listmii@\listmi@
\let\listmiii@\listmi@
\let\listmiv@\listmi@
\let\listmv@\listmi@
\def\itemi@#1{\noindent@@\llap{#1\hskip5\p@}}
\let\itemii@\itemi@
\let\itemiii@\itemi@
\let\itemiv@\itemi@
\let\itemv@\itemi@
\def\liste@{\penalty-50 \medskip}
\def\listei@{\penalty-100 \smallskip}
\let\listeii@\relax
\let\listeiii@\relax
\let\listeiv@\relax
\expandafter\newcount\csname list@C1\endcsname
\csname list@C1\endcsname\z@
\expandafter\newcount\csname list@C2\endcsname
\csname list@C2\endcsname\z@
\expandafter\newcount\csname list@C3\endcsname
\csname list@C3\endcsname\z@
\expandafter\newcount\csname list@C4\endcsname
\csname list@C4\endcsname\z@
\expandafter\newcount\csname list@C5\endcsname
\csname list@C5\endcsname\z@
\expandafter\let\csname list@P1\endcsname\empty
\expandafter\let\csname list@P2\endcsname\empty
\expandafter\let\csname list@P3\endcsname\empty
\expandafter\let\csname list@P4\endcsname\empty
\expandafter\let\csname list@P5\endcsname\empty
\expandafter\let\csname list@Q1\endcsname\empty
\expandafter\let\csname list@Q2\endcsname\empty
\expandafter\let\csname list@Q3\endcsname\empty
\expandafter\let\csname list@Q4\endcsname\empty
\expandafter\let\csname list@Q5\endcsname\empty
\expandafter\def\csname list@S1\endcsname#1{{\rm(}{#1\/}{\rm)}}
\expandafter\def\csname list@S2\endcsname#1{{\rm(}{#1\/}{\rm)}}
\expandafter\def\csname list@S3\endcsname#1{{\rm(}{#1\/}{\rm)}}
\expandafter\def\csname list@S4\endcsname#1{{\rm(}{#1\/}{\rm)}}
\expandafter\def\csname list@S5\endcsname#1{{\rm(}{#1\/}{\rm)}}
\expandafter\let\csname list@N1\endcsname\arabic
\expandafter\let\csname list@N2\endcsname\arabic
\expandafter\let\csname list@N3\endcsname\arabic
\expandafter\let\csname list@N4\endcsname\arabic
\expandafter\let\csname list@N5\endcsname\arabic
\expandafter\def\csname list@F1\endcsname{\rm}
\expandafter\def\csname list@F2\endcsname{\rm}
\expandafter\def\csname list@F3\endcsname{\rm}
\expandafter\def\csname list@F4\endcsname{\rm}
\expandafter\def\csname list@F5\endcsname{\rm}
\newcount\listlevel@
\listlevel@\z@
\def\list@@C{\csname list@C\number\listlevel@\endcsname}
\def\list@@P{\csname list@P\number\listlevel@\endcsname}
\def\list@@Q{\csname list@Q\number\listlevel@\endcsname}
\def\list@@S{\csname list@S\number\listlevel@\endcsname}
\def\list@@N{\csname list@N\number\listlevel@\endcsname}
\def\list@@F{\csname list@F\number\listlevel@\endcsname}
\newif\iffirstitemi@
\newif\iffirstitemii@
\newif\iffirstitemiii@
\newif\iffirstitemiv@
\newif\iffirstitemv@
\def\Firstitem@true{\csname firstitem\romannumeral\listlevel@
 @true\endcsname}
\def\Firstitem@false{\csname firstitem\romannumeral\listlevel@
 @false\endcsname}
\def\Listm@{\csname listm\romannumeral\listlevel@ @\endcsname}
\def\Item@{\csname item\romannumeral\listlevel@ @\endcsname}
\def\Liste@{\csname liste\romannumeral\listlevel@ @\endcsname}
\newif\iflistcontinue@
\def\keepitem{\listcontinue@true}
\newcount\list@C@
\def\list{%
 \iflistcontinue@\csname list@C1\endcsname\csname list@C@\endcsname\fi
 \global\csname list@C2\endcsname\z@
 \global\csname list@C3\endcsname\z@
 \global\csname list@C4\endcsname\z@
 \global\csname list@C5\endcsname\z@
 \begingroup
 \firstitemi@true
 \listlevel@\@ne
 \def\item{\FN@\item@}%
 \FN@\list@}
\Invalid@\runinitem
\def\list@{\ifx\next\par
 \DN@\par{\FN@\list@}\else
 \ifx\next\runinitem
  \DN@\runinitem{\FN@\runinitem@}\else
  \DN@{\par\dimen@\parskip\parskip\dimen@}\fi\fi\next@}
\newif\ifoutlevel@
\newif\ifrunin@
\def\item@{%
 \ifoutlevel@\Liste@\outlevel@false\fi
 \ifrunin@\runin@false\par
  \dimen@\parskip\parskip\dimen@
  \Listm@\fi
 \iffirstitemi@\listbi@\listmi@\firstitemi@false\else\par\fi
 \iffirstitemii@\listbii@\listmii@\firstitemii@false\else\par\fi
 \iffirstitemiii@\listbiii@\listmiii@\firstitemiii@false\else\par\fi
 \iffirstitemiv@\listbiv@\listmiv@\firstitemiv@false\else\par\fi
 \iffirstitemv@\listbv@\listmv@\firstitemv@false\else\par\fi
 \DN@"##1"{{\let\pre\list@@P\let\post\list@@Q
  \let\style\list@@S\let\numstyle\list@@N
  \vskip-\parskip
  \Item@{\list@@F##1}%
  \noexpands@
  \Qlabel@{##1}}%
  \locallabel@
  \FNSSP@}%
 \DNii@{\global\advance\list@@C\@ne
  {\noexpands@
   \xdef\Thelabel@@@{\number\list@@C}%
   \xdefThelabel@\list@@N
   \xdef\Thelabel@@@@{\list@@P\Thelabel@\list@@Q}%
   \xdefThelabel@@\list@@S
  }%
  \locallabel@
  \vskip-\parskip
  \Item@{\list@@F\thelabel@@}%
  \FN@\pretendspace@}%
 \ifx\next"\expandafter\next@\else\expandafter\nextii@\fi}
\def\runinitem@{%
  \runin@true
  \Firstitem@false
  \DN@"##1"{{\let\pre\list@@P\let\post\list@@Q
   \let\style\list@@S\let\numstyle\list@@N
   \unskip\space{\list@@F##1} %
   \noexpands@
   \Qlabel@{##1}}%
   \locallabel@
   \ignorespaces}%
  \DNii@{\global\advance\list@@C\@ne
   {\noexpands@
    \xdef\Thelabel@@@{\number\list@@C}%
    \xdefThelabel@\list@@N
    \xdef\Thelabel@@@@{\list@@P\Thelabel@\list@@Q}%
    \xdefThelabel@@\list@@S
   }%
   \locallabel@
   \unskip\space{\list@@F\thelabel@@} }%
  \ifx\next"\expandafter\next@\else\expandafter\nextii@\fi}
\def\inlevel{\ifnum\listlevel@=5
 \DN@{\Err@{Already 5 levels down}}\else
 \DN@{\begingroup\advance\listlevel@\@ne
 \Firstitem@true\FN@\inlevel@}\fi\next@}
\def\inlevel@{\ifx\next\par
 \DN@\par{\FN@\inlevel@}\else
 \ifx\next\runinitem
  \DN@\runinitem{\FN@\runinitem@}\else
  \let\next@\relax\fi\fi\next@}
\def\outlevel{\ifnum\listlevel@=\@ne
 \Err@{At top level}\else
 \par\global\list@@C\z@\endgroup\outlevel@true\fi}
\def\endlist{%
 \expandafter\global\csname list@C@\endcsname\csname list@C1\endcsname
 \par
 \global\toks\@ne{}\count@\listlevel@
 {\loop
  \ifnum\count@>\z@\global\toks\@ne\expandafter{\the\toks\@ne\endgroup}%
  \advance\count@\m@ne
  \repeat}%
 \the\toks\@ne
 \liste@
 \listcontinue@false\global\csname list@C1\endcsname\z@
 \vskip-\parskip
 \noindent@@
 \FN@\pretendspace@}
\newif\iffirstdescribe@
\def\describe{\par
 \begingroup\firstdescribe@true
 \def\item##1{%
  \iffirstdescribe@\penalty50 \medskip\vskip-\parskip
  \firstdescribe@false\else\par\fi
  \noindent@@\hangindent2pc\hangafter\@ne
  {\bf##1}\hskip.5em}}

\Invalid@\pullin
\Invalid@\pullinmore
\newif\iffirstpull@
\def\margins{\par\begingroup\firstpull@true
 \def\pullin##1##2{\par
  \iffirstpull@\firstpull@false\else\endgroup\fi
  \begingroup\DN@{##1}%
  \ifx\next@\empty\leftskip\z@\else\ifx\next@\space\leftskip\z@
  \else\leftskip##1\fi\fi
  \DN@{##2}\ifx\next@\empty\rightskip\z@\else\ifx\next@\space
  \rightskip\z@\else\rightskip##2\fi\fi\ignorespaces}%
 \def\pullinmore##1##2{\par
  \xdef\Next@{\leftskip\the\leftskip\relax\rightskip\the\rightskip\relax}%
  \iffirstpull@\firstpull@false\else\endgroup\fi
  \begingroup\Next@
  \DN@{##1}%
  \ifx\next@\empty\else\ifx\next@\space\else\advance\leftskip##1\fi\fi
  \DN@{##2}\ifx\next@\empty\else\ifx\next@\space\else
  \advance\rightskip##2\fi\fi\ignorespaces}}

\newif\ifnopunct@
\newif\ifnospace@
\newif\ifoverlong@
\let\nofrillslist@\empty
\let\overlonglist@\empty
\def\nopunct{\nopunct@true\FN@\nopunct@}
\def\nospace{\nospace@true\FN@\nospace@}
\def\overlong{\overlong@true\FN@\overlong@}
\def\nopunct@{\ifx\next\nospace
 \DN@\nospace{\nospace@true\FN@\nopnos@}\else\ifx\next\overlong
 \DN@\overlong{\overlong@true\FN@\nopol@}\else
 \let\next@\nopunct@@\fi\fi\next@}
\def\nopunct@@#1{\ismember@\nofrillslist@#1%
 \iftest@\let\next@#1\else
 \DN@{\nopunct@false\Err@{\noexpand\nopunct can't be used with
 \string#1}#1}\fi\next@}
\def\nospace@{\ifx\next\nopunct
 \DN@\nopunct{\nopunct@true\FN@\nopnos@}\else\ifx\next\overlong
 \DN@\overlong{\overlong@true\FN@\nosol@}\else
 \let\next@\nospace@@\fi\fi\next@}
\def\nospace@@#1{\ismember@\nofrillslist@#1%
 \iftest@\let\next@#1\else
 \DN@{\nospace@false\Err@{\noexpand\nospace can't be used with
 \string#1}#1}\fi\next@}
\def\overlong@{\ifx\next\nopunct
 \DN@\nopunct{\nopunct@true\FN@\nopol@}\else\ifx\next\nospace
 \DN@\nospace{\nospace@true\FN@\nosol@}\else
 \let\next@\overlong@@\fi\fi\next@}
\def\overlong@@#1{\ismember@\overlonglist@#1%
 \iftest@\let\next@#1\else
 \DN@{\overlong@false\Err@{\noexpand\overlong can't be used with
 \string#1}#1}\fi\next@}
\def\nopnos@{\ifx\next\overlong
 \DN@\overlong{\overlong@true\nopnosol@}\else
 \let\next@\nopnos@@\fi\next@}
\def\nopol@{\ifx\next\nospace
 \DN@\nospace{\nospace@true\nopnosol@}\else
 \let\next@\nopol@@\fi\next@}
\def\nosol@{\ifx\next\nopunct
 \DN@\nopunct{\nopunct@true\nopnosol@}\else
 \let\next@\nosol@@\fi\next@}
\def\nopnos@@#1{\ismember@\nofrillslist@#1%
 \iftest@\let\next@#1\else
 \DN@{\nopunct@false\nospace@false
  \Err@{\noexpand\nopunct\noexpand\nospace
   can't be used with \string#1}#1}\fi\next@}
\def\testii@#1{\ismember@\nofrillslist@#1%
 \iftest@\let\nextiii@ T\else\let\nextiii@ F\fi
 \ismember@\overlonglist@#1%
 \iftest@\let\nextiv@ T\else\let\nextiv@ F\fi
 \test@false\if\nextiii@ T\if\nextiv@ T\test@true\fi\fi}
\def\nopol@@#1{\testii@{#1}%
 \iftest@\let\next@#1%
 \else\DN@{\if\nextiii@ T\else\nopunct@false\fi
  \if\nextiv@ T\else\overlong@false\fi
  \Err@{\if\nextiii@ T\else\noexpand\nopunct\fi
  \if\nextiv@ T\else\noexpand\overlong\fi can't be used
  with \string#1}#1}\fi\next@}
\def\nosol@@#1{\testii@{#1}%
 \iftest@\let\next@#1%
 \else\DN@{\if\nextiii@ T\else\nospace@false\fi
  \if\nextiv@ T\else\overlong@false\fi
  \Err@{\if\nextiii@ T\else\noexpand\nospace\fi
  \if\nextiv@ T\else\noexpand\overlong\fi can't be used
  with \string#1}#1}\fi\next@}
\def\nopnosol@#1{\testii@{#1}%
 \iftest@\let\next@#1%
 \else\DN@{\if\nextiii@ T\else\nopunct@false\nospace@false\fi
  \if\nextiv@ T\else\overlong@false\fi
  \Err@{\if\nextiii@ T\else\noexpand\nopunct\noexpand\nospace\fi
  \if\nextiv@ T\else\noexpand\overlong\fi can't be used
  with \string#1}#1}\fi\next@}
\def\punct@#1{\ifnopunct@\else#1\fi}
\def\addspace@#1{\ifnospace@\else#1\fi}
\def\hss@{\ifoverlong@\z@ plus\@m\p@ minus\@m\p@
 \else \z@ plus\@m\p@\fi}
\rightadd@\demo\to\nofrillslist@
\newif\ifclaim@
\def\exxx@{\expandafter\expandafter\expandafter\eat@\expandafter\string}
\let\colon@:
\def\demo#1{\ifclaim@
 \Err@{Previous \expandafter\noexpand\claimtype@ has
  no matching \string\end\exxx@\claimtype@}%
 \let\next@\relax
 \else
  \par
  \ifdim\lastskip<\smallskipamount\removelastskip\smallskip\fi
  \begingroup
  \noindent@@{\smc\ignorespaces#1\unskip
   \punct@{\null\colon@}\addspace@\enspace}%
  \nopunct@false\nospace@false
  \rm
  \DN@{\FNSSP@}%
 \fi
 \next@}
\def\enddemo{\par\endgroup\nopunct@false\nospace@false\smallskip}
\rightadd@\claim\to\nofrillslist@
\def\claim@F{\smc}
\def\claim@@@F{\csname\exxx@\claimtype@ @F\endcsname}
\def\claimformat@#1#2#3{%
 \medbreak\noindent@@{\smc#1 {\claim@@@F#2} #3%
 \punct@{\null.}\addspace@\enspace}\sl}
\def\claimformat@@#1#2{\claimformat@{\ignorespaces#1\unskip}%
 {\ifx\thelabel@@\empty\unskip\else\thelabel@@\fi}%
 {\ignorespaces#2\unskip}%
 \let\Claimformat@@\claimformat@@\FNSSP@}
\let\Claimformat@@\claimformat@@
\def\claim@@@P{\csname\exxx@\claimtype@ @P\endcsname}
\def\claim@@@Q{\csname\exxx@\claimtype@ @Q\endcsname}
\def\claim@@@S{\csname\exxx@\claimtype@ @S\endcsname}
\def\claim@@@N{\csname\exxx@\claimtype@ @N\endcsname}
\def\claim@@@C{\csname claim@C\claimclass@\endcsname}
\newcount\claim@C
\claim@C\z@
\let\claim@P\empty
\let\claim@Q\empty
\def\claim@S#1{#1\/}
\let\claim@N\arabic
\def\claim{\claim@true\let\claimclass@\empty
 \def\claimtype@{\claim}\FN@\claim@}
\def\claim@{%
 \ifx\next\c
  \let\next@\claim@c
 \else
  \ifx\next"%
   \let\next@\claim@q
  \else
   \begingroup\global\advance\claim@C\@ne
   {\noexpands@
    \xdef\Thelabel@@@{\number\claim@C}%
    \xdefThelabel@\claim@N
    \xdef\Thelabel@@@@{\claim@P\Thelabel@\claim@Q}%
    \xdefThelabel@@\claim@S
   }%
   \locallabel@
   \let\next@\Claimformat@@
  \fi
 \fi
 \next@}
\def\claim@c\c#1{\claim@true\begingroup
 \expandafter
 \ifx\csname claim@C#1\endcsname\relax
  \expandafter\newcount@\csname claim@C#1\endcsname
  \global\csname claim@C#1\endcsname\@ne
 \else
  \global\advance\csname claim@C#1\endcsname\@ne
 \fi
 \def\claimclass@{#1}%
 {\noexpands@
  \xdef\Thelabel@@@{\number\claim@@@C}%
  \xdefThelabel@\claim@@@N
  \xdef\Thelabel@@@@{\claim@@@P\Thelabel@\claim@@@Q}%
  \xdefThelabel@@\claim@@@S
 }%
 \locallabel@
 \FNSS@\claim@c@}
\def\claim@q"#1"{\begingroup
 {\let\pre\claim@@@P\let\post\claim@@@Q
  \let\style\claim@@@S\let\numstyle\claim@@@N
  \noexpands@
  \Qlabel@{#1}}%
 \locallabel@
 \FNSS@\claim@q@}
\def\claim@c@{\ifx\next"%
 \global\advance\claim@@@C\m@ne\let\next@\claim@cq
 \else\let\next@\Claimformat@@\fi\next@}
\def\claim@cq"#1"{{\let\pre\claim@@@P\let\post\claim@@@Q
 \let\style\claim@@@S\let\numstyle\claim@@@N
 \noexpands@
 \Qlabel@{#1}}%
 \locallabel@
 \FNSS@\Claimformat@@}
\def\claim@q@{\ifx\next\c\expandafter\claim@qc
 \else\expandafter\Claimformat@@\fi}
\def\claim@qc\c#1{\expandafter\ifx\csname claim@C#1\endcsname\relax
 \expandafter\newcount@\csname claim@C#1\endcsname
 \global\csname claim@C#1\endcsname\z@\fi
 \FNSS@\Claimformat@@}
\def\endclaim{\endgroup\claim@false\nopunct@false\nospace@false
 \let\Claimformat@@\claimformat@@\medbreak}
\Invalid@\claimclause
\def\newclaim{\FN@\newclaim@}
\def\newclaim@{\ifx\next\claimclause
 \DN@\claimclause##1{\newclaim@@{##1}}\else
 \DN@{\newclaim@@\relax}\fi\next@}
\def\claimlist@{\\\claim}
\newtoks\claim@i
\newtoks\claim@v
\let\noclaimclause@=F
\def\newclaim@@#1#2#3\c#4#5{\define#2{}%
 \rightadd@#2\to\claimlist@\rightadd@#2\to\nofrillslist@%
 \expandafter\def\csname\exstring@#2@P\endcsname{\claim@P}%
 \expandafter\def\csname\exstring@#2@Q\endcsname{\claim@Q}%
 \expandafter\def\csname\exstring@#2@S\endcsname{\claim@S}%
 \expandafter\def\csname\exstring@#2@N\endcsname{\claim@N}%
 \expandafter\def\csname\exstring@#2@F\endcsname{\claim@F}%
 \expandafter\def\csname end\exstring@#2\endcsname{\endclaim}%
 \expandafter\ifx\csname claim@C#4\endcsname\relax
  \expandafter\newcount@\csname claim@C#4\endcsname
  \global\csname claim@C#4\endcsname\z@\fi
 \edef\next@{\let\csname\exstring@#2@C\endcsname
   \csname claim@C#4\endcsname}\next@
 \def#2{\ifx\noclaimclause@ T\else#1\fi
  \global\claim@i{#1}\gdef\claim@iv{#4}\global\claim@v{#5}%
  \def\claimtype@{#2}\def\Claimformat@@{\claimformat@@{#5}}\claim@c\c{#4}}}
\def\shortenclaim#1#2{\define#2{}%
 \ismember@\claimlist@#1%
 \iftest@
  \rightadd@#2\to\nofrillslist@%
  \expandafter\def\csname\exstring@#2@P\endcsname
   {\csname\exstring@#1@P\endcsname}%
  \expandafter\def\csname\exstring@#2@Q\endcsname
   {\csname\exstring@#1@Q\endcsname}%
  \expandafter\def\csname\exstring@#2@S\endcsname
   {\csname\exstring@#1@S\endcsname}%
  \expandafter\def\csname\exstring@#2@N\endcsname
   {\csname\exstring@#1@N\endcsname}%
  \expandafter\def\csname\exstring@#2@F\endcsname
   {\csname\exstring@#1@F\endcsname}%
  \expandafter\def\csname end\exstring@#2\endcsname{\endclaim}%
  \edef\next@{\let\csname\exstring@#2@C\endcsname
    \csname claim\exstring@#1C\endcsname}\next@
  \setbox\z@\vbox{\let\noclaimclause@ T#1""\relax\endgroup}%
  \edef#2{\the\claim@i
   \def\noexpand\claimtype@{\noexpand#2}%
   \def\noexpand\Claimformat@@{\noexpand\claimformat@@{\the\claim@v}\relax}%
   \noexpand\claim@c\noexpand\c{\claim@iv}}%
 \else
  \Err@{\noexpand#1not yet created by \string\newclaim}%
 \fi}
\def\classtest@#1{\DN@{#1}\ifx\next@\claimclass@
 \test@true\else\test@false\fi}
\def\typetest@#1{\DN@{#1}\ifx\next@\claimtype@\test@true\else
  \test@false\fi}
\newif\iftoc@
\def\tocfile{\iftoc@\else\alloc@@7\write\chardef\sixt@@n\toc@
 \immediate\openout\toc@=\jobname.toc
 \alloc@@7\write\chardef\sixt@@n\tic@
 \immediate\openout\tic@=\jobname.tic
 \global\toc@true\fi}
\rightadd@\hl\to\nofrillslist@
\rightadd@\HL\to\overlonglist@
\def\HL@@C{\csname HL@C\HLlevel@\endcsname}
\def\HL@@P{\csname HL@P\HLlevel@\endcsname}
\def\HL@@Q{\csname HL@Q\HLlevel@\endcsname}
\def\HL@@S{\csname HL@S\HLlevel@\endcsname}
\def\HL@@N{\csname HL@N\HLlevel@\endcsname}
\def\HL@@F{\csname HL@F\HLlevel@\endcsname}
\def\HL@@@C{\csname\exxx@\HLtype@ @C\endcsname}
\def\HL@@@P{\csname\exxx@\HLtype@ @P\endcsname}
\def\HL@@@Q{\csname\exxx@\HLtype@ @Q\endcsname}
\def\HL@@@S{\csname\exxx@\HLtype@ @S\endcsname}
\def\HL@@@N{\csname\exxx@\HLtype@ @N\endcsname}
\def\HL#1{\expandafter
 \ifx\csname HL@C#1\endcsname\relax
  \DN@{\Err@{\string\HL#1 not defined in this style}}%
 \else
  \DN@{\gdef\HLlevel@{#1}\def\HLname@{\HL{#1}}\let\HLtype@\relax\FNSS@\HL@}%
 \fi
 \next@}%
\newif\ifquoted@
\let\aftertoc@\relax
\def\HL@{%
 \DN@"##1"##2\endHL{\def\entry@{##2}\quoted@true
  {\noexpands@
  \ifx\HLtype@\relax
   \let\pre\HL@@P\let\post\HL@@Q\let\style\HL@@S\let\numstyle\HL@@N
  \else
   \let\pre\HL@@@P\let\post\HL@@@Q\let\style\HL@@@S\let\numstyle\HL@@@N
  \fi
  \Qlabel@{##1}\let\style\relax\xdef\Qlabel@@@@{##1}%
  \xdef\Thepref@{\Thelabel@@@@}}%
  \csname HL@\HLlevel@\endcsname##2\endHL
  \let\pref\Thepref@
  \csname HL@I\HLlevel@\endcsname
  \csname HL@J\HLlevel@\endcsname
  \let\pref\pref@
  \HLtoc@
  \aftertoc@
  \let\aftertoc@\relax\overlong@false}%
 \DNii@##1\endHL{\def\entry@{##1}\quoted@false
  {\noexpands@
  \ifx\HLtype@\relax
   \global\advance\HL@@C\@ne
   \xdef\Thelabel@@@{\number\HL@@C}%
   \xdefThelabel@{\HL@@N}%
   \xdef\Thelabel@@@@{\HL@@P\Thelabel@\HL@@Q}%
   \xdefThelabel@@{\HL@@S}%
  \else
   \global\advance\HL@@@C\@ne
   \xdef\Thelabel@@@{\number\HL@@@C}%
   \xdefThelabel@{\HL@@@N}%
   \xdef\Thelabel@@@@{\HL@@@P\Thelabel@\HL@@@Q}%
   \xdefThelabel@@{\HL@@@S}%
  \fi
  \xdef\Thepref@{\Thelabel@@@@}}%
  \csname HL@\HLlevel@\endcsname##1\endHL
  \let\pref\Thepref@
  \csname HL@I\HLlevel@\endcsname
  \csname HL@J\HLlevel@\endcsname
  \let\pref\pref@
  \HLtoc@
  \aftertoc@
  \let\aftertoc@\relax\overlong@false}%
 \ifx\next"\expandafter\next@\else\expandafter\nextii@\fi}%
\Invalid@\endHL
\def\hl@@C{\csname hl@C\hllevel@\endcsname}
\def\hl@@P{\csname hl@P\hllevel@\endcsname}
\def\hl@@Q{\csname hl@Q\hllevel@\endcsname}
\def\hl@@S{\csname hl@S\hllevel@\endcsname}
\def\hl@@N{\csname hl@N\hllevel@\endcsname}
\def\hl@@F{\csname hl@F\hllevel@\endcsname}
\def\hl@@@C{\csname\exxx@\hltype@ @C\endcsname}
\def\hl@@@P{\csname\exxx@\hltype@ @P\endcsname}
\def\hl@@@Q{\csname\exxx@\hltype@ @Q\endcsname}
\def\hl@@@S{\csname\exxx@\hltype@ @S\endcsname}
\def\hl@@@N{\csname\exxx@\hltype@ @N\endcsname}
\def\hl#1{\expandafter
 \ifx\csname hl@C#1\endcsname\relax
  \DN@{\Err@{\string\hl#1 not defined in this style}}%
 \else
  \DN@{\gdef\hllevel@{#1}\def\hlname@{\hl{#1}}\let\hltype@\relax\FNSS@\hl@}%
 \fi
 \next@}
\def\hl@{%
 \DN@"##1"##2{\def\entry@{##2}\quoted@true
  {\noexpands@
  \ifx\hltype@\relax
   \let\pre\hl@@P\let\post\hl@@Q\let\style\hl@@S\let\numstyle\hl@@N
  \else
   \let\pre\hl@@@P\let\post\hl@@@Q\let\style\hl@@@S\let\numstyle\hl@@@N
  \fi
  \Qlabel@{##1}\let\style\relax\xdef\Qlabel@@@@{##1}%
  \xdef\Thepref@{\Thelabel@@@@}}%
  \csname hl@\hllevel@\endcsname{##2}%
  \let\pref\Thepref@
  \csname hl@I\hllevel@\endcsname
  \csname hl@J\hllevel@\endcsname
  \let\pref\pref@
  \hltoc@
  \aftertoc@
  \let\aftertoc@\relax\nopunct@false\nospace@false\FNSSP@}%
 \DNii@##1{\def\entry@{##1}\quoted@false
  {\noexpands@
  \ifx\hltype@\relax
   \global\advance\hl@@C\@ne
   \xdef\Thelabel@@@{\number\hl@@C}%
   \xdefThelabel@{\hl@@N}%
   \xdef\Thelabel@@@@{\hl@@P\Thelabel@\hl@@Q}%
   \xdefThelabel@@{\hl@@S}%
  \else
   \global\advance\hl@@@C\@ne
   \xdef\Thelabel@@@{\number\hl@@@C}%
   \xdefThelabel@{\hl@@@N}%
   \xdef\Thelabel@@@@{\hl@@@P\Thelabel@\hl@@@Q}%
   \xdefThelabel@@{\hl@@@S}%
  \fi
  \xdef\Thepref@{\Thelabel@@@@}}%
  \csname hl@\hllevel@\endcsname{##1}%
  \let\pref\Thepref@
  \csname hl@I\hllevel@\endcsname
  \csname hl@J\hllevel@\endcsname
  \let\pref\pref@
  \hltoc@
  \aftertoc@
  \let\aftertoc@\relax\nopunct@false\nospace@false\FNSSP@}%
 \ifx\next"\expandafter\next@\else\expandafter\nextii@\fi}%
\def\six@#1#2 #3 #4 #5 #6 #7 {\DN@{#2}\ifx\next@\empty
 \DN@##1\six@{}\else
 \write#1{ #2 #3 #4 #5 #6 #7}\DN@{\six@#1}\fi
 \next@}
\def\Sixtoc@{\ifx\macdef@\empty\else
 \DN@##1##2\next@{\def\macdef@{##1##2}}%
 \expandafter\next@\macdef@\next@
 \edef\next@
  {\noexpand\six@\toc@\macdef@
  \space\space\space\space\space\space\space\space\space\space\space\space
  \noexpand\six@}%
 \next@\let\macdef@\relax\fi}
\def\QorThelabel@@@@{\ifquoted@
 \noexpand\noexpand\noexpand"\Qlabel@@@@\noexpand\noexpand\noexpand"\else
 \Thelabel@@@@\fi}
\def\HLtoc@{%
 \iftoc@
 \expandafter\expandafter\expandafter\unmacro@
  \expandafter\meaning\csname HL@W\HLlevel@\endcsname\unmacro@
  {\noexpands@\let\style\relax
   \edef\next@{\write\toc@{\noexpand\noexpand\expandafter\noexpand\HLname@
   {\macdef@}{\QorThelabel@@@@}}}%
  \next@}%
  \expandafter\unmacro@\meaning\entry@\unmacro@
  \Sixtoc@
  \write\toc@{\noexpand\Page{\number\pageno}{\page@N}%
   {\page@P}{\page@Q}^^J}%
 \fi}
\def\hltoc@{%
 \iftoc@
 \expandafter\expandafter\expandafter\unmacro@
  \expandafter\meaning\csname hl@W\hllevel@\endcsname\unmacro@
  {\noexpands@\let\style\relax
  \edef\next@{\write\toc@{%
   \ifnopunct@\noexpand\noexpand\noexpand\nopunct\fi
   \ifnospace@\noexpand\noexpand\noexpand\nospace\fi
   \noexpand\noexpand\expandafter\noexpand\hlname@
   {\macdef@}{\QorThelabel@@@@}}}%
  \next@}%
  \expandafter\unmacro@\meaning\entry@\unmacro@
  \Sixtoc@
  \write\toc@{\noexpand\Page{\number\pageno}{\page@N}%
   {\page@P}{\page@Q}^^J}%
 \fi}
\def\mainfile#1{\def\mainfile@{#1}}
\def\checkmainfile@{\ifx\mainfile@\undefined
 \Err@{No \noexpand\mainfile specified}\fi}
\expandafter\newcount@\csname HL@C1\endcsname
\csname HL@C1\endcsname\z@
\expandafter\def\csname HL@S1\endcsname#1{#1\null.}
\expandafter\let\csname HL@N1\endcsname\arabic
\expandafter\let\csname HL@P1\endcsname\empty
\expandafter\let\csname HL@Q1\endcsname\empty
\expandafter\def\csname HL@F1\endcsname{\bf}
\expandafter\let\csname HL@W1\endcsname\empty
\expandafter\newcount@\csname hl@C1\endcsname
\csname hl@C1\endcsname\z@
\expandafter\def\csname hl@S1\endcsname#1{#1\/}
\expandafter\let\csname hl@N1\endcsname\arabic
\expandafter\let\csname hl@P1\endcsname\empty
\expandafter\let\csname hl@Q1\endcsname\empty
\expandafter\def\csname hl@F1\endcsname{\bf}
\expandafter\let\csname hl@W1\endcsname\empty
\expandafter\def\csname HL@1\endcsname#1\endHL{\bigbreak
 {\locallabel@
  \global\setbox\@ne\vbox{\Let@\tabskip\hss@
  \halign to\hsize{\bf\hfil\ignorespaces##\unskip\hfil\cr
  \expandafter\ifx\csname HL@W1\endcsname\empty\else
   \csname HL@W1\endcsname\space\fi
  {\HL@@F\ifx\thelabel@@\empty\else\thelabel@@\space\fi}%
  \ignorespaces#1\crcr}}%
  }%
 \unvbox\@ne\nobreak\medskip}
\expandafter\def\csname hl@1\endcsname#1{\medbreak\noindent@@
 {\locallabel@
 \bf{\hl@@F\ifx\thelabel@@\empty\else\thelabel@@\space\fi}%
 \ignorespaces#1\unskip\punct@{\null.}\addspace@\enspace}}
\expandafter\def\csname HL@I1\endcsname{\Reset\hl1{1}%
 \ifx\pref\empty\newpre\hl1{}\else\newpre\hl1{\pref.}\fi}
\def\NameHL#1#2{\define#2{}%
 \expandafter\ifx\csname HL@R#1\endcsname\relax
 \else
  \def\nextiv@{\let\nextiii@}%
  \expandafter\nextiv@\csname HL@R#1\endcsname
  \expandafter\let\nextiii@\undefined
  \expandafter\let\csname\exxx@\nextiii@ @C\endcsname\relax
  \expandafter\let\csname\exxx@\nextiii@ @P\endcsname\relax
  \expandafter\let\csname\exxx@\nextiii@ @Q\endcsname\relax
  \expandafter\let\csname\exxx@\nextiii@ @S\endcsname\relax
  \expandafter\let\csname\exxx@\nextiii@ @N\endcsname\relax
  \expandafter\let\csname\exxx@\nextiii@ @F\endcsname\relax
  \expandafter\let\csname\exxx@\nextiii@ @W\endcsname\relax
  \expandafter\let\csname end\exxx@\nextiii@\endcsname\undefined
 \fi
 \expandafter\gdef\csname HL@R#1\endcsname{#2}%
 \expandafter\gdef\csname\exstring@#2@R\endcsname{{HL}{#1}}%
 \iftoc@\write\toc@{\noexpand\NameHL#1\noexpand#2^^J}\fi
 \rightadd@#2\to\overlonglist@
 \edef\next@{\let\csname\exstring@#2@C\endcsname\expandafter\noexpand
  \csname HL@C#1\endcsname}\next@
 \edef\next@{\let\csname\exstring@#2@P\endcsname\expandafter\noexpand
  \csname HL@P#1\endcsname}\next@
 \edef\next@{\let\csname\exstring@#2@Q\endcsname\expandafter\noexpand
  \csname HL@Q#1\endcsname}\next@
 \edef\next@{\let\csname\exstring@#2@S\endcsname\expandafter\noexpand
  \csname HL@S#1\endcsname}\next@
 \edef\next@{\let\csname\exstring@#2@N\endcsname\expandafter\noexpand
  \csname HL@N#1\endcsname}\next@
 \edef\next@{\let\csname\exstring@#2@F\endcsname\expandafter\noexpand
  \csname HL@F#1\endcsname}\next@
 \edef\next@{\let\csname\exstring@#2@W\endcsname\expandafter\noexpand
  \csname HL@W#1\endcsname}\next@
 \edef\next@{\def\noexpand#2####1\expandafter\noexpand
  \csname end\exstring@#2\endcsname
  {\def\noexpand\HLtype@{\noexpand#2}%
   \def\noexpand\HLname@{\noexpand#2}%
   \gdef\noexpand\HLlevel@{#1}%
   \noexpand\FNSS@\noexpand\HL@####1\noexpand\endHL}}%
  \next@
 \edef\next@{\noexpand\Invalid@\expandafter\noexpand
  \csname end\exstring@#2\endcsname}%
 \next@}
\def\Namehl#1#2{\define#2{}%
 \expandafter\ifx\csname hl@R#1\endcsname\relax
 \else
  \def\nextiv@{\let\nextiii@}%
  \expandafter\nextiv@\csname hl@R#1\endcsname
  \expandafter\let\nextiii@\undefined
  \expandafter\let\csname\exxx@\nextiii@ @C\endcsname\relax
  \expandafter\let\csname\exxx@\nextiii@ @P\endcsname\relax
  \expandafter\let\csname\exxx@\nextiii@ @Q\endcsname\relax
  \expandafter\let\csname\exxx@\nextiii@ @S\endcsname\relax
  \expandafter\let\csname\exxx@\nextiii@ @N\endcsname\relax
  \expandafter\let\csname\exxx@\nextiii@ @F\endcsname\relax
  \expandafter\let\csname\exxx@\nextiii@ @W\endcsname\relax
 \fi
 \expandafter\gdef\csname hl@R#1\endcsname{#2}%
 \expandafter\gdef\csname\exstring@#2@R\endcsname{{hl}{#1}}%
 \iftoc@\write\toc@{\noexpand\Namehl#1\noexpand#2^^J}\fi
 \rightadd@#2\to\nofrillslist@%
 \edef\next@{\let\csname\exstring@#2@C\endcsname\expandafter\noexpand
  \csname hl@C#1\endcsname}\next@
 \edef\next@{\let\csname\exstring@#2@P\endcsname\expandafter\noexpand
  \csname hl@P#1\endcsname}\next@
 \edef\next@{\let\csname\exstring@#2@Q\endcsname\expandafter\noexpand
  \csname hl@Q#1\endcsname}\next@
 \edef\next@{\let\csname\exstring@#2@S\endcsname\expandafter\noexpand
  \csname hl@S#1\endcsname}\next@
 \edef\next@{\let\csname\exstring@#2@N\endcsname\expandafter\noexpand
  \csname hl@N#1\endcsname}\next@
 \edef\next@{\let\csname\exstring@#2@F\endcsname\expandafter\noexpand
  \csname hl@F#1\endcsname}\next@
 \edef\next@{\let\csname\exstring@#2@W\endcsname\expandafter\noexpand
  \csname hl@W#1\endcsname}\next@
 \edef\next@{\def\noexpand#2{%
  \def\noexpand\hltype@{\noexpand#2}%
  \def\noexpand\hlname@{\noexpand#2}%
  \gdef\noexpand\hllevel@{#1}%
  \noexpand\FNSS@\noexpand\hl@}}%
 \next@}%
\def\Initialize{\FN@\Init@}
\def\Init@{\ifx\next\HL\let\next@\InitH@\else\ifx\next\hl\let\next@\InitH@
  \else\let\next@\InitS@\fi\fi\next@}
\def\InitH@#1#2{\expandafter\ifx\csname\exstring@#1@C#2\endcsname\relax
 \DN@{\Err@{\noexpand#1level #2 not defined in this style}}\else
 \DN@{\expandafter\gdef\csname\exstring@#1@J#2\endcsname}\fi\next@}
\def\InitC@#1#2{\edef\nextii@{\expandafter\noexpand\csname#1\endcsname{#2}}}
\def\InitS@#1{\expandafter\ifx\csname\exstring@#1@R\endcsname\relax
 \Err@{\noexpand#1not defined in this style}\let\next@\relax\else
 \DN@{\let\next@}\expandafter\next@\csname\exstring@#1@R\endcsname
 \expandafter\InitC@\next@
 \DN@{\expandafter\InitH@\nextii@}\fi\next@}
\def\value#1{\expandafter
 \ifx\csname\exstring@#1@C\endcsname\relax
  \expandafter\ifx\csname\exstring@#1@C1\endcsname\relax
   \DN@{\Err@{\noexpand\value can't be used with \string#1}}%
  \else
   \DN@{\value@#1}%
  \fi
 \else
  \DN@{\number\csname\exstring@#1@C\endcsname\relax}%
 \fi
 \next@}
\def\value@#1#2{\expandafter
 \ifx\csname\exstring@#1@C#2\endcsname\relax
  \DN@{\Err@{\string\value\string#1 can't be followed by \string#2}}%
 \else
  \DN@{\number\csname\exstring@#1@C#2\endcsname\relax}%
 \fi
 \next@}
\newcount\Value
\def\Evaluate#1{\expandafter
 \ifx\csname\exstring@#1@C\endcsname\relax
  \expandafter\ifx\csname\exstring@#1@C1\endcsname\relax
   \DN@{\Err@{\noexpand\Evaluate can't be used with \string#1}}%
  \else
   \DN@{\Evaluate@#1}%
  \fi
 \else
  \DN@{\global\Value\csname\exstring@#1@C\endcsname}%
 \fi
 \next@}
\def\Evaluate@#1#2{\expandafter
 \ifx\csname\exstring@#1@C#2\endcsname\relax
  \DN@{\Err@{\string\Evaluate\string#1 can't be followed by \string#2}}%
 \else
  \DN@{\global\Value\csname\exstring@#1@C#2\endcsname}%
 \fi\next@}
\def\pre#1{\expandafter
 \ifx\csname\exstring@#1@P\endcsname\relax
  \expandafter\ifx\csname\exstring@#1@P1\endcsname\relax
   \DN@{\Err@{\noexpand\pre can't be used with \string#1}}%
  \else
   \DN@{\pre@#1}%
  \fi
 \else
  \DN@{{\csname\exstring@#1@P\endcsname}}%
 \fi
 \next@}
\def\pre@#1#2{\expandafter
 \ifx\csname\exstring@#1@P#2\endcsname\relax
  \DN@{\Err@{\string\pre\string#1 can't be followed by \string#2}}%
 \else
  \DN@{{\csname\exstring@#1@P#2\endcsname}}%
 \fi
 \next@}
\def\post#1{\expandafter
 \ifx\csname\exstring@#1@Q\endcsname\relax
  \expandafter\ifx\csname\exstring@#1@Q1\endcsname\relax
   \DN@{\Err@{\noexpand\post can't be used with \string#1}}%
  \else
   \DN@{\post@#1}%
  \fi
 \else
  \DN@{{\csname\exstring@#1@Q\endcsname}}%
 \fi
 \next@}
\def\post@#1#2{\expandafter
 \ifx\csname\exstring@#1@Q#2\endcsname\relax
  \DN@{\Err@{\string\post\string#1 can't be followed by \string#2}}%
 \else
  \DN@{{\csname\exstring@#1@Q#2\endcsname}}%
 \fi
 \next@}
\def\style#1{\expandafter
 \ifx\csname\exstring@#1@S\endcsname\relax
  \expandafter\ifx\csname\exstring@#1@S1\endcsname\relax
   \DN@{\Err@{\noexpand\style can't be used with \string#1}}%
  \else
   \DN@{\style@#1}%
  \fi
 \else
  \DN@{\csname\exstring@#1@S\endcsname}%
 \fi
 \next@}
\def\style@#1#2{\expandafter
 \ifx\csname\exstring@#1@S#2\endcsname\relax
  \DN@{\Err@{\string\style\string#1 can't be followed by \string#2}}%
 \else
  \DN@{\csname\exstring@#1@S#2\endcsname}%
 \fi
 \next@}
\def\fontstyle#1{\expandafter
 \ifx\csname\exstring@#1@F\endcsname\relax
  \expandafter\ifx\csname\exstring@#1@F1\endcsname\relax
   \DN@{\Err@{\noexpand\fontstyle can't be used with \string#1}}%
  \else
   \DN@{\fontstyle@#1}%
  \fi
 \else
  \DN@##1{{\csname\exstring@#1@F\endcsname##1}}%
 \fi
 \next@}
\def\fontstyle@#1#2{\expandafter
 \ifx\csname\exstring@#1@F#2\endcsname\relax
  \DN@{\Err@{\string\fontstyle\string#1 can't be followed by \string#2}}%
 \else
  \DN@##1{{\csname\exstring@#1@F#2\endcsname##1}}%
 \fi
 \next@}
\def\Reset#1{\expandafter
 \ifx\csname\exstring@#1@C\endcsname\relax
  \expandafter\ifx\csname\exstring@#1@C1\endcsname\relax
   \DN@{\Err@{\noexpand\Reset can't be used with \string#1}}%
  \else
   \DN@{\Reset@#1}%
  \fi
 \else
  \DN@##1{\count@##1\relax\ifx#1\page\else\advance\count@\m@ne\fi
   \global\csname\exstring@#1@C\endcsname\count@}%
 \fi
 \next@}
\def\Reset@#1#2{\expandafter
 \ifx\csname\exstring@#1@C#2\endcsname\relax
  \DN@{\Err@{\string\Reset\string#1 can't be followed by \string#2}}%
 \else
  \DN@##1{\count@##1\relax\advance\count@\m@ne
   \global\csname\exstring@#1@C#2\endcsname\count@}%
 \fi
 \next@}
\def\Offset#1{\expandafter
 \ifx\csname\exstring@#1@C\endcsname\relax
  \expandafter\ifx\csname\exstring@#1@C1\endcsname\relax
   \DN@{\Err@{\noexpand\Offset can't be used with \string#1}}%
  \else
   \DN@{\Offset@#1}%
  \fi
 \else
  \DN@##1{\count@##1\relax\advance\count@\m@ne\global\advance
   \csname\exstring@#1@C\endcsname\count@}%
 \fi
 \next@}
\def\Offset@#1#2{\expandafter
 \ifx\csname\exstring@#1@C#2\endcsname\relax
  \DN@{\Err@{\string\Offset\string#1 can't be followed by \string#2}}%
 \else
  \DN@##1{\count@##1\relax\advance\count@\m@ne
   \global\advance\csname\exstring@#1@C#2\endcsname\count@}%
 \fi
 \next@}
\def\getR@#1#2{\def\nextiv@{\let\nextiii@}\expandafter\nextiv@
 \csname\exstring@#1@R#2\endcsname}
\def\letR@#1#2#3{\expandafter\let\csname#1@#3#2\endcsname\Next@}
\def\letR@@#1#2{\expandafter\let\csname\exstring@#1@#2\endcsname\Next@}
\def\newpre#1{\expandafter
 \ifx\csname\exstring@#1@P\endcsname\relax
  \expandafter\ifx\csname\exstring@#1@P1\endcsname\relax
   \DN@{\Err@{\noexpand\newpre can't be used with \string#1}}%
  \else
   \DN@{\newpre@#1}%
  \fi
 \else
  \DN@{%
   \DNii@{%
    \endgroup
    \expandafter\let\csname\exstring@#1@P\endcsname\Next@
    \expandafter\ifx\csname\exstring@#1@R\endcsname\relax\else
    \getR@#1{}\expandafter\letR@\nextiii@ P\fi
    }%
   \begingroup\noexpands@\afterassignment\nextii@\xdef\Next@}%
 \fi
 \next@}
\def\newpre@#1#2{\expandafter
 \ifx\csname\exstring@#1@P#2\endcsname\relax
  \DN@{\Err@{\string\newpre\string#1 can't be followed by \string#2}}%
 \else
  \DN@{%
   \DNii@{%
    \endgroup
    \expandafter\let\csname\exstring@#1@P#2\endcsname\Next@
    \expandafter\ifx\csname\exstring@#1@R#2\endcsname\relax\else
    \getR@#1{#2}\expandafter\letR@@\nextiii@ P\fi
    }%
   \begingroup\noexpands@\afterassignment\nextii@\xdef\Next@}%
 \fi
 \next@}
\def\newpost#1{\expandafter
 \ifx\csname\exstring@#1@Q\endcsname\relax
  \expandafter\ifx\csname\exstring@#1@Q1\endcsname\relax
   \DN@{\Err@{\noexpand\newpost can't be used with \string#1}}%
  \else
   \DN@{\newpost@#1}%
  \fi
 \else
  \DN@{%
   \DNii@{%
    \endgroup
    \expandafter\let\csname\exstring@#1@Q\endcsname\Next@
    \expandafter\ifx\csname\exstring@#1@R\endcsname\relax\else
    \getR@#1{}\expandafter\letR@\nextiii@ Q\fi
    }%
   \begingroup\noexpands@\afterassignment\nextii@\xdef\Next@}%
 \fi
 \next@}
\def\newpost@#1#2{\expandafter
 \ifx\csname\exstring@#1@Q#2\endcsname\relax
  \DN@{\Err@{\string\newpost\string#1 can't be followed by \string#2}}%
 \else
  \DN@{%
   \DNii@{%
    \endgroup
    \expandafter\let\csname\exstring@#1@Q#2\endcsname\Next@
    \expandafter\ifx\csname\exstring@#1@R#2\endcsname\relax\else
    \getR@#1{#2}\expandafter\letR@@\nextiii@ Q\fi
    }%
   \begingroup\noexpands@\afterassignment\nextii@\xdef\Next@}%
 \fi
 \next@}
\def\newstyle#1{\expandafter
 \ifx\csname\exstring@#1@S\endcsname\relax
  \expandafter\ifx\csname\exstring@#1@S1\endcsname\relax
   \DN@{\Err@{\noexpand\newstyle can't be used
    with \string#1}}%
  \else
   \DN@{\newstyle@#1}%
  \fi
 \else
  \DN@{%
   \DNii@{%
    \expandafter\let\csname\exstring@#1@S\endcsname\Next@
    \expandafter\ifx\csname\exstring@#1@R\endcsname\relax\else
    \getR@#1{}\expandafter\letR@\nextiii@ S\fi
    }%
   \afterassignment\nextii@\gdef\Next@}%
 \fi
 \next@}
\def\newstyle@#1#2{\expandafter
 \ifx\csname\exstring@#1@S#2\endcsname\relax
  \DN@{\Err@{\string\newstyle\string#1 can't be followed by
   \string#2}}%
 \else
  \DN@{%
   \DNii@{%
    \expandafter\let\csname\exstring@#1@S#2\endcsname\Next@
    \expandafter\ifx\csname\exstring@#1@R#2\endcsname\relax\else
    \getR@#1{#2}\expandafter\letR@@\nextiii@ S\fi
    }%
   \afterassignment\nextii@\gdef\Next@}%
 \fi
 \next@}
\def\newnumstyle#1{\expandafter
 \ifx\csname\exstring@#1@N\endcsname\relax
  \expandafter\ifx\csname\exstring@#1@N1\endcsname\relax
   \DN@{\Err@{\noexpand\newnumstyle can't be used with
    \string#1}}%
  \else
   \DN@{\newnumstyle@#1}%
  \fi
 \else
  \DN@##1{%
   \gdef\Next@{##1}%
    \expandafter\let\csname\exstring@#1@N\endcsname\Next@
    \expandafter\ifx\csname\exstring@#1@R\endcsname\relax\else
    \getR@#1{}\expandafter\letR@\nextiii@ N\fi
    }%
 \fi
 \next@}
\def\newnumstyle@#1#2{\expandafter
 \ifx\csname\exstring@#1@N#2\endcsname\relax
  \DN@{\Err@{\string\newnumstyle\string#1 can't be followed by
   \string#2}}%
 \else
  \DN@##1{%
   \gdef\Next@{##1}%
    \expandafter\let\csname\exstring@#1@N#2\endcsname\Next@
    \expandafter\ifx\csname\exstring@#1@R#2\endcsname\relax\else
    \getR@#1{#2}\expandafter\letR@@\nextiii@ N\fi
    }%
  \fi
 \next@}
\def\newfontstyle#1{\expandafter
 \ifx\csname\exstring@#1@F\endcsname\relax
  \expandafter\ifx\csname\exstring@#1@F1\endcsname\relax
   \DN@{\Err@{\noexpand\newfontstyle can't be used with
    \string#1}}%
  \else
   \DN@{\newfontstyle@#1}%
  \fi
 \else
  \DN@##1{%
   \gdef\Next@{##1}%
    \expandafter\let\csname\exstring@#1@F\endcsname\Next@
    \expandafter\ifx\csname\exstring@#1@R\endcsname\relax\else
    \getR@#1{}\expandafter\letR@\nextiii@ F\fi
    }%
 \fi
 \next@}
\def\newfontstyle@#1#2{\expandafter
 \ifx\csname\exstring@#1@F#2\endcsname\relax
  \DN@{\Err@{\string\newfontstyle\string#1 can't be followed by
   \string#2}}%
 \else
  \DN@##1{%
   \gdef\Next@{##1}%
    \expandafter\let\csname\exstring@#1@F#2\endcsname\Next@
    \expandafter\ifx\csname\exstring@#1@R#2\endcsname\relax\else
    \getR@#1{#2}\expandafter\letR@@\nextiii@ F\fi
    }%
 \fi
 \next@}
\def\word#1{\expandafter
 \ifx\csname\exstring@#1@W\endcsname\relax
  \expandafter\ifx\csname\exstring@#1@W1\endcsname\relax
   \DN@{\Err@{\noexpand\word can't be used with \string#1}}%
  \else
   \DN@{\word@#1}%
  \fi
 \else
  \DN@{{\csname\exstring@#1@W\endcsname}}%
 \fi
 \next@}
\def\word@#1#2{\expandafter
 \ifx\csname\exstring@#1@W#2\endcsname\relax
  \DN@{\Err@{\string\word\noexpand#1can't be followed by \string#2}}%
 \else
  \DN@{{\csname\exstring@#1@W#2\endcsname}}%
 \fi
 \next@}
\def\newword#1{\expandafter
 \ifx\csname\exstring@#1@W\endcsname\relax
  \expandafter\ifx\csname\exstring@#1@W1\endcsname\relax
   \DN@{\Err@{\noexpand\newword can't be used  with \string#1}}%
  \else
   \DN@{\newword@#1}%
  \fi
 \else
  \DN@{%
   \DNii@{%
    \expandafter\let\csname\exstring@#1@W\endcsname\Next@
    \expandafter\ifx\csname\exstring@#1@R\endcsname\relax\else
     \getR@#1{}\expandafter\letR@\nextiii@ W\fi
    }%
   \afterassignment\nextii@\gdef\Next@}%
 \fi
 \next@}
\def\newword@#1#2{\expandafter
 \ifx\csname\exstring@#1@W#2\endcsname\relax
  \DN@{\Err@{\string\newword\noexpand#1can't be followed by \string#2}}%
 \else
  \DN@{%
   \DNii@{%
    \expandafter\let\csname\exstring@#1@W#2\endcsname\Next@
    \expandafter\ifx\csname\exstring@#1@R#2\endcsname\relax\else
     \getR@#1{#2}\expandafter\letR@@\nextiii@ W\fi
    }%
   \afterassignment\nextii@\gdef\Next@}%
 \fi
 \next@}
\newif\iffn@
\newcount\footmark@C
\footmark@C\z@
\def\footmark@S#1{$^{#1}$}
\let\footmark@N\arabic
\def\footmark@F{\rm}
\def\foottext@S#1{$^{#1}$}
\def\foottext@F{\rm}
\let\modifyfootnote@\relax
\def\modifyfootnote#1{\def\modifyfootnote@{#1}}
\def\vfootnote@#1{\insert\footins
 \bgroup
 \floatingpenalty\@MM\interlinepenalty\interfootnotelinepenalty
 \leftskip\z@\rightskip\z@\spaceskip\z@\xspaceskip\z@
 \rm\splittopskip\ht\strutbox\splitmaxdepth\dp\strutbox
 \locallabel@\noindent@@{\foottext@F#1}\modifyfootnote@
 \footstrut\FN@\fo@t}
\def\fo@t{\ifcat\bgroup\noexpand\next\expandafter\f@@t\else
 \expandafter\f@t\fi}
\def\f@t#1{#1\@foot}
\def\f@@t{\bgroup\aftergroup\@foot\afterassignment\FNSSP@\let\next@}
\def\@foot{\unskip\lower\dp\strutbox\vbox to\dp\strutbox{}\egroup
 \iffn@\expandafter\fn@false\else
 \expandafter\postvanish@\fi}
\newif\ifplainfn@
\plainfn@true
\def\fancyfootnotes{\plainfn@false}
\newcount\fancyfootmarkcount@
\fancyfootmarkcount@\z@
\newcount\lastfnpage@
\lastfnpage@-\@M
\let\justfootmarklist@\empty
\def\footmark{\let\@sf\empty
 \ifhmode\edef\@sf{\spacefactor\the\spacefactor}\/\fi
 \DN@{\ifx"\next\expandafter\nextii@\else\expandafter\footmark@\fi}%
 \DNii@"##1"{%
  \iffirstchoice@
   {\let\style\footmark@S\let\numstyle\footmark@N
   \footmark@F##1%
   \noexpands@
   \let\style\foottext@S
   \Qlabel@{##1}%
   }%
   \iffn@\else
    {\noexpands@
    \xdef\Next@{{\Thelabel@}{\Thelabel@@}{\Thelabel@@@}{\Thelabel@@@@}}%
    }%
    \expandafter\rightappend@\Next@\to\justfootmarklist@
   \fi
  \fi
  \@sf\relax}%
 \FN@\next@}
\def\footmark@{%
 \iffirstchoice@
  \global\advance\footmark@C\@ne
  \ifplainfn@
   \xdef\adjustedfootmark@{\number\footmark@C}%
  \else
   {\let\\\or\xdef\Next@{\ifcase\number\footmark@C\fnpages@\else
     -\@M\fi}}%
   \ifnum\Next@=-\@M
    \xdef\adjustedfootmark@{\number\footmark@C}%
   \else
    \ifnum\Next@=\lastfnpage@
     \global\advance\fancyfootmarkcount@\@ne
    \else
     \global\fancyfootmarkcount@\@ne
     \global\lastfnpage@\Next@
    \fi
    \xdef\adjustedfootmark@{\number\fancyfootmarkcount@}%
   \fi
  \fi
  {\noexpands@
  \xdef\Thelabel@@@{\adjustedfootmark@}%
  \xdefThelabel@\footmark@N
  \xdef\Thelabel@@@@{\Thelabel@}%
  \xdefThelabel@@\foottext@S
  }%
  \iffn@\else
   {\noexpands@
   \xdef\Next@{{\Thelabel@}{\Thelabel@@}{\Thelabel@@@}{\Thelabel@@@@}}%
   }%
   \expandafter\rightappend@\Next@\to\justfootmarklist@
  \fi
  \ifplainfn@
  \else
   \edef\next@{\write\laxwrite@{F\noexpand\the\pageno}}\next@
  \fi
 \fi
 \footmark@S{\footmark@N{\adjustedfootmark@}}%
 \@sf\relax}
\def\foottext{\prevanish@
 \ifx\justfootmarklist@\empty
  \Err@{There is no \noexpand\footmark for this \string\foottext}\fi
 \DN@\\##1##2\next@{\DN@{##1}\gdef\justfootmarklist@{##2}}%
 \expandafter\next@\justfootmarklist@\next@
 \expandafter\foottext@\next@}
\def\foottext@#1#2#3#4{{\noexpands@
  \xdef\Thelabel@{#1}\xdef\Thelabel@@{#2}%
  \xdef\Thelabel@@@{#3}\xdef\Thelabel@@@@{#4}}%
  \vfootnote@{\thelabel@@}}
\rightadd@\foottext\to\vanishlist@
\def\footnote{\fn@true
 \let\@sf\empty
 \ifhmode\edef\@sf{\spacefactor\the\spacefactor}\/\fi
 \DN@{\ifx"\next\expandafter\nextii@\else\expandafter\nextiii@\fi}%
 \DNii@"##1"{\footmark"##1"\vfootnote@{\let\style\foottext@S
  \let\numstyle\footmark@N##1}}%
 \def\nextiii@{\footmark\vfootnote@{\foottext@S{\footmark@N
  {\adjustedfootmark@}}}}%
 \FN@\next@}
\newdimen\litindent
\litindent20\p@
\newbox\litbox@
\newbox\Litbox@
\newcount\interlitpenalty@
\interlitpenalty@\@M
\newcount\litlines@
{\obeyspaces\gdef\defspace@{\def {\allowbreak\hskip.5emminus.15em}}}
{\obeylines\gdef\letM@{\let^^M\CtrlM@}}
\def\CtrlM@{\egroup
 \ifcase\litlines@\advance\litlines@\@ne\or
 \box\litbox@\advance\litlines@\@ne\else
 \penalty\interlitpenalty@\box\litbox@\fi
 \Lit@}
\def\Lit@{\setbox\litbox@\hbox\bgroup\litdefs@\hskip\litindent}
\newcount\littab@
\littab@8
\def\littab#1{\littab@#1\relax}
{\catcode`\^^I=\active\gdef\letTAB@{\let^^I\TAB@}}
\def\TAB@{\egroup
 \dimen@\wd\litbox@
 \advance\dimen@-\litindent
 \setboxz@h{\tt0}%
 \dimen@ii\littab@\wdz@
 \divide\dimen@\dimen@ii
 \multiply\dimen@\dimen@ii
 \advance\dimen@\littab@\wdz@
 \advance\dimen@\litindent
 \setbox\litbox@\hbox\bgroup\litdefs@\hbox to\dimen@{\unhbox\litbox@\hfil}}
{\catcode`\`=\active\gdef`{\relax\lq}}
\let\litbs@\relax
\let\litbs@@\relax
\def\litbackslash#1{%
 \edef\litbs@{\catcode`\string#1=\z@
 \def\noexpand\litbs@@{\def\expandafter\noexpand\csname\string#1\endcsname
  {\char`\string#1}}}}
\def\litcodes@{\catcode`\\=12
 \catcode`\{=12 \catcode`\}=12
 \catcode`\$=12 \catcode`\&=12
 \catcode`\#=12
 \catcode`\^=12 \catcode`\_=12
 \catcode`\@=12 \catcode`\~=12 \catcode`\"=12
 \catcode`\;=12 \catcode`\:=12 \catcode`\!=12 \catcode`\?=12
 \catcode`\%=12 \litbs@\catcode`\`=\active\obeyspaces\defspace@}
\def\activate@#1#2{{\lccode`\~=`#2%
 \lowercase{%
  \if0#1%
  \gdef\Next@{\def~{\egroup\endgroup\bigskip\vskip-\parskip
   \def\next@{\noindent@@\FN@\pretendspace@}\FNSS@\next@}}\else
  \gdef\Next@{\def~{\egroup\egroup\endgroup}}\fi
  }%
 }}
\def\litdefs@{\let\0\empty\let\1\litdelim@\def\ {\char32 }\litbs@@}%
\def\litdelimiter#1{%
 \edef\litdelim@{\char`#1}%
 \def\lit#1{\leavevmode\begingroup\litcodes@\litdefs@
  \tt\hyphenchar\tentt\m@ne\lit@}%
 \def\lit@##1#1{##1\endgroup\null}%
 \def\Lit#1{\ifhmode$$\abovedisplayskip\bigskipamount
  \abovedisplayshortskip\bigskipamount
  \belowdisplayskip\z@\belowdisplayshortskip\z@
  \postdisplaypenalty\@M
  $$\vskip-\baselineskip\else\bigskip\fi
  \begingroup\litlines@\z@
  \catcode`#1=\active\activate@0#1\Next@
  \def\displaybreak{\egroup\break\litlines@\z@\Lit@}%
  \def\allowdisplaybreak{\egroup\allowbreak\litlines@\z@\Lit@}%
  \def\allowdisplaybreaks{\egroup\allowbreak\interlitpenalty@\z@
   \litlines@\z@\Lit@}%
  \litcodes@\tt\catcode`\^^I=\active\letTAB@
  \obeylines\letM@\Lit@}%
 \def\Litbox##1=#1{\begingroup\ifodd##1\relax\aftergroup\global\fi
  \aftergroup\setbox\aftergroup##1\aftergroup\box\aftergroup\Litbox@
  \def\allowdisplaybreak{\egroup\allowbreak\litlines@\z@\Lit@}%
  \def\allowdisplaybreaks{\egroup\allowbreak\interlitpenalty@\z@
   \litlines@\z@\Lit@}%
  \catcode`#1=\active\activate@1#1\Next@
  \litcodes@\tt\catcode`\^^I=\active\letTAB@
  \obeylines\letM@\global\setbox\Litbox@\vbox\bgroup\litindent\z@%
  \litlines@\z@\Lit@}%
}
\newbox\titlebox@
\setbox\titlebox@\vbox{}
\rightadd@\title\to\overlonglist@
\def\title{\begingroup\Let@
 \global\setbox\titlebox@\vbox\bgroup\tabskip\hss@
 \halign to\hsize\bgroup\bf\hfil\ignorespaces##\unskip\hfil\cr}
\def\endtitle{\crcr\egroup\egroup\endgroup\overlong@false}
\newbox\authorbox@
\rightadd@\author\to\overlonglist@
\def\author{\begingroup\Let@
 \global\setbox\authorbox@\vbox\bgroup\tabskip\hss@
 \halign to\hsize\bgroup\rm\hfil\ignorespaces##\unskip\hfil\cr}
\def\endauthor{\crcr\egroup\egroup\endgroup\overlong@false}
\newbox\affilbox@
\def\affil{\begingroup\Let@
 \global\setbox\affilbox@\vbox\bgroup\tabskip\hss@
 \halign to\hsize\bgroup\rm\hfil\ignorespaces##\unskip\hfil\cr}%
\def\endaffil{\crcr\egroup\egroup\endgroup\overlong@false}
\let\date@\relax
\def\date#1{\gdef\date@{\ignorespaces#1\unskip}}
\def\today{\ifcase\month\or January\or February\or March\or April\or May\or
 June\or July\or August\or September\or October\or November\or December\fi
 \space\number\day, \number\year}
\def\maketitle{\hrule\height\z@\vskip-\topskip
 \vskip24\p@ plus12\p@ minus12\p@
 \unvbox\titlebox@
 \ifvoid\authorbox@\else\vskip12\p@ plus6\p@ minus3\p@\unvbox\authorbox@\fi
 \ifvoid\affilbox@\else\vskip10\p@ plus5\p@ minus2\p@\unvbox\affilbox@\fi
 \ifx\date@\relax\else\vskip6\p@ plus2\p@ minus\p@\centerline{\rm\date@}\fi
 \vskip18\p@ plus12\p@ minus6\p@}
\def\cite{%
 \DNii@(##1)##2{{\rm[}{##2}, {##1\/}{\rm]}}%
 \def\nextiii@##1{{\rm[}{##1\/}{\rm]}}%
 \DN@{\ifx\next(\expandafter\nextii@\else\expandafter\nextiii@\fi}%
 \FN@\next@}
\def\makebib@W{Bibliography}
\def\makebibp@W{References}
\def\makebib{\begingroup\rm\bigbreak\centerline{\smc\makebib@W}%
 \nobreak\medskip
 \sfcode`\.=\@m\everypar{}\parindent\z@
 \def\nopunct{\nopunct@true}\def\nospace{\nospace@true}%
 \nopunct@false\nospace@false
 \def\lkerns@{\null\kern\m@ne sp\kern\@ne sp}%
 \def\nkerns@{\null\kern-\tw@ sp\kern\tw@ sp}%
}

\newif\ifnoprepunct@
\newif\ifnoprespace@
\newif\ifnoquotes@
\def\noprepunct{\noprepunct@true}
\def\noprespace{\noprespace@true}
\def\noquotes{\noquotes@true}
\newbox\nobox@
\newbox\keybox@
\newbox\bybox@
\newbox\paperbox@
\newbox\paperinfobox@
\newbox\jourbox@
\newbox\volbox@
\newbox\issuebox@
\newbox\yrbox@
\newbox\pgbox@
\newbox\ppbox@
\newbox\bookbox@
\newbox\inbookbox@
\newbox\bookinfobox@
\newbox\publbox@
\newbox\publaddrbox@
\newbox\edbox@
\newbox\edsbox@
\newbox\langbox@
\newbox\translbox@
\newbox\finalinfobox@
\def\setbibinfo@#1{\edef\next@{\ifnopunct@1\else0\fi
 \ifnospace@1\else0\fi\ifnoprepunct@1\else0\fi\ifnoprespace@1\else0\fi
 \ifnoquotes@1\else0\fi}%
 \DNii@{00000}%
 \ifx\next@\nextii@\else\xdef\bibinfo@{\bibinfo@\the#1,\next@}%
 \fi}
\def\getbibinfo@#1{\ifx\bibinfo@\empty
 \let\next@0\let\nextii@0\let\nextiii@0\let\nextiv@0\let\nextv@0\else
 \edef\next@{\def
  \noexpand\next@####1\the#1,####2####3####4####5####6####7\noexpand\next@
  {\let\noexpand\next@####2\let\noexpand\nextii@####3%
  \let\noexpand\nextiii@####4\let\noexpand\nextiv@####5%
  \let\noexpand\nextv@####6}%
  \noexpand\next@\bibinfo@\the#1,00000\noexpand\next@}\next@
 \fi}
\newif\ifbookinquotes@
\def\bookinquotes{\bookinquotes@true}
\newif\ifpaperinquotes@
\def\paperinquotes{\paperinquotes@true}
\newif\ifininbook@
\def\ininbook{\ininbook@true}
\newif\ifopenquotes@
\def\closequotes@{\ifopenquotes@''\openquotes@false\fi}
\newif\ifbeginbib@
\newif\ifendbib@
\newif\ifprevjour@
\newif\ifprevbook@
\newdimen\bibindent@
\bibindent@20\p@
\def\bib{\global\let\bibinfo@\empty\global\let\translinfo@\relax\beginbib@true
 \begingroup\noindent@
 \hangindent\bibindent@\hangafter\@ne\bib@}
\def\v@id#1{\setbox#1\box\voidb@x}
\def\bib@{\v@id\nobox@\v@id\keybox@\v@id\bybox@\v@id\paperbox@
 \v@id\paperinfobox@\v@id\jourbox@\v@id\volbox@\v@id\issuebox@
 \v@id\yrbox@\v@id\pgbox@\v@id\ppbox@\v@id\bookbox@\v@id\inbookbox@
 \v@id\bookinfobox@\v@id\publbox@\v@id\publaddrbox@\v@id\edbox@
 \v@id\edsbox@\v@id\langbox@\v@id\translbox@\v@id\finalinfobox@
 \bgroup}
\def\Setnonemptybox@#1#2{\unskip\setbibinfo@#1\egroup#2%
 \def\aftergroup@{\ifdim\wd#1=\z@\setbox#1\box\voidb@x\fi}%
 \setbox#1\vbox\bgroup\aftergroup\aftergroup@\hsize\maxdimen\leftskip\z@
 \rightskip\z@\hbadness\@M\hfuzz\maxdimen\noindent}
\def\setnonemptybox@#1{\Setnonemptybox@#1\relax}
\def\no{\setnonemptybox@\nobox@}
\def\key{\setnonemptybox@\keybox@\bf}
\def\by{\setnonemptybox@\bybox@}
\def\bysame{\setnonemptybox@\bybox@\leaders\hrule\hskip3em\null}
\def\paper{\setnonemptybox@\paperbox@
 \ifpaperinquotes@\getbibinfo@\paperbox@
 \if\nextv@1\else``\fi\else\it\fi}
\def\paperinfo{\setnonemptybox@\paperinfobox@}
\def\jour{\Setnonemptybox@\jourbox@\prevjour@true}
\def\vol{\setnonemptybox@\volbox@\bf}
\def\issue{\setnonemptybox@\issuebox@}
\def\yr{\setnonemptybox@\yrbox@}

\def\pg{\setnonemptybox@\pgbox@}
\def\pp{\setnonemptybox@\ppbox@}
\def\book{\Setnonemptybox@\bookbox@\prevbook@true
 \ifbookinquotes@\getbibinfo@\bookbox@
 \if\nextv@1\else``\fi\else\it\fi}
\def\inbook{\Setnonemptybox@\inbookbox@\prevbook@true
 \ifininbook@ in \fi\ifbookinquotes@\getbibinfo@\inbookbox@
 \if\nextv@1\else``\fi\fi}
\def\bookinfo{\setnonemptybox@\bookinfobox@}
\def\publ{\setnonemptybox@\publbox@}
\def\publaddr{\setnonemptybox@\publaddrbox@}
\def\ed{\setnonemptybox@\edbox@}
\def\eds{\setnonemptybox@\edsbox@}
\def\lang{\setnonemptybox@\langbox@}
\def\finalinfo{\setnonemptybox@\finalinfobox@}
\def\setboxzl@{\setbox\z@\lastbox}
\def\getbox@#1{\setbox\z@\vbox{\vskip-\@M\p@
 \unvbox#1%
 \setboxzl@
 \global\setbox\@ne\hbox{\unhbox\z@\unskip\unskip\unpenalty}%
 \ifdim\lastskip=-\@M\p@\else
 \loop\ifdim\lastskip=-\@M\p@
 \else\unskip\unpenalty\setboxzl@
 \global\setbox\@ne\hbox{\unhbox\z@\unhbox\@ne}%
 \repeat\fi}%
 \unhbox\@ne}
\def\adjustpunct@#1{\count@\lastkern
 \ifnum\count@=\z@#1\closequotes@\else
 \ifnum\count@>\tw@#1\closequotes@\else
 \ifnum\count@<-\tw@#1\closequotes@\else
  \unkern\unkern\setboxzl@
  \skip@\lastskip\unskip
  \count@@\lastpenalty\unpenalty
  \ifnum\count@=\tw@\unskip\setboxzl@\fi
  \ifdim\skip@=\z@\else\hskip\skip@\fi
  #1\closequotes@
  \ifnum\count@=\tw@\null\hfill\fi
  \penalty\count@@
 \fi\fi\fi}
\def\prepunct@#1#2{\getbibinfo@#2%
 \ifnopunct@
 \else
  \if\nextiii@0\adjustpunct@#1\fi
 \fi
 \closequotes@
 \ifnospace@
 \else
  \if\nextiv@0\space\else\fi
 \fi
 \nopunct@false\nospace@false
 \if\next@1\nopunct@true\fi
 \if\nextii@1\nospace@true\fi}
\def\ppunbox@#1#2{\prepunct@{#1}#2%
 \getbox@#2}
\let\semicolon@;
\def\endbib@{%
 \ifbeginbib@
  \ifvoid\nobox@
   \ifvoid\keybox@\else\hbox to\bibindent@{[\getbox@\keybox@]\hss}\fi
  \else\hbox to\bibindent@{\hss\getbox@\nobox@. }\fi
  \ifvoid\bybox@\else\getbox@\bybox@\fi
 \else
  \nopunct@true
  \ifvoid\bybox@\else\ppunbox@\relax\bybox@\fi
 \fi
 \ifvoid\translbox@\else\ppunbox@,\translbox@\fi
 \ifvoid\paperbox@\else\ppunbox@,\paperbox@\ifpaperinquotes@
  \if\nextv@1\else\openquotes@true\fi\fi
 \fi
 \ifvoid\paperinfobox@\else\ppunbox@,\paperinfobox@\fi
 \test@false
 \ifvoid\jourbox@\else\test@true\ppunbox@,\jourbox@\fi
 \ifprevjour@\test@true\fi
 \iftest@
  \ifvoid\volbox@\else\ppunbox@\relax\volbox@\fi
  \ifvoid\issuebox@
   \else\prepunct@\relax\issuebox@ no.~\getbox@\issuebox@\fi
  \ifvoid\yrbox@\else\prepunct@\relax\yrbox@(\getbox@\yrbox@)\fi
  \ifvoid\ppbox@\else\ppunbox@,\ppbox@\fi
  \ifvoid\pgbox@\else\prepunct@,\pgbox@ p.~\getbox@\pgbox@\fi
 \fi
 \test@false
 \ifvoid\bookbox@\else\test@true\ppunbox@,\bookbox@\ifbookinquotes@
  \if\nextv@1\else\openquotes@true\fi\fi\fi
 \ifvoid\inbookbox@\else\test@true\ppunbox@,\inbookbox@\ifbookinquotes@
  \if\nextv@1\else\openquotes@true\fi\fi\fi
 \ifprevbook@\test@true\fi
 \iftest@
  \ifvoid\edbox@\else\prepunct@\relax\edbox@(\getbox@\edbox@, ed.)\fi
  \ifvoid\edsbox@\else\prepunct@\relax\edsbox@(\getbox@\edsbox@, eds.)\fi
  \ifvoid\bookinfobox@\else\ppunbox@,\bookinfobox@\fi
  \ifvoid\publbox@\else\ppunbox@,\publbox@\fi
  \ifvoid\publaddrbox@\else\ppunbox@,\publaddrbox@\fi
  \ifvoid\yrbox@\else\ppunbox@,\yrbox@\fi
  \ifvoid\ppbox@\else\prepunct@,\ppbox@ pp.~\getbox@\ppbox@\fi
  \ifvoid\pgbox@\else\prepunct@,\pgbox@ p.~\getbox@\pgbox@\fi
 \fi
 \ifvoid\finalinfobox@
  \ifendbib@
   \ifnopunct@\else.\closequotes@\fi
  \else
  \ifvoid\langbox@\else\space(\getbox@\langbox@)\fi
   \/\semicolon@\closequotes@
  \fi
 \else
  \ifendbib@
   \ppunbox@{.\spacefactor3000\relax}\finalinfobox@
    \ifnopunct@\else.\fi
  \else
   \ppunbox@,\finalinfobox@\/\semicolon@\fi
 \fi
 \ifvoid\langbox@\else\space(\getbox@\langbox@)\fi
}
\def\endbib{\unskip\egroup\endbib@true\endbib@\par\endgroup}
\def\morebib{\unskip\egroup
 \endbib@false\endbib@
 \global\let\bibinfo@\empty\beginbib@false
 \bib@}
\def\anotherbib{\unskip\egroup
 \endbib@false\endbib@
 \global\let\bibinfo@\empty\beginbib@false
 \prevjour@false\prevbook@false\bib@}
\def\transl{\unskip
 \xdef\translinfo@{\the\translbox@,\ifnopunct@1\else0\fi
 \ifnospace@1\else0\fi\ifnoprepunct@1\else0\fi\ifnoprespace@1\else0\fi0}%
 \egroup\endbib@false\endbib@
 \global\let\bibinfo@\translinfo@\beginbib@false
 \bib@
 \egroup
 \def\aftergroup@{\ifdim\wd\translbox@=\z@\setbox\translbox@\box\voidb@x\fi}%
 \setbox\translbox@\vbox\bgroup\aftergroup\aftergroup@
 \hsize\maxdimen\leftskip\z@\rightskip\z@\hbadness\@M\hfuzz\maxdimen
 \noindent}
\newwrite\auxwrite@
\newread\bbl@
\def\UseBibTeX{\immediate\openout\auxwrite@=\jobname.aux
 \let\cite\BTcite@
 \def\nocite##1{\immediate\write\auxwrite@{\string\citation{##1}}}%
 \def\bibliographystyle##1{\immediate\write\auxwrite@{\string
  \bibstyle{##1}}}%
 \def\bibliography@W{Bibliography}%
 \def\bibliography##1{\immediate\write\auxwrite@{\string\bibdata{##1}}%
  \immediate\openin\bbl@=\jobname.bbl
  \ifeof\bbl@
   \W@{No .bbl file}%
  \else
   \immediate\closein\bbl@
   \begingroup\input bibtex \input\jobname.bbl \endgroup
  \fi}%
 }
\def\BTcite@{%
 \DNii@(##1)##2{{\rm[}\BTcite@@##2,\BTcite@@{\rm, }{##1\/}{\rm]}%
  \immediate\write\auxwrite@{\string\citation{##2}}}%
 \def\nextiii@##1{{\rm[}\BTcite@@##1,\BTcite@@\/{\rm]}%
  \immediate\write\auxwrite@{\string\citation{##1}}}%
 \DN@{\ifx\next(\expandafter\nextii@\else\expandafter\nextiii@\fi}%
 \FN@\next@}%
\def\BTcite@@#1,{\BTcite@@@{#1}\FN@\BTcite@@@@}
\def\BTcite@@@@{\ifx\next\BTcite@@
 \expandafter\eat@\else{\rm, }\expandafter\BTcite@@\fi}
\catcode`\~=11
\def\BTcite@@@#1{\nolabel@\cite{#1}\relax
 \DNii@##1~##2\nextii@{##1}%
 \csL@{#1}\expandafter\nextii@\Next@\nextii@\fi}
\catcode`\~=\active

\def\beginthebibliography@#1{\rm\setboxz@h{#1\ }\bibindent@\wdz@
 \bigbreak\centerline{\smc\bibliography@W}\nobreak\medskip
 \sfcode`\.=\@m\everypar{}\parindent\z@}

\newif\iffigproofing@
\def\Figureproofing{\figproofing@true}
\def\noFigureproofing{\figproofing@false}
\newif\ifHby@
\def\Hbyw#1{\global\Hby@true\hbyw\vsize{#1}}
\def\hbyw#1#2{%
 \hbox{%
  \ifHby@
  \else
   \iffigproofing@
    \setbox\z@\vbox{\hrule\width5\p@}\ht\z@\z@
    \vbox to#1{\hrule\height5\p@\width.4\p@\vfil\hrule\height5\p@\width.4\p@}%
    \kern-.4\p@\rlap{\copy\z@}\raise#1\hbox{\rlap{\copy\z@}}%
   \fi
  \fi
  \vbox to#1{\hbox to#2{}\vfil}%
  \ifHby@
  \else
   \iffigproofing@
    \vbox to#1{\hrule\height5\p@\width.4\p@\vfil\hrule\height5\p@\width.4\p@}%
    \kern-.4\p@\llap{\copy\z@}\raise#1\hbox{\llap{\boxz@}}%
   \fi
  \fi}}
\newcount\island@C
\let\island@P\empty
\let\island@Q\empty
\def\island@S#1{#1\null.}
\let\island@N\arabic
\def\island@F{\rm}
\def\island@@@P{\csname\exxx@\islandtype@ @P\endcsname}
\def\island@@@Q{\csname\exxx@\islandtype@ @Q\endcsname}
\def\island@@@S{\csname\exxx@\islandtype@ @S\endcsname}
\def\island@@@N{\csname\exxx@\islandtype@ @N\endcsname}
\def\island@@@F{\csname\exxx@\islandtype@ @F\endcsname}
\def\island@@@C{\csname island@C\islandclass@\endcsname}
\newif\ifplace@
\newif\ifisland@
\def\island{%
 \ifplace@
  \DN@{\let\islandclass@\empty\def\islandtype@{\island}\FN@\island@}%
 \else
  \long\DN@##1\endisland{\Err@{\noexpand\island must be used after some
   type of \string\...place}}%
 \fi
 \next@}
\def\island@{\ifx\next\c\let\next@\island@c\else
 \DN@{\FN@\island@@}\fi\next@}
\def\island@@{\ifcat\bgroup\noexpand\next\let\next@\island@@@\else
 \DN@{\Err@{\noexpand\island must be followed by a {prefix} for
 \string\caption's}}\fi\next@}
\newbox\islandbox@
\newcount\captioncount@
\def\island@@@#1{\def\captionprefix@{#1}\captioncount@\z@
 \global\setbox\islandbox@\vbox\bgroup}
\def\island@c\c#1{%
 \ifplace@
 \DN@{\def\islandclass@{#1}%
  \expandafter\ifx\csname island@C#1\endcsname\relax
  \expandafter\newcount@\csname island@C#1\endcsname
   \global\csname island@C#1\endcsname\z@\fi
  \FNSS@\island@c@}%
 \else
 \DN@{\edef\next@{\long\def\noexpand\next@########1\expandafter\noexpand
  \csname end\exxx@\islandtype@\endcsname{\noexpand\Err@{\noexpand\noexpand
  \expandafter\noexpand
  \islandtype@ must be used after some type of \noexpand\string
   \noexpand\...place}}}\next@\next@}%
 \fi
 \next@}
\def\island@c@{%
 \ifcat\bgroup\noexpand\next
  \let\next@\island@c@@
 \else
  \DN@{\Err@{\noexpand\island\string\c{\expandafter\string\islandclass@} must
   be followed by a {prefix} for \string\caption's}}%
 \fi\next@}
\def\island@c@@#1{\def\captionprefix@{#1}%
 \captioncount@\z@\global\setbox\islandbox@\vbox\bgroup}
\rightadd@\caption\to\nofrillslist@
\newbox\captionbox@
\newbox\Captionbox@
\def\caption{%
 \ifnum\captioncount@=\z@
  \ifnopunct@
   \DN@{\egroup\nopunct@true}%
  \else
   \let\next@\egroup
  \fi
 \else
  \let\next@\relax
 \fi
 \next@
 \advance\captioncount@\@ne
 \FN@\caption@}
\def\caption@{\ifx\next"\expandafter\caption@q\else\expandafter\caption@@\fi}
\def\caption@q"#1"{\quoted@true
 {\noexpands@
 \let\pre\island@@@P\let\post\island@@@Q
 \let\style\island@@@S\let\numstyle\island@@@N
 \Qlabel@{#1}\let\style\relax\xdef\Qlabel@@@@{#1}}%
 \finishcaption@}
\def\caption@@{\quoted@false
 \global\advance\island@@@C\@ne
 {\noexpands@
 \xdef\Thelabel@@@{\number\island@@@C}%
 \xdefThelabel@\island@@@N
 \xdef\Thelabel@@@@{\island@@@P\Thelabel@\island@@@Q}%
 \xdefThelabel@@\island@@@S
 \xdef\Thepref@{\Thelabel@@@@}}%
 \finishcaption@}
\long\def\captionformat@#1#2#3{\rm\strut#1 {\island@@@F#2} #3%
 \punct@.\strut}
\long\def\widerthanisland@#1#2#3{\test@true\setbox\z@\vbox{\hsize\maxdimen
 \noindent@@\captionformat@{#1}{#2}{#3}\par\setboxzl@}%
 \ifdim\wdz@=\z@
  \global\setbox\captionbox@\hbox{\noset@\unlabel@
   \captionformat@{#1}{#2}{#3}}%
  \ifdim\wd\captionbox@>\wd\islandbox@\else\test@false\fi
 \fi}
\long\def\captionformat@@#1#2#3{\widerthanisland@{#1}{#2}{#3}%
 \iftest@
  \global\setbox\captionbox@\vbox{\hsize\wd\islandbox@
   \vskip-\parskip\noindent@@\noset@\unlabel@
   \captionformat@{#1}{#2}{#3}\par}%
 \else
  \global\setbox\captionbox@
   \hbox to\wd\islandbox@{\hfil\box\captionbox@\hfil}%
 \fi}
\long\def\finishcaption@#1{\def\entry@{#1}%
 {\locallabel@
 \captionformat@@
  {\expandafter\ignorespaces\captionprefix@\unskip}%
  {\ifx\thelabel@@\empty\unskip\else\thelabel@@\fi}%
  {\ignorespaces#1\unskip}%
 \ifnum\captioncount@=\@ne
  \global\setbox\islandbox@\vbox{\ticwrite@\vbox{\box\islandbox@}}%
  \global\setbox\Captionbox@\vbox{\box\captionbox@}%
 \else
  \global\setbox\islandbox@\vbox{\unvbox\islandbox@\setboxzl@
   \ticwrite@\boxz@}%
  \global\setbox\Captionbox@\vbox{\unvbox\Captionbox@
   \smallskip\box\captionbox@}%
 \fi}%
 \nopunct@false\nospace@false\ignorespaces}
\def\Sixtic@{\ifx\macdef@\empty\else
 \DN@##1##2\next@{\def\macdef@{##1##2}}%
 \expandafter\next@\macdef@\next@
 \edef\next@
  {\noexpand\six@\tic@\macdef@
  \space\space\space\space\space\space\space\space\space\space\space\space
  \noexpand\six@}%
 \next@\let\macdef@\relax\fi}
\def\ticwrite@{%
 \iftoc@
  {\noexpands@\let\style\relax
  \DN@{\island}%
  \edef\next@{\write\tic@{%
   \ifnopunct@\noexpand\noexpand\noexpand\nopunct\fi
   \ifx\islandtype@\next@\noexpand\noexpand\noexpand\island
    \noexpand\string\noexpand\c{\islandclass@}{\captionprefix@}%
     {\QorThelabel@@@@}\else\noexpand\noexpand\expandafter\noexpand
     \islandtype@{\QorThelabel@@@@}}\fi}%
  \next@}%
  \expandafter\unmacro@\meaning\entry@\unmacro@
  \Sixtic@
  \write\tic@{\noexpand\Page{\number\pageno}{\page@N}{\page@P}{\page@Q}^^J}%
 \fi}
\def\Htrim@#1{%
 \ifHby@
  \dimen@\vsize
  \ifnum\captioncount@=\z@
  \else
   \advance\dimen@-\ht\Captionbox@
   \advance\dimen@-#1%
  \fi
  \global\Hby@false
  \dimen@ii\wd\islandbox@
  \global\setbox\islandbox@\vbox
   {\unvbox\islandbox@\setboxzl@
   \vbox to\z@{\vss\boxz@}\nointerlineskip\hbyw\dimen@\dimen@ii}%
  \global\Hby@true
 \fi}
\newif\ifdata@
\def\iclasstest@#1{\DN@{#1}\ifx\next@\islandclass@
 \test@true\else\test@false\fi}
\skipdef\skipi@=1
\def\endisland{\ifnum\captioncount@=\z@\expandafter\egroup\fi
 \ifdata@
 \else
  \iclasstest@{T}%
  \iftest@
   {\rm\global\skipi@-\dp\strutbox}\global\advance\skipi@\bigskipamount
   \Htrim@\skipi@
   \global\setbox\islandbox@\vbox
    {\ifnum\captioncount@=\z@\else
     \box\Captionbox@
     \nointerlineskip
     \vskip\skipi@\fi
     \box\islandbox@}%
  \else
   {\rm\global\skipi@\dp\strutbox}\global\advance\skipi@\medskipamount
   \Htrim@\skipi@
   \global\setbox\islandbox@\vbox
    {\box\islandbox@
     \ifnum\captioncount@=\z@\else
     \nointerlineskip
     \vskip\skipi@
     \box\Captionbox@
     \fi}%
  \fi
  \ifHby@
  \else
   \dimen@\ht\islandbox@\advance\dimen@\dp\islandbox@
   \ifdim\dimen@>\vsize
    \DN@{\island}%
    \Err@{%
     \ifx\islandtype@\next@\noexpand\island\else
      \expandafter\noexpand\islandtype@\fi
     \ifnum\captioncount@=\z@\else
       with \noexpand\caption\fi
      is larger than page}%
     \ht\islandbox@=\vsize
   \fi
  \fi
 \fi
 \global\Hby@false\island@true}
\def\newisland#1\c#2#3{\define#1{}%
 \iftoc@\immediate\write\tic@{\noexpand\newisland\noexpand#1%
  \string\c{#2}{#3}^^J}\fi
 \expandafter\def\csname\exstring@#1@S\endcsname{\island@S}%
 \expandafter\def\csname\exstring@#1@N\endcsname{\island@N}%
 \expandafter\def\csname\exstring@#1@P\endcsname{\island@P}%
 \expandafter\def\csname\exstring@#1@Q\endcsname{\island@Q}%
 \expandafter\def\csname\exstring@#1@F\endcsname{\island@F}%
 \expandafter\def\csname end\exstring@#1\endcsname{\endisland}%
 \expandafter
 \ifx\csname island@C#2\endcsname\relax
  \expandafter\newcount@\csname island@C#2\endcsname
  \global\csname island@C#2\endcsname\z@
 \fi
 \edef\next@{\noexpand\expandafter\noexpand\let\noexpand
  \csname\exstring@#1@C\noexpand\endcsname
  \csname island@C#2\endcsname}%
 \next@
 \def#1{\def\islandtype@{#1}\island@c\c{#2}{#3}}}
\newisland\Figure\c{F}{Figure}
\newisland\Table\c{T}{Table}
\newbox\islandboxi
\newbox\islandboxii
\newbox\islandboxiii
\newbox\captionboxi
\newbox\captionboxii
\newbox\captionboxiii
\long\def\islandpairdata#1#2{{\data@true
 \place@true
 #1%
 \global\setbox\islandboxi\box\islandbox@
 \global\setbox\captionboxi\box\Captionbox@
 #2%
 \global\setbox\islandboxii\box\islandbox@
 \global\setbox\captionboxii\box\Captionbox@
 }}
\long\def\islandpairbox#1#2{\islandpairdata{#1}{#2}%
 \dimen@\ht\captionboxi
 \ifdim\ht\captionboxii>\dimen@\dimen@\ht\captionboxii\fi
 \ifdim\dimen@>\z@
  \ifdim\ht\captionboxi<\dimen@
   \global\setbox\captionboxi\vbox to\dimen@{\unvbox\captionboxi\vfil}\fi
  \ifdim\ht\captionboxii<\dimen@
   \global\setbox\captionboxii\vbox to\dimen@{\unvbox\captionboxii\vfil}\fi
 \fi
 \global\setbox\islandbox@\vbox
 {\hbox to\hsize{\hfil\box\islandboxi\hfil\box\islandboxii\hfil}%
 \ifdim\dimen@>\z@\nointerlineskip
 {\rm\global\skipi@\dp\strutbox}\global\advance\skipi@\medskipamount
  \vskip\skipi@
  \hbox to\hsize{\hfil\box\captionboxi\hfil\box\captionboxii\hfil}\fi}}
\long\def\islandpairboxa#1#2{\islandpairdata{#1}{#2}%
 \dimen@\ht\captionboxi
 \ifdim\ht\captionboxii>\dimen@\dimen@\ht\captionboxii\fi
 \ifdim\dimen@>\z@
  \ifdim\ht\captionboxi<\dimen@
   \global\setbox\captionboxi\vbox to\dimen@{\vfil\unvbox\captionboxi}\fi
  \ifdim\ht\captionboxii<\dimen@
   \global\setbox\captionboxii\vbox to\dimen@{\vfil\unvbox\captionboxii}\fi
 \fi
 \dimen@ii\ht\islandboxi
 \ifdim\ht\islandboxii>\dimen@ii \dimen@ii\ht\islandboxii\fi
 \ifdim\dimen@ii>\z@
  \ifdim\ht\islandboxi<\dimen@ii
   \global\setbox\islandboxi\vbox to\dimen@ii{\box\islandboxi\vfil}\fi
  \ifdim\ht\islandboxii<\dimen@ii
   \global\setbox\islandboxii\vbox to\dimen@ii{\box\islandboxii\vfil}\fi
 \fi
 \global\setbox\islandbox@\vbox{\ifdim\dimen@>\z@
  \hbox to\hsize{\hfil\box\captionboxi\hfil\box\captionboxii\hfil}%
  \nointerlineskip{\rm\global\skipi@-\dp\strutbox}%
  \global\advance\skipi@\bigskipamount\vskip\skipi@\fi
  \hbox to\hsize{\hfil\box\islandboxi\hfil\box\islandboxii\hfil}}}
\long\def\islandtripledata#1#2#3{{\data@true\place@true
 #1%
 \global\setbox\islandboxi\box\islandbox@
 \global\setbox\captionboxi\box\Captionbox@
 #2%
 \global\setbox\islandboxii\box\islandbox@
 \global\setbox\captionboxii\box\Captionbox@
 #3%
 \global\setbox\islandboxiii\box\islandbox@
 \global\setbox\captionboxiii\box\Captionbox@
 }}
\long\def\islandtriplebox#1#2#3{\islandtripledata{#1}{#2}{#3}%
 \dimen@\ht\captionboxi
 \ifdim\ht\captionboxii>\dimen@ \dimen@\ht\captionboxii\fi
 \ifdim\ht\captionboxiii>\dimen@ \dimen@\ht\captionboxiii\fi
 \ifdim\dimen@>\z@
  \ifdim\ht\captionboxi<\dimen@
   \global\setbox\captionboxi\vbox to\dimen@{\unvbox\captionboxi\vfil}\fi
  \ifdim\ht\captionboxii<\dimen@
   \global\setbox\captionboxii\vbox to\dimen@{\unvbox\captionboxii\vfil}\fi
  \ifdim\ht\captionboxiii<\dimen@
   \global\setbox\captionboxiii\vbox to\dimen@{\unvbox\captionboxiii\vfil}\fi
 \fi
 \global\setbox\islandbox@\vbox
  {\hbox to\hsize{\hfil\box\islandboxi\hfil\box\islandboxii\hfil
   \box\islandboxiii\hfil}%
 \ifdim\dimen@>\z@\nointerlineskip
  {\rm\global\skipi@\dp\strutbox}\global\advance\skipi@\medskipamount
  \vskip\skipi@
  \hbox to\hsize{\hfil\box\captionboxi\hfil\box\captionboxii\hfil
   \box\captionboxiii\hfil}\fi}}
\def\islandtripleboxa#1#2#3{\islandtripledata{#1}{#2}{#3}%
 \dimen@\ht\captionboxi
 \ifdim\ht\captionboxii>\dimen@ \dimen@\ht\captionboxii\fi
 \ifdim\ht\captionboxiii>\dimen@ \dimen@\ht\captionboxiii\fi
 \ifdim\dimen@>\z@
  \ifdim\ht\captionboxi<\dimen@
   \global\setbox\captionboxi\vbox to\dimen@{\vfil\unvbox\captionboxi}\fi
  \ifdim\ht\captionboxii<\dimen@
   \global\setbox\captionboxii\vbox to\dimen@{\vfil\unvbox\captionboxii}\fi
  \ifdim\ht\captionboxiii<\dimen@
   \global\setbox\captionboxiii\vbox to\dimen@{\vfil\unvbox\captionboxiii}\fi
 \fi
 \dimen@ii\ht\islandboxi
 \ifdim\ht\islandboxii>\dimen@ii \dimen@ii\ht\islandboxii\fi
 \ifdim\ht\islandboxiii>\dimen@ii \dimen@ii\ht\islandboxiii\fi
 \ifdim\dimen@ii>\z@
  \ifdim\ht\islandboxi<\dimen@ii
   \global\setbox\islandboxi\vbox to\dimen@ii{\box\islandboxi\vfil}\fi
  \ifdim\ht\islandboxii<\dimen@ii
   \global\setbox\islandboxii\vbox to\dimen@ii{\box\islandboxii\vfil}\fi
  \ifdim\ht\islandboxiii<\dimen@ii
   \global\setbox\islandboxiii\vbox to\dimen@ii{\box\islandboxiii\vfil}\fi
 \fi
 \global\setbox\islandbox@\vbox
  {\ifdim\dimen@>\z@
  \hbox to\hsize{\hfil\box\captionboxi\hfil\box\captionboxii\hfil
   \box\captionboxiii\hfil}%
  \nointerlineskip{\rm\global\skipi@-\dp\strutbox}%
  \global\advance\skipi@\bigskipamount\vskip\skipi@\fi
  \hbox to\hsize{\hfil\box\islandboxi\hfil\box\islandboxii\hfil
   \box\islandboxiii\hfil}}}
\def\Figurepair#1\and#2\endFigurepair{\island@true
 \islandpairbox{\Figure#1\endFigure}{\Figure#2\endFigure}}
\def\Figuretriple#1\and#2\and#3\endFiguretriple{\island@true
 \islandtriplebox{\Figure#1\endFigure}{\Figure#2\endFigure}%
  {\Figure#3\endFigure}}
\def\Tablepair#1\and#2\endTablepair{\island@true
 \islandpairboxa{\Table#1\endTable}{\Table#2\endTable}}
\def\Tabletriple#1\and#2\and#3\endTabletriple{\island@true
 \islandtripleboxa{\Table#1\endTable}{\Table#2\endTable}%
 {\Table#3\endTable}}
\def\place#1{\place@true\island@false
 #1%
 \ifisland@
  \box\islandbox@
 \else
  \Err@{Whoa ... there's no \string\Figure, \string\Table,
   etc., here}%
 \fi
 \place@false}
\newskip\belowtopfigskip
\belowtopfigskip 15\p@ plus 5\p@ minus5\p@
\newskip\abovebotfigskip
\abovebotfigskip 18\p@ plus 6\p@ minus6\p@
\newdimen\minpagesize
\minpagesize 5pc
\dimen@\belowtopfigskip
\advance\dimen@-\abovebotfigskip
\skip\topins\dimen@
\dimen\topins\z@
\newcount\topinscount@
\newbox\topinsdims@
\def\storedim@{\global\setbox\topinsdims@
 \vbox{\hbox to\dimen@{}\unvbox\topinsdims@}}
\def\advancedimtopins@{%
 \ifnum\pageno=\@ne
 \else
   \advance\dimen@\dimen\topins
   \global\dimen\topins\dimen@
 \fi}
\newcount\flipcount@
\def\fliptopins@{%
 \global\flipcount@\z@
 \ifvoid\topins\else
 \setbox\z@\vbox
  {\vskip\p@
   \unvbox\topins
   \global\setbox\topins\vbox{}%
   \loop
    \test@false
    \ifdim\lastskip=\z@\unskip
     \ifdim\lastskip=\z@
      \test@true\fi\fi
    \iftest@
    \global\advance\flipcount@\@ne
    \setboxzl@
    \global\setbox\topins\vbox{\unvbox\topins\boxz@}%
    \unpenalty
   \repeat}\fi}
\newif\ifPar@
\newcount\Parcount@
\newbox\Parbox@
\expandafter\newbox\csname Parfigbox1\endcsname
\expandafter\newbox\csname Parfigbox2\endcsname
\expandafter\newbox\csname Parfigbox3\endcsname
\expandafter\newbox\csname Parfigbox4\endcsname
\expandafter\newbox\csname Parfigbox5\endcsname
\expandafter\newdimen\csname Parprev1\endcsname
\expandafter\newdimen\csname Parprev2\endcsname
\expandafter\newdimen\csname Parprev3\endcsname
\expandafter\newdimen\csname Parprev4\endcsname
\expandafter\newdimen\csname Parprev5\endcsname
\expandafter\newdimen\csname Parprev6\endcsname
\def\Par{\par\global\csname Parprev1\endcsname\prevdepth
 \global\Parcount@\@ne
 \global\Par@true\global\let\Parlist@\empty
 \global\setbox\Parbox@\vbox\bgroup\break}
\def\place@#1#2{%
 \ifisland@
  \ifhmode
   \ifPar@
    \ifnum\Parcount@>5
     \Err@{Only 5 \string\place's allowed per
      \string\Par...\noexpand\endPar paragraph}%
    \else
     \expandafter\expandafter\expandafter
      \global\expandafter\setbox
       \csname Parfigbox\number\Parcount@\endcsname\box\islandbox@
     \global\advance\Parcount@\@ne
     \xdef\Parlist@{\Parlist@#1}%
    \fi
   \else
    \vadjust{#2}%
   \fi
  \else
   #2%
  \fi
 \else
  \Err@{Whoa ... there's no \string\Figure,
   \string\Table, etc., here}%
 \fi
 \place@false}
\long\def\Aplace#1{\prevanish@
 \place@true\island@false
 #1%
 \place@ a\Aplace@
 \postvanish@}
\long\def\AAplace#1{\prevanish@\place@true\island@false
 #1%
 \place@ A\AAplace@
 \postvanish@}
\newif\ifAA@
\def\AAplace@{\AA@true\Aplace@\AA@false}
\let\AAlist@\empty
\def\Aplace@{\allowbreak
 \dimen@=\ht\islandbox@
 \advance\dimen@\abovebotfigskip
 \ht\islandbox@\dimen@
 \advance\dimen@\dp\islandbox@
 \storedim@
 \ifAA@
  \xdef\AAlist@{\AAlist@1}%
  \advancedimtopins@
 \else
  \xdef\AAlist@{\AAlist@0}%
  \ifnum\topinscount@>\@ne\else\advancedimtopins@\fi
 \fi
 \insert\topins{\penalty\z@\splittopskip\z@\floatingpenalty\z@
  \box\islandbox@}%
 \global\advance\topinscount@\@ne}
\long\def\Bplace#1{\prevanish@\place@true\island@false
 #1%
 \place@ b\Bplace@
 \postvanish@}
\def\Bplace@{\allowbreak
 \ifnum\topinscount@=\z@
  \setbox\z@\vbox{\vbox to-\belowtopfigskip{}}%
  \dimen@-\skip\topins
  \ht\z@\dimen@
  \storedim@
  \advancedimtopins@
  \insert\topins{\boxz@}%
  \global\advance\topinscount@\@ne
  \xdef\AAlist@{\AAlist@0}%
 \fi
 \dimen@\ht\islandbox@
 \advance\dimen@\abovebotfigskip
 \ht\islandbox@\dimen@
 \advance\dimen@\dp\islandbox@
 \storedim@
 \xdef\AAlist@{\AAlist@0}%
 \ifnum\topinscount@>\@ne\else\advancedimtopins@\fi
 \insert\topins{\penalty\z@\splittopskip\z@
  \floatingpenalty\z@
  \box\islandbox@}%
 \global\advance\topinscount@\@ne}
\def\breakisland@{\global\setbox\@ne\lastbox\global\skipi@\lastskip\unskip
 \global\setbox\thr@@\lastbox}%
\def\printisland@{\centerline{\box\thr@@}\nobreak\nointerlineskip
 \vskip\skipi@
 \ifdim\ht\@ne<\z@\box\@ne\else\centerline{\box\@ne}\fi}
\def\bottomfigs@{%
 \count@\@ne
 \loop
  \ifnum\count@<\flipcount@
  \nointerlineskip
  \vskip\abovebotfigskip
  \global\setbox\topins\vbox{\unvbox\topins\setboxzl@
   \unvbox\z@
   \breakisland@}%
  \printisland@
  \advance\count@\@ne
  \repeat}
\def\resetdimtopins@{%
 \global\advance\topinscount@-\flipcount@
 \global\setbox\topinsdims@\vbox
  {\unvbox\topinsdims@
   \count@\z@
   \DN@##1##2\next@{\gdef\AAlist@{##2}}%
   \loop
    \ifnum\count@<\flipcount@\setboxzl@
    \expandafter\next@\AAlist@\next@
    \advance\count@\@ne
    \repeat
   \dimen@\z@
   \count@\z@
   \setbox\tw@\vbox{}%
   \edef\nextiii@{\AAlist@}%
   \DN@##1##2\next@{\DNii@{##1}\def\nextiii@{##2}}%
   \loop
    \test@false
    \ifnum\count@<\topinscount@
    \expandafter\next@\nextiii@\next@
     \ifnum\count@<\tw@
      \test@true
     \else
      \if\nextii@ 1\test@true\fi
     \fi
    \fi
    \iftest@
     \setboxzl@
     \advance\dimen@\wdz@
     \setbox\tw@\vbox{\boxz@\unvbox\tw@}%
     \advance\count@\@ne
    \repeat
    \unvbox\tw@
    \global\dimen\topins\dimen@}}
\def\Place@#1#2{%
 \ifisland@
  \ifhmode
   \ifPar@
    \ifnum\Parcount@>5
     \Err@{Only 5 \string\place's allowed per
       \string\Par...\noexpand\endPar paragraph}%
    \else
     \expandafter\expandafter\expandafter\global\expandafter\setbox
      \csname Parfigbox\number\Parcount@\endcsname\box\islandbox@
     \global\advance\Parcount@\@ne
     \xdef\Parlist@{\Parlist@#1}%
     \vadjust{\break}%
    \fi
   \else
    \Err@{\noexpand#2allowed only in a \string\Par...\noexpand\endPar
     paragraph}%
   \fi
  \else
   #2%
  \fi
 \else
  \Err@{Who ... there's no \string\Figure, \string\Table,
   etc., here}%
 \fi
 \place@false}
\newif\ifC@
\newdimen\Cdim@
\long\def\Cplace#1{\prevanish@\place@true\island@false
 #1%
 \Place@ c\Cplace@
 \postvanish@}
\def\Cplace@{\allowbreak
 \ifnum\topinscount@>\z@\else
  \global\C@true\global\Cdim@\pagetotal\fi
 \Aplace@}
\long\def\Mplace#1{\prevanish@\place@true\island@false
 #1%
 \Place@ m\Mplace@
 \postvanish@}
\long\def\MXplace#1{\prevanish@\place@true\island@false
 #1%
 \Place@ M\MXplace@
 \postvanish@}
\newif\ifMX@
\def\MXplace@{\MX@true\Mplace@\MX@false}
\def\Mplace@{\allowbreak
 \dimen@\ht\islandbox@\advance\dimen@\dp\islandbox@
 \ifdim\pagetotal=\z@\else
  \ifdim\lastskip<\abovebotfigskip\advance\dimen@\abovebotfigskip
  \advance\dimen@-\lastskip\fi
 \fi
 \advance\dimen@\pagetotal
 \ifdim\dimen@>\pagegoal
  \Aplace@
 \else
  \nointerlineskip
  \ifdim\lastskip<\abovebotfigskip\removelastskip\vskip\abovebotfigskip\fi
  \setbox\z@\vbox{\unvbox\islandbox@
   \breakisland@}%
  \printisland@
  \ifnum\topinscount@=\z@
   \setbox\z@\vbox{\vbox to-\belowtopfigskip{}}%
   \dimen@-\skip\topins
   \ht\z@\dimen@
   \storedim@
   \advancedimtopins@
   \insert\topins{\boxz@}%
   \global\advance\topinscount@\@ne
   \xdef\AAlist@{\AAlist@0}%
  \fi
  \ifMX@
   \ifnum\topinscount@=\@ne
    \setbox\z@\vbox{\vbox to-\abovebotfigskip{}}%
    \ht\z@\z@
    \dimen@\z@
    \storedim@
    \advancedimtopins@
    \insert\topins{\boxz@}%
    \global\advance\topinscount@\@ne
    \xdef\AAlist@{\AAlist@0}%
   \fi
  \fi
  \nointerlineskip
  \vskip\belowtopfigskip
 \fi}
\expandafter\newbox\csname Parbox1\endcsname
\expandafter\newbox\csname Parbox2\endcsname
\expandafter\newbox\csname Parbox3\endcsname
\expandafter\newbox\csname Parbox4\endcsname
\expandafter\newbox\csname Parbox5\endcsname
\def\endPar{\egroup
 \count@\@ne
 {\vbadness\@M\vfuzz\maxdimen\splitmaxdepth\maxdimen\splittopskip\ht\strutbox
 \setbox\z@\vsplit\Parbox@ to\ht\Parbox@
 \loop
  \ifnum\count@<\Parcount@
  \expandafter\expandafter\expandafter\global\expandafter\setbox
   \csname Parbox\number\count@\endcsname\vsplit\Parbox@ to\ht\Parbox@
  \count@@\count@\advance\count@@\@ne
  \global\csname Parprev\number\count@@\endcsname
   \dp\csname Parbox\number\count@\endcsname
  \advance\count@\@ne
  \repeat}%
 \vskip\parskip
 \count@\@ne
 \def\nextv@##1##2\nextv@{\DN@{##1}\gdef\Parlist@{##2}}%
 \loop
  \ifnum\count@<\Parcount@
   \dimen@\csname Parprev\number\count@\endcsname
   \advance\dimen@\ht\strutbox
   \ifdim\dimen@<\baselineskip
    \advance\dimen@-\baselineskip\vskip-\dimen@
   \else
    \vskip\lineskip
   \fi
   \unvbox\csname Parbox\number\count@\endcsname
   \global\setbox\islandbox@\box\csname Parfigbox\number\count@\endcsname
   \expandafter\nextv@\Parlist@\nextv@
   \if a\next@\Aplace@\else
   \if A\next@\AAplace@\else
   \if b\next@\Bplace@\else
   \if c\next@\Cplace@\else
   \if m\next@\Mplace@\else
   \if M\next@\MXplace@\fi\fi\fi\fi\fi\fi
  \advance\count@\@ne
  \repeat
 \global\Par@false
 \ifvoid\Parbox@
  \prevdepth\csname Parprev\number\count@\endcsname
 \else
  \dimen@\csname Parprev\number\count@\endcsname\advance\dimen@\ht\strutbox
  \ifdim\dimen@<\baselineskip
    \advance\dimen@-\baselineskip\vskip-\dimen@
  \else
    \vskip\lineskip
  \fi
  \dimen@\dp\Parbox@
  \unvbox\Parbox@
  \prevdepth\dimen@
 \fi}
\def\folio{{\page@F\page@S{\page@P\page@N{\number\page@C}\page@Q}}}
\def\advancepageno{\global\advance\pageno\@ne}
\newif\ifspecialsplit@
\newbox\outbox@
\let\shipout@\shipout
\def\plainoutput{\specialsplit@false\ifvoid\topins\else\ifdim\ht\topins=\z@
 \specialsplit@true\advance\minpagesize-\skip\topins\fi\fi
 \fliptopins@
 \setbox\outbox@\vbox{\makeheadline\pagebody\makefootline}%
 {\noexpands@\let\style\relax
 \shipout@\box\outbox@}%
 \advancepageno
 \resetdimtopins@
 \ifvoid\@cclv\else\unvbox\@cclv\penalty\outputpenalty\fi
 \ifnum\outputpenalty>-\@MM\else\dosupereject\fi}
\def\pagebody{\vbox to\vsize{\boxmaxdepth\maxdepth
 \ifvoid\margin@\else
 \rlap{\kern\hsize\vbox to\z@{\kern4\p@\box\margin@\vss}}\fi
 \pagecontents}}
\newif\ifonlytop@
\def\pagecontents{%
 \onlytop@false
 \ifdim\ht\@cclv<\minpagesize\ifnum\flipcount@<\tw@\ifvoid\footins
  \onlytop@true\fi\fi\fi
 \test@false
 \ifC@
  \ifnum\flipcount@=\@ne
   \global\multiply\Cdim@\tw@
   \ifdim\Cdim@>\ht\@cclv
    \test@true
   \fi
  \fi
 \fi
 \global\C@false
 \iftest@
  \dimen@\ht\@cclv
  \advance\dimen@\skip\topins
  {\vfuzz\maxdimen\vbadness\@M
  \splitmaxdepth\maxdepth\splittopskip\topskip
  \setbox\z@\vsplit\@cclv to\dimen@
  \unvbox\z@}%
  \global\setbox\topins\vbox{\unvbox\topins
   \global\setbox\@ne\lastbox}%
  \setbox\z@\vbox{\unvbox\@ne
   \breakisland@}%
  \nointerlineskip
  \vskip\abovebotfigskip
  \printisland@
 \else
  \ifnum\flipcount@>\z@
   \global\setbox\topins\vbox{\unvbox\topins\global\setbox\@ne\lastbox}%
   \setbox\z@\vbox{\unvbox\@ne
    \breakisland@}%
   \printisland@
   \ifonlytop@\kern-\prevdepth\vfill\else\vskip\belowtopfigskip\fi
  \fi
 \fi
 \ifdim\ht\@cclv<\minpagesize
  \ifonlytop@\else\vfill\fi
 \else
  \ifspecialsplit@
   {\vfuzz\maxdimen\vbadness\@M
   \splitmaxdepth\maxdepth\splittopskip\topskip
   \dimen@ii\ht\@cclv \advance\dimen@ii\skip\topins
   \setbox\z@\vsplit\@cclv to\dimen@ii
   \unvbox\z@}%
  \else
   \unvbox\@cclv
  \fi
 \fi
 \bottomfigs@
 \ifvoid\footins\else\vskip\skip\footins\footnoterule\unvbox\footins\fi}
\newread\readdata@
\def\readthedata@#1{\expandafter
 \ifx\csname#1@D\endcsname\relax
  \immediate\openin\readdata@=#1.dat
  \ifeof\readdata@
   \Err@{No file #1.dat}%
  \else
   {\endlinechar\m@ne\gdef\Next@{}%
   \DNii@##1 ##2 ##3pt{\global\data@ht##1\global\data@dp##2%
    \global\data@wd##3pt}%
   \loop
    \ifeof\readdata@
    \else
    \read\readdata@ to\next@
    \ifx\next@\empty\else
     \edef\next@{\expandafter\nextii@\next@}%
     \expandafter\rightadd@\next@\to\Next@
    \fi
    \repeat}%
   \immediate\closein\readdata@
   \expandafter\expandafter\expandafter\global\expandafter
    \let\csname#1@D\endcsname\Next@\global\let\Next@\relax
  \fi
 \fi}
\newdimen\data@ht
\newdimen\data@dp
\newdimen\data@wd
\newif\ifgetdata@
\def\getdata@#1#2{\global\getdata@true\count@#2\relax
 {\let\\\or\xdef\Next@{\ifcase\number\count@#1\else
 \global\noexpand\getdata@false\fi}}\Next@}
\def\paste#1#2{\readthedata@{#1}%
 \getdata@{\csname#1@D\endcsname}{#2}%
 \ifgetdata@
 \dimen@\data@ht \advance\dimen@\data@dp
  \hbox{\special{dvipaste: #1 #2}%
   \lower\data@dp\vbox to\dimen@{\hbox to\data@wd{}\vfil}}%
 \else
  {\lccode`\Z=`\#\lccode`\N=`\N\lccode`\F=`\F%
   \lowercase{\Err@{No data for File [#1], Z#2}}}%
 \fi}
\newdimen\httable
\newdimen\dptable
\newdimen\wdtable
\def\measuretable#1#2{\readthedata@{#1}%
 \getdata@{\csname#1@D\endcsname}{#2}%
 \ifgetdata@
  \httable\data@ht \dptable\data@dp \wdtable\data@wd
 \else
  {\lccode`\Z=`\#\lccode`\N=`\N\lccode`\F=`\F%
  \lowercase{\Err@{No data for File [#1], Z#2}}}%
 \fi}
\def\East#1#2{\setboxz@h{$\m@th\ssize\;{#1}\;\;$}%
 \setbox\tw@\hbox{$\m@th\ssize\;{#2}\;\;$}\setbox4=\hbox{$\m@th#2$}%
 \dimen@\minaw@
 \ifdim\wdz@>\dimen@\dimen@\wdz@\fi\ifdim\wd\tw@>\dimen@\dimen@\wd\tw@\fi
 \ifdim\wd4 >\z@
  \mathrel{\mathop{\hbox to\dimen@{\rightarrowfill}}\limits^{#1}_{#2}}%
 \else
  \mathrel{\mathop{\hbox to\dimen@{\rightarrowfill}}\limits^{#1}}%
 \fi}
\def\West#1#2{\setboxz@h{$\m@th\ssize\;\;{#1}\;$}%
 \setbox\tw@\hbox{$\m@th\ssize\;\;{#2}\;$}\setbox4=\hbox{$\m@th#2$}%
 \dimen@\minaw@
 \ifdim\wdz@>\dimen@\dimen@\wdz@\fi\ifdim\wd\tw@>\dimen@\dimen@\wd\tw@\fi
 \ifdim\wd4 >\z@
  \mathrel{\mathop{\hbox to\dimen@{\leftarrowfill}}\limits^{#1}_{#2}}%
 \else
  \mathrel{\mathop{\hbox to\dimen@{\leftarrowfill}}\limits^{#1}}%
 \fi}
\font\arrow@i=lams1
\font\arrow@ii=lams2
\font\arrow@iii=lams3
\font\arrow@iv=lams4
\font\arrow@v=lams5
\newdimen\standardcgap
\standardcgap40\p@
\newdimen\hunit
\hunit\tw@\p@
\newdimen\standardrgap
\standardrgap32\p@
\newdimen\vunit
\vunit1.6\p@
\def\Cgaps#1{\RIfM@
 \standardcgap#1\standardcgap\relax\hunit#1\hunit\relax
 \else\nonmatherr@\Cgaps\fi}
\def\Rgaps#1{\RIfM@
 \standardrgap#1\standardrgap\relax\vunit#1\vunit\relax
 \else\nonmatherr@\Rgaps\fi}
\newdimen\getdim@
\def\getcgap@#1{\ifcase#1\or\getdim@\z@\else\getdim@\standardcgap\fi}
\def\getrgap@#1{\ifcase#1\getdim@\z@\else\getdim@\standardrgap\fi}
\def\cgaps{\RIfM@\expandafter\cgaps@\else\expandafter\nonmatherr@
 \expandafter\cgaps\fi}
\def\cgaps@{\ifnum\catcode`\;=\active\expandafter\cgapsA@\else
 \expandafter\cgapsO@\fi}
\def\cgapsO@#1{\toks@{\ifcase\i@\or\getdim@=\z@}%
 \gaps@@\standardcgap#1;\gaps@@\gaps@@
 \edef\next@{\the\toks@\noexpand\else\noexpand\getdim@\noexpand\standardcgap
  \noexpand\fi}%
 \toks@=\expandafter{\next@}%
 \edef\getcgap@##1{\i@##1\relax\the\toks@}\toks@{}}
{\catcode`\;=\active
 \gdef\cgapsA@#1{\toks@{\ifcase\i@\or\getdim@=\z@}%
 \gaps@@\standardcgap#1;\gaps@@\gaps@@
 \edef\next@{\the\toks@\noexpand\else\noexpand\getdim@\noexpand\standardcgap
  \noexpand\fi}%
 \toks@=\expandafter{\next@}%
 \edef\getcgap@##1{\i@##1\relax\the\toks@}\toks@{}}
}
\def\Gaps@@{\gaps@@}
\def\gaps@@#1#2;#3{\mgaps@#1#2\mgaps@
 \edef\next@{\the\toks@\noexpand\or\noexpand\getdim@
  \noexpand#1\the\mgapstoks@@}%
 \toks@\expandafter{\next@}%
 \DN@{#3}%
 \ifx\next@\Gaps@@\def\next@##1\gaps@@{}\else
  \def\next@{\gaps@@#1#3}\fi\next@}
{\catcode`\;=\active
 \gdef\rgaps#1{\RIfM@{\ifnum\catcode`\;=\active\def;{\string;}\fi
   \xdef\Next@{\noexpand\rgaps@{#1}}}%
  \Next@\edef\getrgap@##1{\i@##1\relax\the\toks@}\toks@{}\else
  \nonmatherr@\rgaps\fi}
}
\def\rgaps@#1{\toks@{\ifcase\i@\getdim@=\z@}%
 \gaps@@\standardrgap#1;\gaps@@\gaps@@
 \edef\next@{\the\toks@\noexpand\else\noexpand\getdim@\noexpand\standardrgap
  \noexpand\fi}%
 \toks@=\expandafter{\next@}}
\newbox\ZER@
\def\mgaps@#1{\let\mgapsnext@#1\FNSS@\mgaps@@}
\def\mgaps@@{\ifx\next\w\expandafter\mgaps@@@\else
 \expandafter\mgaps@@@@\fi}
\newtoks\mgapstoks@@
\def\mgaps@@@@#1\mgaps@{\getdim@\mgapsnext@\getdim@#1\getdim@
 \edef\next@{\noexpand\getdim@\the\getdim@}%
 \mgapstoks@@\expandafter{\next@}}
\def\mgaps@@@\w#1#2\mgaps@{\mgaps@@@@#2\mgaps@
 \setbox\ZER@\hbox{$\m@th\hskip15\p@\tsize@#1$}%
 \dimen@\wd\ZER@
 \ifdim\dimen@>\getdim@\getdim@\dimen@\fi
 \edef\next@{\noexpand\getdim@\the\getdim@}%
 \mgapstoks@@\expandafter{\next@}}
\def\changewidth#1#2{\setbox\ZER@{$\m@th#2}%
 \hbox to\wd\ZER@{\hss$\m@th#1$\hss}}
\atdef@({\FN@\ARROW@}
\def\ARROW@{\ifx\next)\let\next@\OPTIONS@\else
 \DN@{\csname\string @(\endcsname}\fi\next@}
\newif\ifoptions@
\def\OPTIONS@){\ifoptions@\let\next@\relax\else
 \DN@{\global\options@true\begingroup\optioncodes@}\fi\next@}
\newif\ifN@
\newif\ifE@
\newif\ifNESW@
\newif\ifH@
\newif\ifV@
\newif\ifHshort@
\expandafter\def\csname\string @(\endcsname #1,#2){%
 \ifoptions@\expandafter\endgroup\fi
 \N@false\E@false\H@false\V@false\Hshort@false
 \ifnum#1>\z@\E@true\fi
 \ifnum#1=\z@\V@true\global\tX@false\global\tY@false\global\a@false\fi
 \ifnum#2>\z@\N@true\fi
 \ifnum#2=\z@\H@true\global\tX@false\global\tY@false\global\a@false
  \ifshort@\Hshort@true\fi\fi
 \NESW@false
 \ifN@\ifE@\NESW@true\fi\else\ifE@\else\NESW@true\fi\fi
 \arrow@{#1}{#2}%
 \global\options@false
 \global\scount@\z@\global\tcount@\z@\global\arrcount@\z@
 \global\s@false\global\sxdimen@\z@\global\sydimen@\z@
 \global\tX@false\global\tXdimen@i\z@\global\tXdimen@ii\z@
 \global\tY@false\global\tYdimen@i\z@\global\tYdimen@ii\z@
 \global\a@false\global\exacount@\z@
 \global\x@false\global\xdimen@\z@
 \global\X@false\global\Xdimen@\z@
 \global\y@false\global\ydimen@\z@
 \global\Y@false\global\Ydimen@\z@
 \global\p@false\global\pdimen@\z@
 \global\label@ifalse\global\label@iifalse
 \global\dl@ifalse\global\ldimen@i\z@
 \global\dl@iifalse\global\ldimen@ii\z@
 \global\short@false\global\unshort@false}
\newif\iflabel@i
\newif\iflabel@ii
\newcount\scount@
\newcount\tcount@
\newcount\arrcount@
\newif\ifs@
\newdimen\sxdimen@
\newdimen\sydimen@
\newif\iftX@
\newdimen\tXdimen@i
\newdimen\tXdimen@ii
\newif\iftY@
\newdimen\tYdimen@i
\newdimen\tYdimen@ii
\newif\ifa@
\newcount\exacount@
\newif\ifx@
\newdimen\xdimen@
\newif\ifX@
\newdimen\Xdimen@
\newif\ify@
\newdimen\ydimen@
\newif\ifY@
\newdimen\Ydimen@
\newif\ifp@
\newdimen\pdimen@
\newif\ifdl@i
\newif\ifdl@ii
\newdimen\ldimen@i
\newdimen\ldimen@ii
\newif\ifshort@
\newif\ifunshort@
\def\zero@#1{\ifnum\scount@=\z@
 \if#1e\global\scount@\m@ne\else
 \if#1t\global\scount@\tw@\else
 \if#1h\global\scount@\thr@@\else
 \if#1'\global\scount@6 \else
 \if#1`\global\scount@7 \else
 \if#1(\global\scount@8 \else
 \if#1)\global\scount@9 \else
 \if#1s\global\scount@12 \else
 \if#1H\global\scount@13 \else
 \Err@{\Invalid@@ option \string\0}\fi\fi\fi\fi\fi\fi\fi\fi\fi
 \fi}
\def\one@#1{\ifnum\tcount@=\z@
 \if#1e\global\tcount@\m@ne\else
 \if#1h\global\tcount@\tw@\else
 \if#1t\global\tcount@\thr@@\else
 \if#1'\global\tcount@4 \else
 \if#1`\global\tcount@5 \else
 \if#1(\global\tcount@\ten@ \else
 \if#1)\global\tcount@11 \else
 \if#1s\global\tcount@12 \else
 \if#1H\global\tcount@13 \else
 \Err@{\Invalid@@ option \string\1}\fi\fi\fi\fi\fi\fi\fi\fi\fi
 \fi}
\def\a@#1{\ifnum\arrcount@=\z@
 \if#10\global\arrcount@\m@ne\else
 \if#1+\global\arrcount@\@ne\else
 \if#1-\global\arrcount@\tw@\else
 \if#1=\global\arrcount@\thr@@\else
 \Err@{\Invalid@@ option \string\a}\fi\fi\fi\fi
 \fi}
\def\ds@{\ifnum\catcode`\;=\active\expandafter\dsA@\else
 \expandafter\dsO@\fi}
\def\dsO@(#1;#2){\ds@@{#1}{#2}}
\def\ds@@#1#2{\ifs@\else
 \global\s@true
 \global\sxdimen@\hunit\global\sxdimen@#1\sxdimen@\relax
 \global\sydimen@\vunit\global\sydimen@#2\sydimen@\relax
 \fi}
\def\dtX@{\ifnum\catcode`\;=\active\expandafter\dtXA@\else
 \expandafter\dtXO@\fi}
\def\dtXO@(#1;#2){\dtX@@{#1}{#2}}
\def\dtX@@#1#2{\iftX@\else
 \global\tX@true
 \global\tXdimen@i\hunit\global\tXdimen@i#1\tXdimen@i\relax
 \global\tXdimen@ii\vunit\global\tXdimen@ii#2\tXdimen@ii\relax
 \fi}
\def\dtY@{\ifnum\catcode`\;=\active\expandafter\dtYA@\else
 \expandafter\dtYO@\fi}
\def\dtYO@(#1;#2){\dtY@@{#1}{#2}}
\def\dtY@@#1#2{\iftY@\else
 \global\tY@true
 \global\tYdimen@i\hunit\global\tYdimen@i#1\tYdimen@i\relax
 \global\tYdimen@ii\vunit\global\tYdimen@ii#2\tYdimen@ii\relax
 \fi}
{\catcode`\;=\active
 \gdef\dsA@(#1;#2){\ds@@{#1}{#2}}
 \gdef\dtXA@(#1;#2){\dtX@@{#1}{#2}}
 \gdef\dtYA@(#1;#2){\dtY@@{#1}{#2}}
}
\def\da@#1{\ifa@\else\global\a@true\global\exacount@#1\relax\fi}
\def\dx@#1{\ifx@\else
 \global\x@true
 \global\xdimen@\hunit\global\xdimen@#1\xdimen@\relax
 \fi}
\def\dX@#1{\ifX@\else
 \global\X@true
 \global\Xdimen@\hunit\global\Xdimen@#1\Xdimen@\relax
 \fi}
\def\dy@#1{\ify@\else
 \global\y@true
 \global\ydimen@\vunit\global\ydimen@#1\ydimen@\relax
 \fi}
\def\dY@#1{\ifY@\else
 \global\Y@true
 \global\Ydimen@\vunit\global\Ydimen@#1\Ydimen@\relax
 \fi}
\def\p@@#1{\ifp@\else
 \global\p@true
 \global\pdimen@\hunit\global\divide\pdimen@\tw@
 \global\pdimen@#1\pdimen@\relax
 \fi}
\def\L@#1{\iflabel@i\else
 \global\label@itrue\gdef\label@i{#1}%
 \fi}
\def\l@#1{\iflabel@ii\else
 \global\label@iitrue\gdef\label@ii{#1}%
 \fi}
\def\dL@#1{\ifdl@i\else
 \global\dl@itrue\global\ldimen@i\hunit\global\ldimen@i#1\ldimen@i\relax
 \fi}
\def\dl@#1{\ifdl@ii\else
 \global\dl@iitrue\global\ldimen@ii\hunit\global\ldimen@ii#1\ldimen@ii\relax
 \fi}
\def\s@{\ifunshort@\else\global\short@true\fi}
\def\uns@{\ifshort@\else\global\unshort@true\global\short@false\fi}
\def\optioncodes@{\let\0\zero@\let\1\one@\let\a\a@\let\ds\ds@\let\dtX\dtX@
 \let\dtY\dtY@\let\da\da@\let\dx\dx@\let\dX\dX@\let\dY\dY@\let\dy\dy@
 \let\p\p@@\let\L\L@\let\l\l@\let\dL\dL@\let\dl\dl@\let\s\s@\let\uns\uns@}
\def\slopes@{\\161\\152\\143\\134\\255\\126\\357\\238\\349\\45{10}\\56{11}%
 \\11{12}\\65{13}\\54{14}\\43{15}\\32{16}\\53{17}\\21{18}\\52{19}\\31{20}%
 \\41{21}\\51{22}\\61{23}}
\newcount\tan@i
\newcount\tan@ip
\newcount\tan@ii
\newcount\tan@iip
\newdimen\slope@i
\newdimen\slope@ip
\newdimen\slope@ii
\newdimen\slope@iip
\newcount\angcount@
\newcount\extracount@
\def\slope@{{\slope@i\secondy@\advance\slope@i-\firsty@
 \ifN@\else\multiply\slope@i\m@ne\fi
 \slope@ii\secondx@\advance\slope@ii-\firstx@
 \ifE@\else\multiply\slope@ii\m@ne\fi
 \ifdim\slope@ii<\z@
  \global\tan@i6 \global\tan@ii\@ne\global\angcount@23
 \else
  \dimen@\slope@i\multiply\dimen@6
  \ifdim\dimen@<\slope@ii
   \global\tan@i\@ne\global\tan@ii6 \global\angcount@\@ne
  \else
   \dimen@\slope@ii\multiply\dimen@6
   \ifdim\dimen@<\slope@i
    \global\tan@i6 \global\tan@ii\@ne\global\angcount@23
   \else
    \global\tan@ip\z@\global\tan@iip\@ne
    \def\\##1##2##3{\global\angcount@##3\relax
     \slope@ip\slope@i\slope@iip\slope@ii
     \multiply\slope@iip##1\relax\multiply\slope@ip##2\relax
     \ifdim\slope@iip<\slope@ip
      \global\tan@ip##1\relax\global\tan@iip##2\relax
     \else
      \global\tan@i##1\relax\global\tan@ii##2\relax
      \def\\####1####2####3{}%
     \fi}%
    \slopes@
    \slope@i\secondy@\advance\slope@i-\firsty@
    \ifN@\else\multiply\slope@i\m@ne\fi
    \multiply\slope@i\tan@ii\multiply\slope@i\tan@iip\multiply\slope@i\tw@
    \count@\tan@i\multiply\count@\tan@iip
    \extracount@\tan@ip\multiply\extracount@\tan@ii
    \advance\count@\extracount@
    \slope@ii\secondx@\advance\slope@ii-\firstx@
    \ifE@\else\multiply\slope@ii\m@ne\fi
    \multiply\slope@ii\count@
    \ifdim\slope@i<\slope@ii
     \global\tan@i\tan@ip\global\tan@ii\tan@iip
     \global\advance\angcount@\m@ne
    \fi
   \fi
  \fi
 \fi}%
}
\def\slope@a#1{{\def\\##1##2##3{\ifnum##3=#1\global\tan@i##1\relax
 \global\tan@ii##2\relax\fi}\slopes@}}
\newcount\i@
\newcount\j@
\newcount\colcount@
\newcount\Colcount@
\newcount\tcolcount@
\newdimen\rowht@
\newdimen\rowdp@
\newcount\rowcount@
\newcount\Rowcount@
\newcount\maxcolrow@
\newtoks\colwidthtoks@
\newtoks\Rowheighttoks@
\newtoks\Rowdepthtoks@
\newtoks\widthtoks@
\newtoks\Widthtoks@
\newtoks\heighttoks@
\newtoks\Heighttoks@
\newtoks\depthtoks@
\newtoks\Depthtoks@
\newif\iffirstCDcr@
\def\dotoks@i{%
 \global\widthtoks@\expandafter{\the\widthtoks@\else\getdim@\z@\fi}%
 \global\heighttoks@\expandafter{\the\heighttoks@\else\getdim@\z@\fi}%
 \global\depthtoks@\expandafter{\the\depthtoks@\else\getdim@\z@\fi}}
\def\dotoks@ii{%
 \global\widthtoks@{\ifcase\j@}%
 \global\heighttoks@{\ifcase\j@}%
 \global\depthtoks@{\ifcase\j@}}
\def\preCD@#1\endCD{\setbox\ZER@
 \vbox{%
  \def\arrow@##1##2{{}}%
  \global\rowcount@\m@ne\global\colcount@\z@\global\Colcount@\z@
  \global\firstCDcr@true\toks@{}%
  \global\widthtoks@{\ifcase\j@}%
  \global\Widthtoks@{\ifcase\i@}%
  \global\heighttoks@{\ifcase\j@}%
  \global\Heighttoks@{\ifcase\i@}%
  \global\depthtoks@{\ifcase\j@}%
  \global\Depthtoks@{\ifcase\i@}%
  \global\Rowheighttoks@{\ifcase\i@}%
  \global\Rowdepthtoks@{\ifcase\i@}%
  \Let@
  \everycr{%
   \noalign{%
    \global\advance\rowcount@\@ne
    \ifnum\colcount@<\Colcount@
    \else
     \global\Colcount@\colcount@\global\maxcolrow@\rowcount@
    \fi
    \global\colcount@\z@
    \iffirstCDcr@
     \global\firstCDcr@false
    \else
     \edef\next@{\the\Rowheighttoks@\noexpand\or\noexpand\getdim@\the\rowht@}%
      \global\Rowheighttoks@\expandafter{\next@}%
     \edef\next@{\the\Rowdepthtoks@\noexpand\or\noexpand\getdim@\the\rowdp@}%
      \global\Rowdepthtoks@\expandafter{\next@}%
     \global\rowht@\z@\global\rowdp@\z@
     \dotoks@i
     \edef\next@{\the\Widthtoks@\noexpand\or\the\widthtoks@}%
      \global\Widthtoks@\expandafter{\next@}%
     \edef\next@{\the\Heighttoks@\noexpand\or\the\heighttoks@}%
      \global\Heighttoks@\expandafter{\next@}%
     \edef\next@{\the\Depthtoks@\noexpand\or\the\depthtoks@}%
      \global\Depthtoks@\expandafter{\next@}%
     \dotoks@ii
    \fi}}%
  \tabskip\z@
  \halign{&\setbox\ZER@\hbox{\vrule\height\ten@\p@\width\z@\depth\z@     
   $\m@th\displaystyle{##}$}\copy\ZER@
   \ifdim\ht\ZER@>\rowht@\global\rowht@\ht\ZER@\fi
   \ifdim\dp\ZER@>\rowdp@\global\rowdp@\dp\ZER@\fi
   \global\advance\colcount@\@ne
   \edef\next@{\the\widthtoks@\noexpand\or\noexpand\getdim@\the\wd\ZER@}%
    \global\widthtoks@\expandafter{\next@}%
   \edef\next@{\the\heighttoks@\noexpand\or\noexpand\getdim@\the\ht\ZER@}%
    \global\heighttoks@\expandafter{\next@}%
   \edef\next@{\the\depthtoks@\noexpand\or\noexpand\getdim@\the\dp\ZER@}%
    \global\depthtoks@\expandafter{\next@}%
   \cr#1\crcr}}%
 \Rowcount@\rowcount@
 \global\Widthtoks@\expandafter{\the\Widthtoks@\fi\relax}%
 \edef\Width@##1##2{\i@##1\relax\j@##2\relax\the\Widthtoks@}%
 \global\Heighttoks@\expandafter{\the\Heighttoks@\fi\relax}%
 \edef\Height@##1##2{\i@##1\relax\j@##2\relax\the\Heighttoks@}%
 \global\Depthtoks@\expandafter{\the\Depthtoks@\fi\relax}%
 \edef\Depth@##1##2{\i@##1\relax\j@##2\relax\the\Depthtoks@}%
 \edef\next@{\the\Rowheighttoks@\noexpand\fi\relax}%
 \global\Rowheighttoks@\expandafter{\next@}%
 \edef\Rowheight@##1{\i@##1\relax\the\Rowheighttoks@}%
 \edef\next@{\the\Rowdepthtoks@\noexpand\fi\relax}%
 \global\Rowdepthtoks@\expandafter{\next@}%
 \edef\Rowdepth@##1{\i@##1\relax\the\Rowdepthtoks@}%
 \global\colwidthtoks@{\fi}%
 \setbox\ZER@\vbox{%
  \unvbox\ZER@
  \count@\rowcount@
  \loop
   \unskip\unpenalty
   \setbox\ZER@\lastbox
   \ifnum\count@>\maxcolrow@\advance\count@\m@ne
   \repeat
  \hbox{%
   \unhbox\ZER@
   \count@\z@
   \loop
    \unskip
    \setbox\ZER@\lastbox
    \edef\next@{\noexpand\or\noexpand\getdim@\the\wd\ZER@\the\colwidthtoks@}%
     \global\colwidthtoks@\expandafter{\next@}%
    \advance\count@\@ne
    \ifnum\count@<\Colcount@
    \repeat}}%
 \edef\next@{\noexpand\ifcase\noexpand\i@\the\colwidthtoks@}%
  \global\colwidthtoks@\expandafter{\next@}%
 \edef\Colwidth@##1{\i@##1\relax\the\colwidthtoks@}%
 \global\colwidthtoks@{}\global\Rowheighttoks@{}\global\Rowdepthtoks@{}%
 \global\widthtoks@{}\global\Widthtoks@{}\global\heighttoks@{}%
 \global\Heighttoks@{}\global\depthtoks@{}\global\Depthtoks@{}%
}
\newcount\xoff@
\newcount\yoff@
\newcount\endcount@
\newcount\rcount@
\newdimen\firstx@
\newdimen\firsty@
\newdimen\secondx@
\newdimen\secondy@
\newdimen\tocenter@
\newdimen\charht@
\newdimen\charwd@
\def\outside@{\Err@{This arrow points outside the \string\CD}}
\newif\ifsvertex@
\newif\iftvertex@
\def\arrow@#1#2{\global\xoff@#1\relax\global\yoff@#2\relax
 \count@\rowcount@\advance\count@-\yoff@
 \ifnum\count@<\@ne\outside@\else\ifnum\count@>\Rowcount@\outside@\fi\fi
 \count@\colcount@\advance\count@\xoff@
 \ifnum\count@<\@ne\outside@\else\ifnum\count@>\Colcount@\outside@\fi\fi
 \tcolcount@\colcount@\advance\tcolcount@\xoff@
 \Width@\rowcount@\colcount@\divide\getdim@\tw@\tocenter@-\getdim@
 \ifdim\getdim@=\z@
  \firstx@\z@\firsty@\mathaxis@\svertex@true
 \else
  \svertex@false
  \ifHshort@
   \Colwidth@\colcount@\divide\getdim@\tw@
   \ifE@ \firstx@\getdim@ \else \firstx@-\getdim@ \fi
  \else
   \ifE@ \firstx@\getdim@ \else \firstx@-\getdim@ \fi
  \fi
  \ifE@
   \ifH@ \advance\firstx@\thr@@\p@ \else \advance\firstx@-\thr@@\p@ \fi  
  \else
   \ifH@ \advance\firstx@-\thr@@\p@ \else \advance\firstx@\thr@@\p@ \fi  
  \fi
  \ifN@
   \Height@\rowcount@\colcount@ \firsty@\getdim@                         
   \ifV@ \advance\firsty@\thr@@\p@ \fi                                   
  \else
   \ifV@
    \Depth@\rowcount@\colcount@ \firsty@-\getdim@                        
    \advance\firsty@-\thr@@\p@                                           
   \else
    \firsty@\z@                                                          
   \fi
  \fi
 \fi
 \ifV@
 \else
  \Colwidth@\colcount@\divide\getdim@\tw@
  \ifE@\secondx@\getdim@\else\secondx@-\getdim@\fi
  \ifE@\else\getcgap@\colcount@\advance\secondx@-\getdim@\fi
  \endcount@\colcount@\advance\endcount@\xoff@
  \count@\colcount@
  \ifE@
   \advance\count@\@ne
   \loop
    \ifnum\count@<\endcount@
    \Colwidth@\count@\advance\secondx@\getdim@
    \getcgap@\count@\advance\secondx@\getdim@
    \advance\count@\@ne
    \repeat
  \else
   \advance\count@\m@ne
   \loop
    \ifnum\count@>\endcount@
    \Colwidth@\count@\advance\secondx@-\getdim@
    \getcgap@\count@\advance\secondx@-\getdim@
    \advance\count@\m@ne
    \repeat
  \fi
  \Colwidth@\count@\divide\getdim@\tw@
  \ifHshort@
  \else
   \ifE@\advance\secondx@\getdim@\else\advance\secondx@-\getdim@\fi
  \fi
  \ifE@\getcgap@\count@\advance\secondx@\getdim@\fi
  \rcount@\rowcount@\advance\rcount@-\yoff@
  \Width@\rcount@\count@\divide\getdim@\tw@
  \tvertex@false
  \ifH@\ifdim\getdim@=\z@\tvertex@true\Hshort@false\fi\fi
  \ifHshort@
  \else
   \ifE@\advance\secondx@-\getdim@\else\advance\secondx@\getdim@\fi
  \fi
  \iftvertex@
   \advance\secondx@.4\p@
  \else
   \ifE@\advance\secondx@-\thr@@\p@\else\advance\secondx@\thr@@\p@\fi    
  \fi
 \fi
 \ifH@
 \else
  \ifN@
   \Rowheight@\rowcount@\secondy@\getdim@
  \else
   \Rowdepth@\rowcount@\secondy@-\getdim@
   \getrgap@\rowcount@\advance\secondy@-\getdim@
  \fi
  \endcount@\rowcount@\advance\endcount@-\yoff@
  \count@\rowcount@
  \ifN@
   \advance\count@\m@ne
   \loop
    \ifnum\count@>\endcount@
    \Rowheight@\count@\advance\secondy@\getdim@
    \Rowdepth@\count@\advance\secondy@\getdim@
    \getrgap@\count@\advance\secondy@\getdim@
    \advance\count@\m@ne
    \repeat
  \else
   \advance\count@\@ne
   \loop
    \ifnum\count@<\endcount@
    \Rowheight@\count@\advance\secondy@-\getdim@
    \Rowdepth@\count@\advance\secondy@-\getdim@
    \getrgap@\count@\advance\secondy@-\getdim@
    \advance\count@\@ne
    \repeat
  \fi
  \tvertex@false
  \ifV@\Width@\count@\colcount@\ifdim\getdim@=\z@\tvertex@true\fi\fi
  \ifN@
   \getrgap@\count@\advance\secondy@\getdim@
   \Rowdepth@\count@\advance\secondy@\getdim@
   \iftvertex@
    \advance\secondy@\mathaxis@
   \else
    \Depth@\count@\tcolcount@\advance\secondy@-\getdim@
    \advance\secondy@-\thr@@\p@                                          
   \fi
  \else
   \Rowheight@\count@\advance\secondy@-\getdim@
   \iftvertex@
    \advance\secondy@\mathaxis@
   \else
    \Height@\count@\tcolcount@\advance\secondy@\getdim@
    \advance\secondy@\thr@@\p@                                           
   \fi
  \fi
 \fi
 \ifV@\else\advance\firstx@\sxdimen@\fi
 \ifH@\else\advance\firsty@\sydimen@\fi
 \iftX@
  \advance\secondy@\tXdimen@ii
  \advance\secondx@\tXdimen@i
  \slope@
 \else
  \iftY@
   \advance\secondy@\tYdimen@ii
   \advance\secondx@\tYdimen@i
   \slope@
   \secondy@\secondx@\advance\secondy@-\firstx@
   \ifNESW@\else\multiply\secondy@\m@ne\fi
   \multiply\secondy@\tan@i\divide\secondy@\tan@ii\advance\secondy@\firsty@
  \else
   \ifa@
    \slope@
    \ifNESW@\global\advance\angcount@\exacount@\else
     \global\advance\angcount@-\exacount@\fi
    \ifnum\angcount@>23 \global\angcount@23 \fi
    \ifnum\angcount@<\@ne\global\angcount@\@ne\fi
    \slope@a\angcount@
    \ifY@
     \advance\secondy@\Ydimen@
    \else
     \ifX@
      \advance\secondx@\Xdimen@
      \dimen@\secondx@\advance\dimen@-\firstx@
      \ifNESW@\else\multiply\dimen@\m@ne\fi
      \multiply\dimen@\tan@i\divide\dimen@\tan@ii
      \advance\dimen@\firsty@\secondy@\dimen@
     \fi
    \fi
   \else
    \ifH@\else\ifV@\else\slope@\fi\fi
   \fi
  \fi
 \fi
 \ifH@\else\ifV@\else\ifsvertex@\else
  \dimen@6\p@\multiply\dimen@\tan@ii
  \count@\tan@i\advance\count@\tan@ii\divide\dimen@\count@
  \ifE@\advance\firstx@\dimen@\else\advance\firstx@-\dimen@\fi
  \multiply\dimen@\tan@i\divide\dimen@\tan@ii
  \ifN@\advance\firsty@\dimen@\else\advance\firsty@-\dimen@\fi
 \fi\fi\fi
 \ifp@
  \ifH@\else\ifV@\else
   \getcos@\pdimen@\advance\firsty@\dimen@\advance\secondy@\dimen@
   \ifNESW@\advance\firstx@-\dimen@ii\else\advance\firstx@\dimen@ii\fi
  \fi\fi
 \fi
 \ifH@\else\ifV@\else
  \ifnum\tan@i>\tan@ii
   \charht@\ten@\p@\charwd@\ten@\p@
   \multiply\charwd@\tan@ii\divide\charwd@\tan@i
  \else
   \charwd@\ten@\p@\charht@\ten@\p@
   \divide\charht@\tan@ii\multiply\charht@\tan@i
  \fi
  \ifnum\tcount@=\thr@@
   \ifN@\advance\secondy@-.3\charht@\else\advance\secondy@.3\charht@\fi
  \fi
  \ifnum\scount@=\tw@
   \ifE@\advance\firstx@.3\charht@\else\advance\firstx@-.3\charht@\fi
  \fi
  \ifnum\tcount@=12
   \ifN@\advance\secondy@-\charht@\else\advance\secondy@\charht@\fi
  \fi
  \iftY@
  \else
   \ifa@
    \ifX@
    \else
     \secondx@\secondy@\advance\secondx@-\firsty@
     \ifNESW@\else\multiply\secondx@\m@ne\fi
     \multiply\secondx@\tan@ii\divide\secondx@\tan@i
     \advance\secondx@\firstx@
    \fi
   \fi
  \fi
 \fi\fi
 \ifH@\harrow@\else\ifV@\varrow@\else\arrow@@\fi\fi}
\newdimen\mathaxis@
\mathaxis@90\p@\divide\mathaxis@36
\def\harrow@b{\ifE@\hskip\tocenter@\hskip\firstx@\fi}
\def\harrow@bb{\ifE@\hskip\xdimen@\else\hskip\Xdimen@\fi}
\def\harrow@e{\ifE@\else\hskip-\firstx@\hskip-\tocenter@\fi}
\def\harrow@ee{\ifE@\hskip-\Xdimen@\else\hskip-\xdimen@\fi}
\def\harrow@{\dimen@\secondx@\advance\dimen@-\firstx@
 \ifE@\let\next@\rlap\else\multiply\dimen@\m@ne\let\next@\llap\fi
 \next@{%
  \harrow@b
  \smash{\raise\pdimen@\hbox to\dimen@
   {\harrow@bb\arrow@ii
    \ifnum\arrcount@=\m@ne\else\ifnum\arrcount@=\thr@@\else
     \ifE@
      \ifnum\scount@=\m@ne
      \else
       \ifcase\scount@\or\or\char118 \or\char117 \or\or\or\char119 \or
       \char120 \or\char121 \or\char122 \or\or\or\arrow@i\char125 \or
       \char117 \hskip\thr@@\p@\char117 \hskip-\thr@@\p@\fi
      \fi
     \else
      \ifnum\tcount@=\m@ne
      \else
       \ifcase\tcount@\char117 \or\or\char117 \or\char118 \or\char119 \or
       \char120 \or\or\or\or\or\char121 \or\char122 \or\arrow@i\char125
       \or\char117 \hskip\thr@@\p@\char117 \hskip-\thr@@\p@\fi
      \fi
     \fi
    \fi\fi
    \dimen@\mathaxis@\advance\dimen@.2\p@
    \dimen@ii\mathaxis@\advance\dimen@ii-.2\p@
    \ifnum\arrcount@=\m@ne
     \let\leads@\null
    \else
     \ifcase\arrcount@
      \def\leads@{\hrule\height\dimen@\depth-\dimen@ii}\or
      \def\leads@{\hrule\height\dimen@\depth-\dimen@ii}\or
      \def\leads@{\hbox to\ten@\p@{%
       \leaders\hrule\height\dimen@\depth-\dimen@ii\hfil
       \hfil
      \leaders\hrule\height\dimen@\depth-\dimen@ii\hskip\z@ plus2fil\relax
       \hfil
       \leaders\hrule\height\dimen@\depth-\dimen@ii\hfil}}\or
     \def\leads@{\hbox{\hbox to\ten@\p@{\dimen@\mathaxis@\advance\dimen@1.2\p@
       \dimen@ii\dimen@\advance\dimen@ii-.4\p@
       \leaders\hrule\height\dimen@\depth-\dimen@ii\hfil}%
       \kern-\ten@\p@
       \hbox to\ten@\p@{\dimen@\mathaxis@\advance\dimen@-1.2\p@
       \dimen@ii\dimen@\advance\dimen@ii-.4\p@
       \leaders\hrule\height\dimen@\depth-\dimen@ii\hfil}}}\fi
    \fi
    \cleaders\leads@\hfil
    \ifnum\arrcount@=\m@ne\else\ifnum\arrcount@=\thr@@\else
     \arrow@i
     \ifE@
      \ifnum\tcount@=\m@ne
      \else
       \ifcase\tcount@\char119 \or\or\char119 \or\char120 \or\char121 \or
       \char122 \or \or\or\or\or\char123 \or\char124 \or
       \char125 \or\char119 \hskip-\thr@@\p@\char119 \hskip\thr@@\p@\fi
      \fi
     \else
      \ifcase\scount@\or\or\char120 \or\char119 \or\or\or\char121 \or\char122
      \or\char123 \or\char124 \or\or\or\char125 \or
      \char119 \hskip-\thr@@\p@\char119 \hskip\thr@@\p@\fi
     \fi
    \fi\fi
    \harrow@ee}}%
  \harrow@e}%
 \iflabel@i
  \dimen@ii\z@\setbox\ZER@\hbox{$\m@th\tsize@@\label@i$}%
  \ifnum\arrcount@=\m@ne
  \else
   \advance\dimen@ii\mathaxis@
   \advance\dimen@ii\dp\ZER@\advance\dimen@ii\tw@\p@
   \ifnum\arrcount@=\thr@@\advance\dimen@ii\tw@\p@\fi
  \fi
  \advance\dimen@ii\pdimen@
  \next@{\harrow@b\smash{\raise\dimen@ii\hbox to\dimen@
   {\harrow@bb\hskip\tw@\ldimen@i\hfil\box\ZER@\hfil\harrow@ee}}\harrow@e}%
 \fi
 \iflabel@ii
  \ifnum\arrcount@=\m@ne
  \else
   \setbox\ZER@\hbox{$\m@th\tsize@\label@ii$}%
   \dimen@ii-\ht\ZER@\advance\dimen@ii-\tw@\p@
   \ifnum\arrcount@=\thr@@\advance\dimen@ii-\tw@\p@\fi
   \advance\dimen@ii\mathaxis@\advance\dimen@ii\pdimen@
   \next@{\harrow@b\smash{\raise\dimen@ii\hbox to\dimen@
    {\harrow@bb\hskip\tw@\ldimen@ii\hfil\box\ZER@\hfil\harrow@ee}}\harrow@e}%
  \fi
 \fi}
\let\tsize@\tsize
\def\tsizeCDlabels{\let\tsize@\tsize}
\def\ssizeCDlabels{\let\tsize@\ssize}
\def\tsize@@{\ifnum\arrcount@=\m@ne\else\tsize@\fi}
\def\varrow@{\dimen@\secondy@\advance\dimen@-\firsty@
 \ifN@\else\multiply\dimen@\m@ne\fi
 \setbox\ZER@\vbox to\dimen@
  {\ifN@\vskip-\Ydimen@\else\vskip\ydimen@\fi
   \ifnum\arrcount@=\m@ne\else\ifnum\arrcount@=\thr@@\else
    \hbox{\arrow@iii
     \ifN@
      \ifnum\tcount@=\m@ne
      \else
       \ifcase\tcount@\char117 \or\or\char117 \or\char118 \or\char119 \or
       \char120 \or\or\or\or\or\char121 \or\char122 \or\char123 \or
       \vbox{\hbox{\char117}\nointerlineskip\vskip\thr@@\p@
       \hbox{\char117}\vskip-\thr@@\p@}\fi
      \fi
     \else
      \ifcase\scount@\or\or\char118 \or\char117 \or\or\or\char119 \or
      \char120 \or\char121 \or\char122 \or\or\or\char123 \or
      \vbox{\hbox{\char117}\nointerlineskip\vskip\thr@@\p@
      \hbox{\char117}\vskip-\thr@@\p@}\fi
     \fi}%
    \nointerlineskip
   \fi\fi
   \ifnum\arrcount@=\m@ne
    \let\leads@\null
   \else
    \ifcase\arrcount@\let\leads@\vrule\or\let\leads@\vrule\or
    \def\leads@{\vbox to\ten@\p@{%
     \hrule\height1.67\p@\depth\z@\width.4\p@
     \vfil
     \hrule\height3.33\p@\depth\z@\width.4\p@
     \vfil
     \hrule\height1.67\p@\depth\z@\width.4\p@}}\or
    \def\leads@{\hbox{\vrule\height\p@\hskip\tw@\p@\vrule}}\fi
   \fi
  \cleaders\leads@\vfill\nointerlineskip
   \ifnum\arrcount@=\m@ne\else\ifnum\arrcount@=\thr@@\else
    \hbox{\arrow@iv
     \ifN@
      \ifcase\scount@\or\or\char118 \or\char117 \or\or\or\char119 \or
      \char120 \or\char121 \or\char122 \or\or\or\arrow@iii\char123 \or
      \vbox{\hbox{\char117}\nointerlineskip\vskip-\thr@@\p@
      \hbox{\char117}\vskip\thr@@\p@}\fi
     \else
      \ifnum\tcount@=\m@ne
      \else
       \ifcase\tcount@\char117 \or\or\char117 \or\char118 \or\char119 \or
       \char120 \or\or\or\or\or\char121 \or\char122 \or\arrow@iii\char123 \or
       \vbox{\hbox{\char117}\nointerlineskip\vskip-\thr@@\p@
       \hbox{\char117}\vskip\thr@@\p@}\fi
      \fi
     \fi}%
   \fi\fi
   \ifN@\vskip\ydimen@\else\vskip-\Ydimen@\fi}%
 \ifN@
  \dimen@ii\firsty@
 \else
  \dimen@ii-\firsty@\advance\dimen@ii\ht\ZER@\multiply\dimen@ii\m@ne
 \fi
 \rlap{\smash{\hskip\tocenter@\hskip\pdimen@\raise\dimen@ii\box\ZER@}}%
 \iflabel@i
  \setbox\ZER@\vbox to\dimen@{\vfil
   \hbox{$\m@th\tsize@@\label@i$}\vskip\tw@\ldimen@i\vfil}%
  \rlap{\smash{\hskip\tocenter@\hskip\pdimen@
  \ifnum\arrcount@=\m@ne\let\next@\relax\else\let\next@\llap\fi
  \next@{\raise\dimen@ii\hbox{\ifnum\arrcount@=\m@ne\hskip-.5\wd\ZER@\fi
   \box\ZER@\ifnum\arrcount@=\m@ne\else\hskip\tw@\p@\fi}}}}%
 \fi
 \iflabel@ii
  \ifnum\arrcount@=\m@ne
  \else
   \setbox\ZER@\vbox to\dimen@{\vfil
    \hbox{$\m@th\tsize@\label@ii$}\vskip\tw@\ldimen@ii\vfil}%
   \rlap{\smash{\hskip\tocenter@\hskip\pdimen@
   \rlap{\raise\dimen@ii\hbox{\ifnum\arrcount@=\thr@@\hskip4.5\p@\else
    \hskip2.5\p@\fi\box\ZER@}}}}%
  \fi
 \fi
}
\newdimen\goal@
\newdimen\shifted@
\newcount\Tcount@
\newcount\Scount@
\newbox\shaft@
\newcount\slcount@
\def\getcos@#1{%
 \ifnum\tan@i<\tan@ii
  \dimen@#1%
  \ifnum\slcount@<8 \count@9 \else \ifnum\slcount@<12 \count@8 \else
   \count@7 \fi\fi
  \multiply\dimen@\count@\divide\dimen@\ten@
  \dimen@ii\dimen@\multiply\dimen@ii\tan@i\divide\dimen@ii\tan@ii
 \else
  \dimen@ii#1%
  \count@-\slcount@\advance\count@24
  \ifnum\count@<8 \count@9 \else \ifnum\count@<12 \count@8
   \else\count@7 \fi\fi
  \multiply\dimen@ii\count@\divide\dimen@ii\ten@
  \dimen@\dimen@ii\multiply\dimen@\tan@ii\divide\dimen@\tan@i
 \fi}
\newdimen\adjust@
\def\Nnext@{\ifN@\let\next@\raise\else\let\next@\lower\fi}
\def\arrow@@{\slcount@\angcount@
 \ifNESW@
  \ifnum\angcount@<\ten@
   \let\arrowfont@\arrow@i\global\advance\angcount@\m@ne
   \global\multiply\angcount@13
  \else
   \ifnum\angcount@<19
    \let\arrowfont@\arrow@ii\global\advance\angcount@-\ten@
    \global\multiply\angcount@13
   \else
    \let\arrowfont@\arrow@iii\global\advance\angcount@-19
    \global\multiply\angcount@13
  \fi\fi
  \Tcount@\angcount@
 \else
  \ifnum\angcount@<5
   \let\arrowfont@\arrow@iii\global\advance\angcount@\m@ne
   \global\multiply\angcount@13 \global\advance\angcount@65
  \else
   \ifnum\angcount@<14
    \let\arrowfont@\arrow@iv\global\advance\angcount@-5
    \global\multiply\angcount@13
   \else
    \ifnum\angcount@<23
     \let\arrowfont@\arrow@v\global\advance\angcount@-14
     \global\multiply\angcount@13
    \else
     \let\arrowfont@\arrow@i\global\angcount@117
  \fi\fi\fi
  \ifnum\angcount@=117 \Tcount@115 \else\Tcount@\angcount@\fi
 \fi
 \Scount@\Tcount@
 \ifE@
  \ifnum\tcount@=\z@\advance\Tcount@\tw@\else\ifnum\tcount@=13
   \advance\Tcount@\tw@\else\advance\Tcount@\tcount@\fi\fi
  \ifnum\scount@=\z@\else\ifnum\scount@=13 \advance\Scount@\thr@@\else
   \advance\Scount@\scount@\fi\fi
 \else
  \ifcase\tcount@\advance\Tcount@\thr@@\or\or\advance\Tcount@\thr@@\or
  \advance\Tcount@\tw@\or\advance\Tcount@6 \or\advance\Tcount@7
  \or\or\or\or\or\advance\Tcount@8 \or\advance\Tcount@9 \or
  \advance\Tcount@12 \or\advance\Tcount@\thr@@\fi
  \ifcase\scount@\or\or\advance\Scount@\thr@@\or\advance\Scount@\tw@\or
  \or\or\advance\Scount@4 \or\advance\Scount@5 \or\advance\Scount@\ten@
  \or\advance\Scount@11 \or\or\or\advance\Scount@12 \or\advance
  \Scount@\tw@\fi
 \fi
 \ifcase\arrcount@\or\or\global\advance\angcount@\@ne\else\fi
 \ifN@\shifted@\firsty@\else\shifted@-\firsty@\fi
 \ifE@\else\advance\shifted@\charht@\fi
 \goal@\secondy@\advance\goal@-\firsty@
 \ifN@\else\multiply\goal@\m@ne\fi
 \setbox\shaft@\hbox{\arrowfont@\char\angcount@}%
 \ifnum\arrcount@=\thr@@
  \getcos@{1.5\p@}%
  \setbox\shaft@\hbox to\wd\shaft@{\arrowfont@
   \rlap{\hskip\dimen@ii
    \smash{\ifNESW@\let\next@\lower\else\let\next@\raise\fi
     \next@\dimen@\hbox{\arrowfont@\char\angcount@}}}%
   \rlap{\hskip-\dimen@ii
    \smash{\ifNESW@\let\next@\raise\else\let\next@\lower\fi
      \next@\dimen@\hbox{\arrowfont@\char\angcount@}}}\hfil}%
 \fi
 \rlap{\smash{\hskip\tocenter@\hskip\firstx@
  \ifnum\arrcount@=\m@ne
  \else
   \ifnum\arrcount@=\thr@@
   \else
    \ifnum\scount@=\m@ne
    \else
     \ifnum\scount@=\z@
     \else
      \setbox\ZER@\hbox{\ifnum\angcount@=117 \arrow@v\else\arrowfont@\fi
       \char\Scount@}%
      \ifNESW@
       \ifnum\scount@=\tw@
        \dimen@\shifted@\advance\dimen@-\charht@
        \ifN@\hskip-\wd\ZER@\fi
        \Nnext@
        \next@\dimen@\copy\ZER@
        \ifN@\else\hskip-\wd\ZER@\fi
       \else
        \Nnext@
        \ifN@\else\hskip-\wd\ZER@\fi
        \next@\shifted@\copy\ZER@
        \ifN@\hskip-\wd\ZER@\fi
       \fi
       \ifnum\scount@=12
        \advance\shifted@\charht@\advance\goal@-\charht@
        \ifN@\hskip\wd\ZER@\else\hskip-\wd\ZER@\fi
       \fi
       \ifnum\scount@=13
        \getcos@{\thr@@\p@}%
        \ifN@\hskip\dimen@\else\hskip-\wd\ZER@\hskip-\dimen@\fi
        \adjust@\shifted@\advance\adjust@\dimen@ii
        \Nnext@
        \next@\adjust@\copy\ZER@
        \ifN@\hskip-\dimen@\hskip-\wd\ZER@\else\hskip\dimen@\fi
       \fi
      \else
       \ifN@\hskip-\wd\ZER@\fi
       \ifnum\scount@=\tw@
        \ifN@\hskip\wd\ZER@\else\hskip-\wd\ZER@\fi
        \dimen@\shifted@\advance\dimen@-\charht@
        \Nnext@
        \next@\dimen@\copy\ZER@
        \ifN@\hskip-\wd\ZER@\fi
       \else
        \Nnext@
        \next@\shifted@\copy\ZER@
        \ifN@\else\hskip-\wd\ZER@\fi
       \fi
       \ifnum\scount@=12
        \advance\shifted@\charht@\advance\goal@-\charht@
        \ifN@\hskip-\wd\ZER@\else\hskip\wd\ZER@\fi
       \fi
       \ifnum\scount@=13
        \getcos@{\thr@@\p@}%
        \ifN@\hskip-\wd\ZER@\hskip-\dimen@\else\hskip\dimen@\fi
        \adjust@\shifted@\advance\adjust@\dimen@ii
        \Nnext@
        \next@\adjust@\copy\ZER@
        \ifN@\hskip\dimen@\else\hskip-\dimen@\hskip-\wd\ZER@\fi
       \fi
      \fi
  \fi\fi\fi\fi
  \ifnum\arrcount@=\m@ne
  \else
   \loop
    \ifdim\goal@>\charht@
    \ifE@\else\hskip-\charwd@\fi
    \Nnext@
    \next@\shifted@\copy\shaft@
    \ifE@\else\hskip-\charwd@\fi
    \advance\shifted@\charht@\advance\goal@-\charht@
    \repeat
   \ifdim\goal@>\z@
    \dimen@\charht@\advance\dimen@-\goal@
    \divide\dimen@\tan@i\multiply\dimen@\tan@ii
    \ifE@\hskip-\dimen@\else\hskip-\charwd@\hskip\dimen@\fi
    \adjust@\shifted@\advance\adjust@-\charht@\advance\adjust@\goal@
    \Nnext@
    \next@\adjust@\copy\shaft@
    \ifE@\else\hskip-\charwd@\fi
   \else
    \adjust@\shifted@\advance\adjust@-\charht@
   \fi
  \fi
  \ifnum\arrcount@=\m@ne
  \else
   \ifnum\arrcount@=\thr@@
   \else
    \ifnum\tcount@=\m@ne
    \else
     \setbox\ZER@
      \hbox{\ifnum\angcount@=117 \arrow@v\else\arrowfont@\fi\char\Tcount@}%
     \ifnum\tcount@=\thr@@
      \advance\adjust@\charht@
      \ifE@\else\ifN@\hskip-\charwd@\else\hskip-\wd\ZER@\fi\fi
     \else
      \ifnum\tcount@=12
       \advance\adjust@\charht@
       \ifE@\else\ifN@\hskip-\charwd@\else\hskip-\wd\ZER@\fi\fi
      \else
       \ifE@\hskip-\wd\ZER@\fi
     \fi\fi
     \Nnext@
     \next@\adjust@\copy\ZER@
     \ifnum\tcount@=13
      \hskip-\wd\ZER@
      \getcos@{\thr@@\p@}%
      \ifE@\hskip-\dimen@\else\hskip\dimen@\fi
      \advance\adjust@-\dimen@ii
      \Nnext@
      \next@\adjust@\box\ZER@
     \fi
  \fi\fi\fi}}%
 \iflabel@i
  \rlap{\hskip\tocenter@
  \dimen@\firstx@\advance\dimen@\secondx@\divide\dimen@\tw@
  \advance\dimen@\ldimen@i
  \dimen@ii\firsty@\advance\dimen@ii\secondy@\divide\dimen@ii\tw@
  \global\multiply\ldimen@i\tan@i\global\divide\ldimen@i\tan@ii
  \ifNESW@\advance\dimen@ii\ldimen@i\else\advance\dimen@ii-\ldimen@i\fi
  \setbox\ZER@\hbox{\ifNESW@\else\ifnum\arrcount@=\thr@@\hskip4\p@\else
   \hskip\tw@\p@\fi\fi
   $\m@th\tsize@@\label@i$\ifNESW@\ifnum\arrcount@=\thr@@\hskip4\p@\else
   \hskip\tw@\p@\fi\fi}%
  \ifnum\arrcount@=\m@ne
   \ifNESW@\advance\dimen@.5\wd\ZER@\advance\dimen@\p@\else
    \advance\dimen@-.5\wd\ZER@\advance\dimen@-\p@\fi
   \advance\dimen@ii-.5\ht\ZER@
  \else
   \advance\dimen@ii\dp\ZER@
   \ifnum\slcount@<6 \advance\dimen@ii\tw@\p@\fi
  \fi
  \hskip\dimen@
  \ifNESW@\let\next@\llap\else\let\next@\rlap\fi
  \next@{\smash{\raise\dimen@ii\box\ZER@}}}%
 \fi
 \iflabel@ii
  \ifnum\arrcount@=\m@ne
  \else
   \rlap{\hskip\tocenter@
   \dimen@\firstx@\advance\dimen@\secondx@\divide\dimen@\tw@
   \ifNESW@\advance\dimen@\ldimen@ii\else\advance\dimen@-\ldimen@ii\fi
   \dimen@ii\firsty@\advance\dimen@ii\secondy@\divide\dimen@ii\tw@
   \global\multiply\ldimen@ii\tan@i\global\divide\ldimen@ii\tan@ii
   \advance\dimen@ii\ldimen@ii
   \setbox\ZER@\hbox{\ifNESW@\ifnum\arrcount@=\thr@@\hskip4\p@\else
    \hskip\tw@\p@\fi\fi
    $\m@th\tsize@\label@ii$\ifNESW@\else\ifnum\arrcount@=\thr@@\hskip4\p@
    \else\hskip\tw@\p@\fi\fi}%
   \advance\dimen@ii-\ht\ZER@
   \ifnum\slcount@<9 \advance\dimen@ii-\thr@@\p@\fi
   \ifNESW@\let\next@\rlap\else\let\next@\llap\fi
   \hskip\dimen@\next@{\smash{\raise\dimen@ii\box\ZER@}}}%
  \fi
 \fi
}
\def\outCD@#1{\def#1{\Err@{\noexpand#1must not be used within \string\CD}}}
\newskip\preCDskip@
\newskip\postCDskip@
\preCDskip@\z@
\postCDskip@\z@
\def\preCDspace#1{\RIfMIfI@
 \onlydmatherr@\preCDspace\else\advance\preCDskip@#1\relax\fi\else
 \onlydmatherr@\preCDspace\fi}
\def\postCDspace#1{\RIfMIfI@
 \onlydmatherr@\postCDspace\else\advance\postCDskip@#1\relax\fi\else
 \onlydmatherr@\postCDspace\fi}
\def\predisplayspace#1{\RIfMIfI@
 \onlydmatherr@\predisplayspace\else
 \advance\abovedisplayskip#1\relax
 \advance\abovedisplayshortskip#1\relax\fi
 \else\onlydmatherr@\preCDspace\fi}
\def\postdisplayspace#1{\RIfMIfI@
 \onlydmatherr@\postdisplayspace\else
 \advance\belowdisplayskip#1\relax
 \advance\belowdisplayshortskip#1\relax\fi
 \else\onlydmatherr@\postdisplayspace\fi}
\def\PreCDSpace#1{\global\preCDskip@#1\relax}
\def\PostCDSpace#1{\global\postCDskip@#1\relax}
\def\CD#1\endCD{%
 \outCD@\cgaps\outCD@\rgaps\outCD@\Cgaps\outCD@\Rgaps
 \preCD@#1\endCD
 \advance\abovedisplayskip\preCDskip@
 \advance\abovedisplayshortskip\preCDskip@
 \advance\belowdisplayskip\postCDskip@
 \advance\belowdisplayshortskip\postCDskip@
 \vcenter{\offinterlineskip
  \vskip\preCDskip@\Let@\global\colcount@\@ne\global\rowcount@\z@
  \everycr{%
   \noalign{%
    \ifnum\rowcount@=\Rowcount@
    \else
     \getrgap@\rowcount@\vskip\getdim@
     \global\advance\rowcount@\@ne\global\colcount@\@ne
    \fi}}%
  \tabskip\z@
  \halign{&\global\xoff@\z@\global\yoff@\z@
   \getcgap@\colcount@\hskip\getdim@
   \hfil\vrule\height\ten@\p@\width\z@\depth\z@
   $\m@th\displaystyle{##}$\hfil
   \global\advance\colcount@\@ne\cr
   #1\crcr}\vskip\postCDskip@}%
 \preCDskip@\z@\postCDskip@\z@
 \def\getcgap@##1{\ifcase##1\or\getdim@\z@\else\getdim@\standardcgap\fi}%
 \def\getrgap@##1{\ifcase##1\getdim@\z@\else\getdim@\standardrgap\fi}%
 \let\Width@\relax\let\Height@\relax\let\Depth@\relax\let\Rowheight@\relax
 \let\Rowdepth@\relax\let\Colwidth@\relax
}

\def\alloc@#1#2#3#4#5{\global\advance\count1#1by\@ne
  \ch@ck#1#4#2%
  \allocationnumber=\count1#1%
  \global#3#5=\allocationnumber
  \wlog{\string#5=\string#2\the\allocationnumber}}
\catcode`\@=\active

\catcode`\"=12
\font\black=cmbx10
\font\sblack=cmbx7
\font\ssblack=cmbx5
\font\blackital=cmmib10  \skewchar\blackital='177
\font\sblackital=cmmib7  \skewchar\sblackital='177
\font\ssblackital=cmmib5  \skewchar\ssblackital='177
\font\sanss=cmss10
\font\ssanss=cmss8 scaled 900
\font\sssanss=cmss8 scaled 600
\font\blackboard=msbm10
\font\sblackboard=msbm7
\font\ssblackboard=msbm5
\font\caligr=eusm10
\font\scaligr=eusm7
\font\sscaligr=eusm5

\font\fraktur=eufm10
\font\sfraktur=eufm7
\font\ssfraktur=eufm5

\font\bsymb=cmsy10 scaled\magstep2
\def\all#1{\setbox0=\hbox{\lower1.5pt\hbox{\bsymb
       \char"38}}\setbox1=\hbox{$_{#1}$} \box0\lower2pt\box1\;}
\def\exi#1{\setbox0=\hbox{\lower1.5pt\hbox{\bsymb \char"39}}
       \setbox1=\hbox{$_{#1}$} \box0\lower2pt\box1\;}

\def\tx#1{{\fam0\relax#1}}

\newfam\bifam
\textfont\bifam=\blackital
\scriptfont\bifam=\sblackital
\scriptscriptfont\bifam=\ssblackital
\def\bi#1{{\fam\bifam\relax#1}}

\newfam\blfam
\textfont\blfam=\black
\scriptfont\blfam=\sblack
\scriptscriptfont\blfam=\ssblack

\newfam\bbfam
\textfont\bbfam=\blackboard
\scriptfont\bbfam=\sblackboard
\scriptscriptfont\bbfam=\ssblackboard
\def\bb#1{{\fam\bbfam\relax#1}}

\newfam\ssfam
\textfont\ssfam=\sanss
\scriptfont\ssfam=\ssanss
\scriptscriptfont\ssfam=\sssanss
\def\ss#1{{\fam\ssfam\relax#1}}

\newfam\clfam
\textfont\clfam=\caligr
\scriptfont\clfam=\scaligr
\scriptscriptfont\clfam=\sscaligr
\def\cl#1{{\fam\clfam\relax#1}}

\newfam\frfam
\textfont\frfam=\fraktur
\scriptfont\frfam=\sfraktur
\scriptscriptfont\frfam=\ssfraktur
\def\fr#1{{\fam\frfam\relax#1}}

\def\hpb#1{\setbox0=\hbox{${#1}$}
    \copy0 \kern-\wd0 \kern.2pt \box0}
\def\vpb#1{\setbox0=\hbox{${#1}$}
    \copy0 \kern-\wd0 \raise.08pt \box0}

\def\pmb#1{\setbox0\hbox{${#1}$} \copy0 \kern-\wd0 \kern.2pt \box0}
\def\pmbb#1{\setbox0\hbox{${#1}$} \copy0 \kern-\wd0
      \kern.2pt \copy0 \kern-\wd0 \kern.2pt \box0}
\def\pmbbb#1{\setbox0\hbox{${#1}$} \copy0 \kern-\wd0
      \kern.2pt \copy0 \kern-\wd0 \kern.2pt
    \copy0 \kern-\wd0 \kern.2pt \box0}
\def\pmxb#1{\setbox0\hbox{${#1}$} \copy0 \kern-\wd0
      \kern.2pt \copy0 \kern-\wd0 \kern.2pt
      \copy0 \kern-\wd0 \kern.2pt \copy0 \kern-\wd0 \kern.2pt \box0}
\def\pmxbb#1{\setbox0\hbox{${#1}$} \copy0 \kern-\wd0 \kern.2pt
      \copy0 \kern-\wd0 \kern.2pt
      \copy0 \kern-\wd0 \kern.2pt \copy0 \kern-\wd0 \kern.2pt
      \copy0 \kern-\wd0 \kern.2pt \box0}

\mathchardef\za="710B  
\mathchardef\zb="710C  
\mathchardef\zg="710D  
\mathchardef\zd="710E  
\mathchardef\zve="710F 
\mathchardef\zz="7110  
\mathchardef\zh="7111  
\mathchardef\zvy="7112 
\mathchardef\zi="7113  
\mathchardef\zk="7114  
\mathchardef\zl="7115  
\mathchardef\zm="7116  
\mathchardef\zn="7117  
\mathchardef\zx="7118  
\mathchardef\zp="7119  
\mathchardef\zr="711A  
\mathchardef\zs="711B  
\mathchardef\zt="711C  
\mathchardef\zu="711D  
\mathchardef\zvf="711E 
\mathchardef\zq="711F  
\mathchardef\zc="7120  
\mathchardef\zw="7121  
\mathchardef\ze="7122  
\mathchardef\zy="7123  
\mathchardef\zvp="7124 
\mathchardef\zvr="7125 
\mathchardef\zvs="7126 
\mathchardef\zf="7127  
\mathchardef\zG="7000  
\mathchardef\zD="7001  
\mathchardef\zY="7002  
\mathchardef\zL="7003  
\mathchardef\zX="7004  
\mathchardef\zP="7005  
\mathchardef\zS="7006  
\mathchardef\zU="7007  
\mathchardef\zF="7008  
\mathchardef\zC="7009  
\mathchardef\zW="700A  

\catcode`\"=\active


\loadmsam
\loadmsbm
\newsymbol\leqslant 1336
\newsymbol\geqslant 133E
\newsymbol\centerdot 1205
\newsymbol\shortmid 2370

\def\leqs{\leqslant}
\def\geqs{\geqslant}
\def\fpr#1{\underset{{#1}}\to\times}

\newcounter\secno
\newcounter\ssecno

\define\sect#1{\Reset\ssecno1\bigpagebreak
	\flushpar {\secno.}\,{\bf #1}\vskip1.2mm}
\newfontstyle\secno{\bf}

\define\ssca#1{\Offset\secno0\bigpagebreak\vskip-4mm
	\flushpar {\secno.\ssecno.}\,{\bf #1}\vskip1.2mm}
\newfontstyle\ssecno{\bf}

\define\sscx#1{\Offset\secno0\bigpagebreak
	\flushpar {\secno.\ssecno.}\,{\bf #1}\vskip1.2mm}
\newfontstyle\ssecno{\bf}

\def\*{{\textstyle *}}

\def\polar{{\textstyle\circ}}
\newsymbol\blacktriangle 104E

\def\proof{\demo{Proof}}
\def\endproof{\hfill \vrule height4pt width6pt depth2pt \enddemo}

\def\N{{\bb N}}
\def\R{{\bb R}}

\def\K{{\bb K}}

\def\ssT{{\scriptscriptstyle {\ss T}}}

\def\by{{\bi y}}

\def\bc{{\bi c}}
\def\bd{{\bi d}}

\def\bL{{\bi L}}
\def\bV{{\bi V}}

\def\bT{{\bi T}}
\def\bC{{\bi C}}

\def\bq{{\bi q}}
\def\bQ{{\bi Q}}

\def\sA{{\ss A}}
\def\sC{{\ss C}}
\def\sD{{\ss D}}
\def\sG{{\ss G}}

\def\sI{{\ss I}}
\def\sJ{{\ss J}}
\def\sK{{\ss K}}
\def\sL{{\ss L}}

\def\sP{{\ss P}}

\def\sT{{\ss T}}
\def\sV{{\ss V}}

\def\st{{\ss t}}
\def\sg{{\ss g}}

\def\sv{{\ss v}}

\def\sQ{{\ss Q}}
\def\sj{{\ss j}}
\def\sq{{\ss q}}

\def\xd{\tx{d}}
\def\xi{\tx{i}}
\def\xD{\tx{D}}

\def\Sup{\operatorname{Sup}}
\def\Dm{\operatorname{Dom}}
\def\dim{\operatorname{dim}}

\def\cD{{\cl D}}
\def\cE{{\cl E}}
\def\cF{{\cl F}}
\def\cG{{\cl G}}

\def\cK{{\cl K}}

\def\cP{{\cl P}}

\def\cT{{\cl T}}
\def\cQ{{\cl Q}}

\input paper.st\relax
\hsize=37pc
\hoffset=-10pt
\vsize=52pc
\voffset=6pt
\TagsOnRight
\document
\input xy
\xyoption{all}

\def\oT{\overset\circ\to\sT}

\def\fc{{\fr c}}

\def\gzmq{{q^\sharp}}

\def\wQ{{\widetilde Q}}

\def\wS{{\widetilde S}}

\def\of{{\bar f}}

\def\oQ{{\overline Q}}
\def\oU{{\overline U}}
\def\oS{{\overline S}}

\def\bc{\bold c}
\def\bd{\bold d}
\def\lq{\bold q}
\def\lg{\bold g}
\def\bg{\pmbb\sg}
\def\bl{\pmbb{\ss l}}
\def\bC{\sC}
\def\bq{\pmbb\sq}
\def\bt{\pmbb\st}

\def\bT{\pmbb\sT}
\def\bV{\pmbb\sV}
\def\bQ{\pmbb\sQ}
\def\bK{\pmbb\sK}
\def\bL{\pmbb\sL}
\def\bG{\pmbb\sG}
\def\bD{\pmbb\sD}
\def\NM{{(N|M)}}
\def\NMc{{(N|M,\bc)}}
\def\bzg{\pmbb\zg}
\def\bzl{\pmbb\zk}

\def\bzp{\pmbb\zp}
\def\bzt{\pmbb\zt}

\def\fc{{\fr c}}

\def\tW{{\tilde\wedge}}
\def\tzp{{\tilde\zp}}

\def\ux{\setbox0\hbox{$\times$} \copy0 \kern-\wd0 \kern.3mm \box0}


    \title
        The origin of variational principles.
    \endtitle

    \author
        W\l odzimierz M. Tulczyjew \\
        Dipartimento di Matematica e Fisica \\
        Universit\`a di Camerino \\
        62032 Camerino, Italy \\
        Istituto Nazionale di Fisica Nucleare,
        Sezione di Napoli, Italy \\
        {\tt tulczy\@libero.it} \\

    \endauthor
        \thanks{Supported by PRIN SINTESI}
    \maketitle

        \sect{Introduction.}
    This note presents an attempt to provide a conceptual framework for variational formulations of classical physics.  It
is based on a study of static systems in collaboration with P. Urbanski of Warsaw.

    Variational principles of physics have all a common source in the {\it principle of virtual work} well known in statics
of machanical systems.  This principle is presented here as the first step in characterizing local stable equilibria of
static systems.  An extended analysis of local equilibria is given for systems with configuration manifolds of finite
dimensions.  Numerous examples of the principle of virtual work and the Legendre transformation applied to static
mechanical systems are provided.  Configuration spaces for the dynamics of autonomous mechanical systems and for statics of
continua are constructed in the final sections.  These configuration spaces are not differential manifolds.

    The final section of the present note is an initial part of a programme of geometric formulations of the foundations of
statics of continua undertaken jointly with F. Cardin of Padova in collaboration with P. Urbanski.

        \sect{Configurations, processes, work.}

    Experimental study of physical systems consists in subjecting a system to external control and observing the results.
A theoretical description of the system should have the power to predict observed results.  The formulation of a
theoretical description must start with assigning to the system parameters characterizing its configuration.  These
parameters do not ever exhaust all characteristics of the system.  The choice is dictated by the interactions to which the
system will be subjected. Statics of mechanical systems assigns to a mechanical system points in a differential manifold
characterizing the geometrical placement of the system in the physical space.  The points are translated in parameters by
means of coordinate systems.  The differential manifold is called the {\it configuration space}.  Control of a static
mechanical system consists in coupling the system with external devices with the same configuration space and observing the
resulting configurations.  At least two configuration spaces are involved.  The {\it control configuration space}
represents degrees of freedom which are being controlled.  The observed configurations are elements of the {\it observed
configuration space}.  The two spaces are not necessarily the same.  The configuration spaces considered represent
configurations of stable local equilibrium.

    It is sometimes necessary to redefine the control configuration space.  It is sometimes found that the same control
arrangements do not result in the same response.  When this happens we must discover and include in the description
additional hidden degrees of freedom participating in the process of control.  The configuration space with these degrees
of freedom included is the {\it internal configuration space}.  We will assume initially that the inclusion of additional
degrees of freedom is not necessary and that the internal configuration space is the same as the control configuration
space.

    An important form of control consists in subjecting a system to a {\it quasi static process} by coupling it to an
external device with slowly changing configuration.  The changes must be slow enough to justify the treatment of a process
as a succession of configurations of static equilibrium configurations rather than a dynamic event. Criteria of equilibrium
are based on the response to quasi static processes.

    Quasistatic processes are represented as oriented embedded arcs in the configuration space.  Let $Q$ be a differential
manifold of dimension $m$ representing the configuration space of a static system.  An {\it embedded arc} is a subset $\bc
\subset  Q$ which is the image of an embedding $\zg \colon [0,a] \rightarrow Q$ of a closed interval $[0,a] \subset \R$
defined as the restriction to $[0,a]$ of an embedding $\widetilde\zg \colon \R \rightarrow Q$.  The embedding $\zg$ is a
representative of an embedded arc called its {\it parameterization}.  Different embeddings may be parameterizations of the
same arc.  A parameterization $\zg \colon [0,a] \rightarrow Q$ of an arc $\bc$ induces an {\it orientation} of the arc. The
arc is oriented from $\zg(0)$ to $\zg(a)$ if the two points are distinct.  If $\zg(0) = \zg(a)$, then the arc is a cycle.
We will exclude cyclic processes from the discussion.  They are of no interest for developing criteria of equilibrium.  Two
parameterizations $\zg$ and $\zg'$ induce the same orientation if $\zg'(0) = \zg(0)$.  The {\it boundary} $\partial\bc$ of
an arc $\bc$ is a set of two points $q$ and $q'$ such that $\zg(0) = q$ and $\zg(a) = q'$ for some parameterization $\zg
\colon [0,a] \rightarrow Q$ of $\bc$. The designation of one of the boundary points as the {\it initial configuration} of
the process represented by an arc specifies an orientation.  An {\it oriented embedded arc} is defined as a pair
$(\bc,q_0)$, where $\bc$ is an embedded arc and $q_0$ is a point in the boundary $\partial\bc$ designated as the initial
configuration.  The remaining boundary point is the {\it terminal configuration} of $(\bc,q_0)$.  A parameterization  $\zg
\colon [0,a] \rightarrow Q$ of an arc $\bc$ such that $\zg(0) = q_0$ is said to be {\it compatible} with the orientation of
the oriented arc $(\bc,q_0)$. Compatible parameterizations will be used for oriented arcs.


    Let $(\bc,q_0)$ be an oriented arc.  We refer to points in $\bc$ as elements of $(\bc,q_0)$.  Elements of an oriented
arc $(\bc,q_0)$ are ordered by the relation $\leqs$ defined in terms of any parameterization $\zg$ compatible with the
orientation.  The relation $\zg(s') \leqs \zg(s)$ is equivalent to $s' \leqs s$.  Relations $<$, $\geqs$, and $>$ are
defined in a similar way.  They reflect the corresponding relations between the values of the parameter.

    Let $(\bc,q_0)$ and $(\bc',q'_0)$ be oriented arcs such that  $\bc' \subset \bc$.  There are two ordering relations in
$(\bc',q'_0)$ since elements of $(\bc',q'_0)$ are also elements of $(\bc,q_0)$.  The arc $(\bc',q'_0)$ is said to be {\it
included} in the arc $(\bc,q_0)$ if the two ordering relations coincide.  The inclusion relation is denoted by $(\bc',q'_0)
\subset (\bc,q_0)$.  The process represented by $(\bc',q'_0)$ is a {\it subprocess} of the process represented by
$(\bc,q_0)$ if $(\bc',q'_0) \subset (\bc,q_0)$.  If $(\bc^1,q^1_0)$ and $(\bc^2,q^2_0)$ are oriented arcs and if one of the
pairs $(\bc^1 \cup \bc^2,q^1_0)$ or $(\bc^1 \cup \bc^2,q^2_0)$ is an oriented arc, then it is considered the {\it union} of
$(\bc^1,q^1_0)$ and $(\bc^2,q^2_0)$ and is denoted by $(\bc^1,q^1_0) \cup (\bc^2,q^2_0)$.  Similarly, if one of the pairs
$(\bc^1 \cap \bc^2,q^1_0)$ or $(\bc^1 \cap \bc^2,q^2_0)$ is an oriented arc, then it is considered the {\it intersection}
of $(\bc^1,q^1_0)$ and $(\bc^2,q^2_0)$ and is denoted by $(\bc^1,q^1_0) \cap (\bc^2,q^2_0)$.

    Let $(\bc,q_0)$ be an oriented arc.  A function $a \colon \bc \rightarrow \R$ is considered a function on $(\bc,q_0)$.
This function is said to be {\it differentiable} if it the restriction to $\bc$ of a differentiable function on $Q$.  Let
$\sA Q$ be the algebra of differentiable functions on $Q$ and let $\sI_0(Q,\bc) \subset \sA Q$ be the ideal of functions
vanishing on $\bc$.  Two functions $f$ and $f'$ on $Q$ have the same restrictions to $\bc$ if and only if $f' - f \in
\sI_0(M,S)$.  It follows that the algebra $\sA(\bc,q_0)$ of differentiable functions on $(\bc,q_0)$ is canonically
isomorphic to the quotient algebra $\sA Q\big/\sI_0(Q,\bc)$.  The orientation of $(\bc,q_0)$ makes it possible to single
out increasing functions.  A function $h \in \sA\bc$ is {\it increasing} if $q' > q$ implies $h(q') > h(q)$.

    An oriented arc represents a quasi static process.  We use these two terms interchangeably.  We denote by $\pmbb\sP Q$
the set of all processes in $Q$.  An essential part of the characterization of a static system is the specification of a
set $\bC \subset \pmbb\sP Q$ of {\it admissible processes} in the configuration space $Q$.  Admissible processes are the
processes that can be actually induced using the control devices at our disposal.  The following conditions are satisfied.
        \list
    \item A subprocess of an admissible process is admissible.
    \item The union of admissible processes is admissible.
    \item If a process can be approximated with admissible processes, then it is admissible.
        \endlist

\noindent The last condition is made more precise by the following statement.  Let \;$\gzmq_1, \gzmq_2, \ldots, \gzmq_i,
\ldots$\; be a sequence of interior configurations in a process $(\bc,q_0)$ converging to the terminal configuration $q_1$.
If processes $(\bc_{\gzmq_i|},q_0)$ are admissible, then $(\bc,q_0)$ is admissible.  The symbol $(\bc_{q|},q_0)$ denotes
the subprocess of $(\bc,q_0)$ with $q \in \bc$ as the terminal configuration.  It is based on the arc
        $$\bc_{q|} = \left\{q' \in \bc;\; q' \leqs q \right\}.
                                                                                                        \tag \label{Fxt1}$$

    Another object characterizing a static system is a function
        $$W \colon \bC \rightarrow \R
                                                                                                        \tag \label{Fxt2}$$
    associating with each admissible process $(\bc,q_0)$ the {\it work} $W(\bc,q_0)$ of the process.  Since subprocesses of
admissible processes are admissible we can associate with each admissible process $(\bc,q_0)$ the function
        $$w_{(\bc,q_0)} \colon \bc \rightarrow \R \colon q \mapsto \cases W(\bc_{q|},q_0) &\text{ if } q \neq q_0 \\ 0
&\text{ if } q = q_0. \endcases
                                                                                                        \tag \label{Fxt3}$$
    The work function will be assumed to satisfy the following conditions.
        \list
    \item Work is additive in the sense that if $(\bc,q_0)$ is the union of admissible processes
        $$(\bc^0,q^0_0), (\bc^1,q^1_0),
\ldots ,(\bc^n,q^n_0)
                                                                                                        \tag \label{Fxt4}$$
     such that for $i = 0, 1, \dots ,n-1$ the terminal configuration of $(\bc^i,q^i_0)$ is the initial configuration of
$(\bc^{i+1},q^{i+1}_0)$, then
        $$W(\bc,q_0) = \sum_{i=0}^n W(\bc^i,q^i_0).
                                                                                                        \tag \label{Fxt5}$$
    \item For each admissible process $(\bc,q_0)$ the function $w_{(\bc,q_0)}$ is differentiable.
        \endlist

        \sect{An alternative definition of orientation.}

    For the purpose of constructing an orientation of an embedded arc $\bc$ we introduce the {\it tangent bundle} $\sT\bc$.
Let $\zg \colon [0,a] \rightarrow Q$ be a parameterization of $\bc$ and let $\widetilde\zg \colon \R \rightarrow Q$ be an
embedding such that $\zg = \widetilde\zg|[0,a]$.  The image $\widetilde\bc = \widetilde\zg(\R)$ is a submanifold of $Q$ and
has a well defined tangent bundle $\sT\widetilde\bc \subset \sT Q$.  The tangent bundle $\sT\bc$ is the restriction
        $$\sT\bc = \sT\widetilde\bc \cap \zt_Q^{-1}(\bc)
                                                                                                        \tag \label{Fxt6}$$
    of $\sT C$ to $\bc$.  The result of this construction is independent of the choice of the parameterization.  An {\it
orientation} of the arc $\bc$ is a subset $o \in \sT\bc$ with the following properties.
        \list
    \item $\zt_Q(o) = \bc$.
    \item If $v \in o$ then $kv \in o$ for each $k > 0$.
    \item If $v \in o$ then $kv \notin o$ for $k \geqs 0$.
    \item There is a parameterization $\zg \colon [0,a] \rightarrow Q$ such that $\st\zg(s) \in o$ for each $s \in ]0,a[$.
        \endlist
    The symbol $\st\zg(s)$ denotes the vector tangent to the curve $\zg$ at $\zg(s)$.  There are obviously two different
orientations of each embedded arc.  An {\it oriented embedded arc} is a pair $(\bc,o)$ of an embedded arc $\bc$ and one of
its orientations $o$.  Parameterizations of an arc satisfying the condition $(4)$ above are said to be {\it compatible}
with the orientation $o$.  A parameterization $\zg \colon [0,a] \rightarrow Q$ can be used to generate an orientation of an
arc. The set
        $$o = \left\{v \in \sT\bc ;\; \exi{s \in [0,a],\, k>0} v = k\st\zg(s) \right\}
                                                                                                        \tag \label{Fxt7}$$
    is an orientation of the arc $\bc = \zg([0,a])$ and the parameterization $\zg$ is compatible with this orientation.
Vectors $\st\zg(0)$ an $\st\zg(a)$ are obtained as limits.  Specifying the initial configuration $q_0$ determines an
orientation $o$ of the arc unless the terminal configuration is coincides with the initial configuration.  If such are
excluded, then the pair $(\bc,q_0)$ is a representation of the oriented arc equivalent to $(\bc,q_0)$.

        \sect{Stable local equilibrium configurations.}

    Let $C^0 \subset Q$ be the set of initial configurations of all admissible processes for a static system.  A point $q_0
\in C^0$ is called a {\it stable local equilibrium configuration} of the system if for each admissible process $(\bc,q_0)$
initiating at $q_0$  the function $w_{(\bc,q_0)}$ has a local minimum at $q_0$.

    The function $w_{(\bc,q_0)}$ will have a local minimum at $q_0$ if there is a point $q \in \bc$ such that the function
$w_{(\bc,q_0)}$ restricted to the arc
        $$\bc_{q|} = \left\{q' \in \bc;\; q' \leqs q \right\}
                                                                                                        \tag \label{Fxt8}$$
    is increasing.  This observation implies that equilibrium at $q_0$ is a property of the germ of $w_{(\bc,q_0)}$ at $q_0$.
We will state the definition of equilibrium in terms of germs.

    In the set $\pmbb\sP Q$ of oriented embedded arcs in $Q$ we introduce an equivalence relation.  Arcs $(\bc,q_0)$ and
$(\bc',q'_0)$ are equivalent if $q'_0 = q_0$ and if there is an open neighbourhood $U \subset Q$ of $q_0$ such that $\bc'
\cap U = \bc \cap U$.  Equivalence classes are called {\it germs} of processes.  The set of germs is denoted by $\sP^\fc
Q$.  The germ of a process $(\bc,q_0)$ is denoted by $\sj^\fc(\bc,q_0)$.  Germs of processes can be defined in terms of
ideals.  In the algebra $\sA Q$ we introduce an ideal $\sI_\fc(Q,q)$ for each $q \in Q$.  A function $f$ is in
$\sI_\fc(Q,q)$ if there is a neighbourhood $U \subset Q$ of $q$ such that $f|U = 0$.  Arcs $(\bc,q_0)$ and $(\bc',q'_0)$
are equivalent if
        $$\sI_\fc(Q,q'_0) + \sI_0(Q,\bc') = \sI_\fc(Q,q_0) + \sI_0(Q,\bc).
                                                                                                        \tag \label{Fxt9}$$
    This is the same equivalence relation as the one defined above. The symbol $C^\fc_{q_0}$ will denote the set of germs of
admissible processes initiating at $q_0$.

    With a process $(\bc,q_0)$ we associate an ideal $\sI_\fc(\bc,q_0)$ in the algebra $\sA(\bc,q_0)$ of differentiable
functions on $(\bc,q_0)$.  A function $h$ is in this ideal if there is a neighbourhood $V \in \bc$ of the initial
configuration $q_0$ such that $h|V = 0$.  The algebra $\sA(\bc,q_0)$ is canonically isomorphic to the quotient algebra $\sA
Q\big/\sI_0(Q,\bc)$ and the ideal $\sI_\fc(\bc,q_0)$ is isomorphic to the quotient
        $$\left(\sI_\fc(Q,q_0) + \sI_0(Q,\bc)\right)\big/\sI_0(Q,\bc).
                                                                                                        \tag \label{Fxt10}$$
    We will adopt the following identification
        $$\sA^\fc(\bc,q_0) = \sA(\bc,q_0)\big/\sI_\fc(\bc,q_0) \;\;\text{ is identified with }\;\;\sA
Q\big/\left(\sI_\fc(Q,q_0) + \sI_0(Q,\bc)\right).
                                                                                                        \tag \label{Fxt11}$$
    As a consequence of this identification the algebra $\sA^\fc(\bc,q_0)$ is associated with the germ $\sj^\fc(\bc,q_0)$
rather than with the process $(\bc,q_0)$ since the algebra $\sA^\fc(\bc',q'_0)$ is the same as the algebra
$\sA^\fc(\bc,q_0)$ if $(\bc',q'_0)$ and $(\bc,q_0)$ are equivalent.

    Each function $h \in \sA(\bc,q_0)$ has a germ $\sj^\fc h \in \sA^\fc(\bc,q_0)$.  The germ $\sj^\fc w_{(\bc,q_0)}$ of
the work function $w_{(\bc,q_0)}$ along an admissible process $(\bc,q_0)$ is the same as the germ $\sj^\fc w_{(\bc',q'_0)}$
if $(\bc',q'_0)$ and $(\bc,q_0)$ are equivalent.  An element of $\sA^\fc(\bc,q_0)$ is said to be {\it increasing} if it has
an increasing representative in $\sA(\bc,q_0)$.

    The criterion of stable local equilibrium is based on response of the system to the initial phase of a process.  It is
therefore correct to adopt the following definition of equilibrium formulated in terms of germs.

    A point $q_0 \in C^0$ is called a {\it stable local equilibrium configuration} if for each germ $\sj^\fc(\bc,q_0) \in
C^\fc_{q_0}$ the germ $\sj^\fc w_{(\bc,q_0)}$ is increasing.

    We have excluded constant processes represented by constant arcs.  In a complete discussion of equilibrium
configurations {\it isolated admissible configurations} and {\it admissible constant processes} should be considered.  Only
constant admissible configurations can initiate at an isolated admissible configuration.  Such configurations ar obviously
stable local equilibrium configurations.  No further discussion is necessary.

        \sect{Differential criteria of local equilibrium.}

        \ssca{Jets of processes and functions.}

    We will denote by $\K$ the set $\N \cup \{\infty, \fc \}$, where $\fc$ stands for the cardinality of $\R$.  The
ordering relations $\leqslant$, $<$, $\geqs$, and $>$ have in $\K$ the usual meaning of inequalities of cardinal numbers.
Let $q$ be a point in a differential manifold $Q$.  In the algebra $\sA Q$ of differentiable functions on $Q$ we introduce
a sequence of ideals
        $$\sI_0(Q,q),\; \sI_1(Q,q),\; \ldots ,\; \sI_{\infty}(Q,q),\; \sI_{\fc}(Q,q).
                                                                                                        \tag \label{Fxt12}$$
    The ideal $\sI_0(Q,q)$ associated with $q$ is maximal in the sense that it is not a proper subset of any ideal except
the trivial ideal $\sA Q$.  It is known that all maximal ideals in $\sA Q$ are associated with points.  For $k \in \N$, the
ideal $\sI_k(Q,q)$ is the power $(\sI_0(Q,q))^{k+1}$ of the ideal $\sI_0(Q,q)$.  The ideal $\sI_\infty(Q,q)$ is the
intersection $\bigcap_{k \in \N}\sI_k(Q,q)$.  The ideal $\sI_\fc(Q,q)$ is the set of functions each vanishing in a
neighbourhood of $q$. This ideal was introduced earlier.  Inclusion relations
        $$\sI_k(Q,q) \subset \sI_{k'}(Q,q)
                                                                                                        \tag \label{Fxt13}$$
    hold for all $k'$ and $k$ in $\K$ such that $k' \leqslant k$.

    For each $k \in \K$ we introduce an equivalence relation in the set $\pmbb\sP Q$ of oriented embedded arcs in $Q$.
Arcs $(\bc,q_0)$ and $(\bc',q'_0)$ are equivalent if
        $$\sI_k(Q,q'_0) + \sI_0(Q,\bc') = \sI_k(Q,q_0) + \sI_0(Q,\bc).
                                                                                                        \tag \label{Fxt14}$$
    Equivalence classes are called $k${\it -jets} of processes.  The $\fc$-jets are the germs introduced earlier.  The set
of $k$-jets will be denoted by $\sP^k Q$.  The $k$-jet of a process $(\bc,q_0)$ will be denoted by $\sj^k(\bc,q_0)$.
Inclusion relations imply the existence of projections
        $$\zp^{k'}{}_k{}_{\,Q} \colon \sP^k Q \rightarrow \sP^{k'} Q \colon \sj^k(\bc,q_0) \mapsto \sj^{k'}(\bc,q_0)
                                                                                                        \tag \label{Fxt15}$$
    for $k' \leqs k$ in addition to
        $$\zp_k{}_{\,Q} \colon \sP^k Q \rightarrow Q \colon \sj^k(\bc,q_0) \mapsto q_0.
                                                                                                        \tag \label{Fxt16}$$

    Corresponding to the sequence \Ref{Fxt12} of ideals we have the sequence
        $$\sA^0(Q,q) = \sA Q\big/\sI_0(Q,q), \ldots ,\; \sA^\infty(Q,q) = \sA Q\big/\sI_{\infty}(Q,q),\; \sA^\fc(Q,q) = \sA
Q\big/\sI_{\fc}(Q,q)
                                                                                                        \tag \label{Fxt17}$$
    of quotient algebras.

    In the algebra $\sA(\bc,q_0)$ of differentiable functions on the oriented arc $(\bc,q_0)$ we have ideals
        $$\sI_0(\bc,q_0),\; \sI_1(\bc,q_0),\; \ldots ,\; \sI_{\infty}(\bc,q_0),\; \sI_{\fc}(\bc,q_0)
                                                                                                        \tag \label{Fxt18}$$
    defined just as the ideals in the sequence \Ref{Fxt12} and the corresponding quotient algebras
        $$\align
        \sA^0(\bc,q_0) = \sA(\bc,q_0)&\big/\sI_0(\bc,q_0),\; \sA^1(\bc,q_0) = \sA(\bc,q_0)\big/\sI_1(\bc,q_0), \ldots \\
    &\ldots ,\; \sA^\infty(\bc,q_0) = \sA(\bc,q_0)\big/\sI_\infty(\bc,q_0),\; \sA^\fc(\bc,q_0) =
\sA(\bc,q_0)\big/\sI_\fc(\bc,q_0).
                                                                                                        \tag \label{Fxt19}\endalign$$
    For each $k \in \K$ the ideal $\sI_k(\bc,q_0)$ is isomorphic to the quotient
        $$\left(\sI_k(Q,q_0) + \sI_0(Q,\bc)\right)\big/\sI_0(Q,\bc).
                                                                                                        \tag \label{Fxt20}$$
    This justifies the following identification
        $$\sA^k(\bc,q_0) = \sA(\bc,q_0)\big/\sI_k(\bc,q_0) \;\;\text{ is identified with }\;\;\sA Q\big/\left(\sI_k(Q,q_0)
+ \sI_0(Q,\bc)\right).
                                                                                                        \tag \label{Fxt21}$$
    The identification implies that the algebra $\sA^k(\bc,q_0)$ is associated with the $k$-jet $\sj^k(\bc,q_0)$
rather than with the process $(\bc,q_0)$ since the algebra $\sA^k(\bc',q'_0)$ is the same as the algebra $\sA^k(\bc,q_0)$
if $(\bc',q'_0)$ and $(\bc,q_0)$ are in the same jet.

    An element of the quotient algebra $\sA^k(\bc,q_0) = \sA(\bc,q_0)\big/\sI_k(\bc,q_0)$ is said to be {\it increasing} if
it has an increasing representative in $\sA(\bc,q_0)$.

        \sscx{Jets of functions on $[0,a] \subset \R$.}

    The algebra $\sA[0,a]$ of differentiable functions on $[0,a] \subset  \R$, its ideals
        $$\sI_0([0,a],0),\; \sI_1([0,a],0),\; \ldots ,\; \sI_{\infty}([0,a],0),\; \sI_{\fc}([0,a],0).
                                                                                                        \tag \label{Fxt22}$$
    and the quotient algebras
        $$\align
        \sA^0([0,a],&0) = \sA[0,a]\big/\sI_0([0,a],0),\; \sA^1([0,a],0) = \sA[0,a]\big/\sI_1([0,a],0), \ldots \\
    &\ldots ,\; \sA^\infty([0,a],0) = \sA[0,a]\big/\sI_\infty([0,a],0),\; \sA^\fc([0,a],0) =
\sA[0,a]\big/\sI_\fc([0,a],0)
                                                                                                        \tag \label{Fxt23}\endalign$$
    deserve special attention.  Elements of an algebra $\sA^k([0,a],0)$ are $k$-jets of functions on $[0,a]$.

    A differentiable function on $[0,a]$ is the restriction to $[0,a]$ of a differentiable function on $\R$.  A function is
in an ideal $\sI_k([0,a],0)$ with $k < \fc$ if and only if it is the restriction to $[0,a]$ of a function in $\sI_k(\R,0)$.
This is not true for $k = \fc$.  As a consequence the quotient algebra $\sA^k([0,a],0) = \sA[0,a]\big/\sI_k([0,a],0)$ is
isomorphic to $\sA^k(\R,0) = \sA\R\big/\sI_k(\R,0)$ for $k < \fc$ but not for $k = \fc$.  In each algebra $\sA^k([0,a],0)$
there is a sequence of ideals
        $$\align
        \sI^k{}_0([0,a],0) = \sI_0([0,a],0)\big/\sI_k([0,a],0),\; &\sI^k{}_1([0,a],0) =
\sI_1([0,a],0)\big/\sI_k([0,a],0),\ldots \\
            &\ldots,\;\sI^k{}_k([0,a],0) = \sI_k([0,a],0)\big/\sI_k([0,a],0).
                                                                                                        \tag \label{Fxt24}\endalign$$
    The quotient algebra $\sA^k([0,a],0)\big/\sI^k{}_{k'}([0,a],0)$ with $k' \leqs k$ is isomorphic to $\sA^{k'}([0,a],0)$.

    A function $g \in \sA[0,a]$ is said to be {\it increasing} if the inequality $s' > s$ implies $q(s') > g(s)$.  An
element of a quotient algebra $\sA^\fc([0,a],0) = \sA[0,a]\big/\sI_\fc([0,a],0)$ is said to be {\it increasing} if it has
an increasing representative in $\sA[0,a]$.  A function $g \in \sA[0,a]$ is in $\sI^\fc([0,a],0)$ if and only if there is a
real number $\zd > 0$ such that $g(s) = 0$ for $s \leqs \zd$.  It follows that the germ $\sj^\fc g(0)$ of a function $g \in
\sA[0,a]$ is increasing if there is a number $\zd > 0$ such that $g$ is increasing in $[0,\zd]$.

        \claim \c{p}{Proposition}{}                                                                              \label{Cxt1}
    For $k \in \N$ a function $g \in \sA[0,a]$ is in $\sI_k([0,a],0)$ if and only if $\xD^i g(0) = 0$ for each $i \leqs k$.
        \endclaim
        \proof
    The derivatives of a function $g \in \sA[0,a]$ at $0$ are well defined and the function can be represented by the
Taylor formula
        $$g = e_0(g) + e_1(g)s + \ldots + e_k(g)s^k + rs^{k+1},
                                                                                                        \tag \label{Fxt25}$$
    where
        $$e_i(g) = \frac{1}{k!} \xD^i g(0),
                                                                                                        \tag \label{Fxt26}$$
    $s \colon [0,a] \rightarrow \R$ is the canonical injection, and $r$ is a differentiable function on $[0,a]$.  The
function $s$ is in $\sI_0([0,a],0)$ and the power $s^{k+1}$ is in $\sI_k([0,a],0)$.  If $\xD^i g(0) = 0$ for each $i \leqs
k$, then $g = rs^{k+1}$ is in $\sI_k([0,a],0)$.

    A function $g \in \sI_l([0,a],0)$ is a combination of products $g_0g_1 \cdots g_l$ of elements of $\sI_0([0,a],0)$. The
derivative $\xD g$ is a combination of products of functions with each product containing at least $l$ factors in
$\sI_0([0,a],0)$.  It follows that the derivative $\xD g$ of a function $g \in \sI_l([0,a],0)$ is in $\sI_{l-1}([0,a],0)$.
If $g \in \sI_k([0,a],0)$, then $\xD^0 g = g \in \sI_k([0,a],0)$, $\xD^1 g = \xD g \in \sI_{k-1}([0,a],0)$, $\xD^2 g \in
\sI_{k-2}([0,a],0)$, $\ldots$, $\xD^k g \in \sI_0([0,a],0)$.  Hence, $\xD^i g(0) = 0$ for each $i \leqs k$.
        \endproof
    A function $g \in \sA[0,a]$ is in $\sI^\infty([0,a],0)$ if and only if $\xD^i g(0) = 0$ for each $i \in \N$.

    It follows from the Proposition \Ref{Cxt1} that for $k < \fc$ the jet $\sj^k g(0) \in \sA^k([0,a],0)$ is fully
represented by the sequence
        $$e_0(g),\; e_1(g), \ldots,\; e_k(g)
                                                                                                        \tag \label{Fxt27}$$
    of derivatives
        $$e_i(g) = \frac{1}{k!} \xD^i g(0)
                                                                                                        \tag \label{Fxt28}$$
    of its representative $g \in \sA[0,a]$.  If $k \in \N$, then the polynomial
        $$e_0(g) + e_1(g)s + \ldots + e_k(g)s^k
                                                                                                        \tag \label{Fxt29}$$
    represents the jet $\sj^k g(0)$ and the product of two jets is represented by the product of the corresponding
polynomials truncated after the first $k+1$ terms.

    For $k < \fc$ an element of the ideal $\sI^k{}_0([0,a],0) \subset \sA^k([0,a],0)$ is represented by the sequence
$e_0,\; e_1, \ldots,\; e_k$ with $e_0 = 0$.  This element is said to be {\it positive} if the first non zero element in the
sequence is positive.  The element is said to be {\it negative} if the first non zero element in the sequence is negative.
Each element of the ideal $\sI^k{}_0([0,a],0)$ is either positive or negative if it is not zero.  There are obvious
relations $>$, $<$, $\geqs$, and $\leqs$ between elements of $\sI^k{}_0([0,a],0)$.

        \claim \c{p}{Proposition}{}                                                                             \label{Cxt2}
    Let $k < \fc$.  If a jet $\sj^k g(0)$ in the ideal $\sI^k{}_0([0,a],0)$ is positive, then the germ $\sj^\fc g(0)$ is
increasing.  If the jet $\sj^k g(0)$ is negative, then the germ $\sj^\fc g(0)$ is decreasing.
        \endclaim
        \proof If $e_l(g)$ is the first non zero element in the sequence \Ref{Fxt27} of derivatives, then the function $g$
is represented by Taylor formula
        $$g(s) = e_l(g)s^l + r(s)s^{l+1}.
                                                                                                        \tag \label{Fxt30}$$
    From
        $$\lim_{s \rightarrow 0}\left((l+1)r(s)s + \xD r(s)s^2\right) = 0
                                                                                                        \tag \label{Fxt31}$$
    it follows that there is a number $\zd > 0$ such that
        $$|(l+1)r(s)s + \xD r(s)s^2| < |le_l(g)|
                                                                                                        \tag \label{Fxt32}$$
    for $|s| < \zd$.  If $e_l(g) > 0$, then the function $g$ is increasing in the interval $[0,\zd]$ since the derivative
        $$\align
    \xD g &= le_l s^{l-1} + (l+1)r(s)s^l + \xD r(s)s^{l+1} \\
            &= \left(le_l + (l+1)r(s)s + \xD r(s)s^2\right)s^{l-1}
                                                                                                        \tag \label{Fxt33}\endalign$$
    is positive for $0 < s < \zd$.  This is a consequence of the Lagrange mean value theorem.  It follows that the germ
$\sj^\fc g(0)$ is increasing.  It is shown in a similar way that if $e_l$ is negative, then the germ is decreasing.
        \endproof

    It follows from the proposition that if the germ $\sj^\fc g(0)$ of a function $g$ is increasing, then the jet $\sj^k
g(0)$ is non negative for each $k < \fc$.

        \sscx{Conditions of equilibrium.}

    Let $h$ be a function in the algebra $\sA(\bc,q_0)$ and let $\zg \colon [0,a] \rightarrow Q$ be a parameterization of
the process $(\bc,q_0)$.  The parameterization induces the mapping $\bc|\zg \colon [0,a] \rightarrow \bc \colon s \mapsto
\zg(s)$. The composition $h \circ \bc|\zg$ is a function on $[0,a]$.  The function $h$ is increasing if and only if $h
\circ \bc|\zg$ is increasing.  Let $h$ be in the ideal $\sI_0(\bc,q_0) \subset  \sA(\bc,q_0)$.  For $k < \fc$, the jet
$\sj^k h \in \sI^k{}_0(\bc,q_0)$ is said to be {\it positive} if the jet $j^k(h \circ \bc|\zg)(0)$ is positive.  The jet
$\sj^k h$ is said to be {\it negative} if the jet $j^k(h \circ \bc|\zg)(0)$ is negative.  Being positive or negative is a
property of the jet $\sj^k h$ independent of the choice of the parameterization.  The following proposition is an
adaptation of Proposition \Ref{Cxt2}.
        \claim \c{p}{Proposition}{}                                                                             \label{Cxt3}
    Let $k < \fc$.  If a jet $\sj^k h$ in the ideal $\sI^k{}_0(\bc,q_0)$ is positive, then the germ $\sj^\fc h$ is
increasing.  If the jet $\sj^k h$ is negative, then the germ $\sj^\fc h$ is decreasing.
        \endclaim

    The proposition implies that if the germ $\sj^\fc h$ of a function $h$ is increasing, then the jet $\sj^k h$ is non
negative for each $k < \fc$.

    Let $q_0$ be a configuration in the set $C^0$ an let $k \in \N \cup \{\infty\}$.  If for each jet $j^k(\bc,q_0)$ of an
admissible process $(\bc,q_0)$ the jet $\sj^k w_{(\bc,q_0)}$ is positive, then the germ $\sj^\fc w_{(\bc,q_0)}$.  Hence,
$q_0$ is a stable equilibrium configuration.  We have obtained a sufficient condition for a point $q_0 \in C^0$ to be a
stable local equilibrium configuration for each $k \in \N \cup \{\infty\}$.

    If $q_0$ is a stable equilibrium configuration, then the germ $\sj^\fc w_{(\bc,q_0)}$ is increasing for each germ
$j^\fc(\bc,q_0)$ of an admissible process $(\bc,q_0)$.  It follows that for each $k \in \N \cup \{\infty\}$ and each jet
$j^k(\bc,q_0)$ of an admissible process $(\bc,q_0)$ the jet $\sj^k w_{(\bc,q_0)}$ is non negative.  This results in a
series of necessary conditions for a point $q_0 \in C^0$ to be a stable local equilibrium configuration.  Neither of these
conditions is sufficient.  Variational principles of classical physics are based on the necessary equilibrium condition of
order $k = 1$.

        \sect{Composition of static systems and control.}

    Static systems can be composed if they share the same control configuration space.  The equality of configuration
spaces is usually the result of a suitable choice.  Let two systems be characterized by work functions $W_1 \colon \bC_1
\rightarrow \R$ and $W_2 \colon \bC_2 \rightarrow \R$ defined on sets $\bC_1$ and $\bC_2$ of admissible processes in the
same configuration manifold $Q$.  The set $\bC = \bC_1 \cap \bC_2$ is the set of admissible processes of the composed
system and the work is the function $W = W_1|\bC + W_2|\bC$.  Coupling the system to other systems with the same
configuration space is a form of control.  The function $W$ is used to find the equilibrium configurations of the
controlled system.

    Composition of constrained systems may encounter difficulties.  Examples will be given.

        \sect{Partial control.}

    We have considered control of static systems through interaction with systems with the same configuration space.  This
is not always the case.  One can in general associate three distinct configuration spaces with a static system: the {\it
internal configuration space} $\oQ$, the {\it control configuration space} $Q$, and the {\it observed configuration space}
$\wQ$.  There are differential relations connecting the three spaces.  We will consider the cases when a static system with
a configuration space $\wQ$ is controlled by external devices in a configuration space $\oQ$.  The relation between the two
spaces is a differential fibration $\zh \colon \oQ \rightarrow Q$.  The configuration space $\oQ$ of the controlled system
is the internal configuration space and the configuration space $Q$ of the controlling devices is the control configuration
space.  We will refer to such situations as cases of {\it partial control}.  The observed configuration space $\wQ$ will
coincide either with $Q$ or with $\oQ$.

    Let a controlled system be represented by a work function $W \colon \bC \rightarrow \R$ defined on a set of admissible
processes $\bC \subset \pmbb\sP \oQ$ and let a controlling device be represented by a work function $W' \colon \bC'
\rightarrow \R$ defined on a set of admissible processes $\bC' \subset \pmbb\sP Q$.  An admissible process of the combined
system is a process $\bc \in \bC$ such that $\bc' = \zh(\bc) \in \bC'$.  Let $\overline\bC \subset \bC$ be the set of such
processes.  The work of the combined system is the function
        $$\overline W \colon \overline\bC \rightarrow \R \colon \bc \mapsto W(\bc) + W'(\zh(\bc)).
                                                                                                        \tag \label{Fxt34}$$

    We denote by $\pmbb\sP_V Q$ the set of processes $(\bc,q_0)$ such that $\zh(\bc)$ is a single point set.  Let $C_V$ be
the intersection $C \cap \pmbb\sP_V Q$.  We introduce the concept of the {\it critical set}.  A configuration $q_0 \in C^0$
is in the critical set $Cr(W,\zh)$ if for each germ $\sj^\fc(\bc,q_0) \in C_V{}^\fc_{q_0}$ the germ $\sj^\fc w_{(\bc,q_0)}$
is increasing.

        \sect{Realistic static systems.}

        \ssca{Constraints determined by differential equations.}

    From the set $\bC \subset \pmbb\sP Q$ of admissible processes of a static system we extract the sequence of sets
        $$C^0 \subset Q,\;C^1 \subset \sP^1 Q,\ldots,\;C^k \subset \sP^k Q,\ldots,\;C^\infty \subset \sP^\infty Q,\;C^\fc
\subset \sP^\fc Q.
                                                                                                        \tag \label{Fxt35}$$
    The $k$-jet of a process is in $C^k$ if it has a representative in $\bC$.  The set $C^0$ was already introduced
earlier.  Relations
        $$\zp^{k'}{}_k{}_{\,Q}(C^k) = C^{k'}
                                                                                                        \tag \label{Fxt36}$$
    hold for $k' \leqs k$ and
        $$\zp_k{}_{\,Q}(C^k) = C^0
                                                                                                        \tag \label{Fxt37}$$
    for $k > 0$.

    Let $(\bc,q_0)$ be a process with the property that for each configuration $q \in (\bc,q_0)$ with the exclusion of the
terminal configuration $q_1$ the jet $\sj^k(\bc_{|q},q)$ of the process $(\bc_{|q},q)$ based on the arc
        $$\bc_{|q} = \left\{q' \in \bc;\; q' \geqs q \right\}.
                                                                                                        \tag \label{Fxt38}$$
    is in $C^k$.  We denote by $\bC^k$ the class of all such processes.  If a process $(\bc,q_0)$ with terminal point $q_1$
is in $\bC^k$, then $\sj^k(\bc_{|q},q) \in C^k$ for each $q \in \bc \setminus \{q_1\}$.  It follows that
$\sj^{k'}(\bc_{|q},q) = \zp^{k'}{}_k{}_{\,Q}(\sj^k(\bc_{|q},q)) \in C^{k'}$ for each $q \in \bc \setminus \{q_1\}$.  Hence,
$(\bc,q_0) \in \bC^{k'}$. We have established the inclusion
        $$\bC^{k'} \supset \bC^k
                                                                                                        \tag \label{Fxt39}$$
    for $k' \leqs k$.

    If $(\bc,q_0)$ is an admissible process, then for each configuration $q \in (\bc,q_0)$ with the exclusion of the
terminal configuration $q_1$ the process $(\bc_{|q},q)$ is admissible and the germ $\sj^\fc(\bc_{|q},q)$ is in $C^\fc$.  It
follows that $\bC \subset \bC^\fc$.  Let the germ $\sj^\fc(\bc_{|q},q)$ be in $C^\fc$ for each configuration $q \in
\bc\setminus\{q_1\}$.  We show that the process $(\bc_\gzmq{}_|,q_0)$ is admissible for each configuration $\gzmq$ in the
interior of $\bc$.  For each $q \in \bc\setminus\{q_1\}$ there is a configuration $q' \in (\bc,q_0)$ such that $q' > q$ and
the process $(\bc_{|qq'|},q)$ based on the arc
        $$\bc_{|qq'|} = \left\{q'' \in \bc;\; q' \geqs q'' \geqs q \right\}
                                                                                                        \tag \label{Fxt40}$$
    is admissible.  This is a consequence of $\sj^\fc(\bc_{|q},q) \in C^\fc$.  The union
        $$\bigcup_{q \in \bc\setminus\{q_1\}}(\bc_{|qq'|},q)
                                                                                                        \tag \label{Fxt41}$$
    is an admissible process and for each $\gzmq$ in the interior of $\bc$ the process $(\bc_\gzmq{}_|,q_0)$ is a
subprocess of the union.  Hence, $(\bc_\gzmq{}_|,q_0)$ is admissible.  The process $(\bc,q_0)$ itself is admissible since
it can be approximated by admissible processes.  We have proved the inclusion $\bC^\fc \subset \bC$.

    It may happen that the equality $\bC = \bC^k$ holds for $k < \fc$.  If it holds for $k$ and not for $k-1$, then the set
$\bC$ is said to represent {\it constraints of order} $k$.  The set $C^k$ is interpreted as a differential equation of
order $k$.  Admissible processes are obtained by solving this equation.  Constraints of order 0 are said to be {\it
holonomic}.  Constraints of order 1 are said to be {\it non holonomic}.  Constraints of higher order are not usually
discussed since their presence is not apparent if only first order criteria of equilibrium are considered.

        \sscx{Jets of processes with volume, the work integral.}

    A $k${\it -jet of a process with volume} is a pair $(\sj^k(\bc,q_0),v)$, where $v$ is a vector belonging to the
orientation $o$ of $(\bc,q_0)$ at $q_0$.  We denote by $\sV^k Q$ the space of $k$-jets with volume.  For $k < \fc$ it make
sense to associate $k$-jets with volume with the terminal configuration of a process.  Spaces $\sV^0 Q$ and $\sV^1$ are
identified with the bundle $\oT Q \subset \sT Q$ of tangent vectors with the zero vectors removed.  This is possible since
in the pair $(\sj^0(\bc,q_0),v)$ or $(\sj^1(\bc,q_0),v)$ the vector $v$ already contains the complete information about the
jet. We denote by $V^k$ the $k$-jets with volume associated with admissible processes.  Note that if $(\sj^k(\bc,q_0),v)
\in V^k$, then  $(\sj^k(\bc,q_0),\zl v) \in V^k$ for each $\zl > 0$.  Sets $V^0$ and $V^1$ are interpreted as a subset $V
\subset \oT Q$.

    The work of a realistic static system is defined in terms of a {\it work form}
        $$\zy \colon V^k \rightarrow \R.
                                                                                                        \tag \label{Fxt42}$$
    The work form is positive homogeneous in its vector argument: if $\zl > 0$, then $\zy(\sj^k(\bc,q_0),\zl v) =
\zl\zy(\sj^k(\bc,q_0),v)$.  The work form is used to define the work function $w_{(\bc,q_0)}$ along an admissible
processes. Let $q_1$ be the terminal point of the process $(\bc,q_0)$.  For each configuration $q \in \bc \setminus
\{q_1\}$ the value of the work function is the integral
        $$w_{(\bc,q_0)}(q) = \int_{(\bc|q,q_0)} \zy
                                                                                                        \tag \label{Fxt43}$$
    of the work form defined as the Riemann integral
        $$\int_0^{\zg^{-1}(q)} \zy(\sj^k(\bc_{|\zg(s)},\zg(s)),\st\zg(s))\xd s
                                                                                                        \tag \label{Fxt44}$$
    in terms of a parameterization $\zg \colon [0,a] \rightarrow Q$.  Homogeneity of $\zy$ makes the integral independent
of the parameterization.  The work of the process is the limit

        $$W(\bc,q_0) = \lim_{q \rightarrow q_0} w_{(\bc,q_0)}(q).
                                                                                                        \tag \label{Fxt45}$$

    The work form of a system of the usually considered type is a positive homogeneous function
        $$\zy \colon V \rightarrow \R.
                                                                                                        \tag \label{Fxt46}$$
    This corresponds to $k = 0$ or $k = 1$.

        \sect{The principle of virtual work.}

    Variational principles of classical physics are versions of what is known in statics as the {\it principle of virtual
work}.  This principle consists in applying the first necessary equilibrium condition to a static system.  It is usually
assumed that the constraints are of order 0 or 1 and that the work is represented by a work form
        $$\zy \colon V \rightarrow \R.
                                                                                                        \tag \label{Fxt47}$$
    Elements of $V$ are the {\it admissible virtual displacements}.  The principle of virtual work states that the
inequality
        $$\zy(v) \geqs 0
                                                                                                        \tag \label{Fxt48}$$
    holds for each virtual displacement $v$ tangent to an admissible process initiating at a configuration of equilibrium.

    The principle of virtual work can be extended to more general realistic systems.  If the system has constraints of
order $l$ and the work is defined in terms of a work form
        $$\zy \colon V^k \rightarrow \R,
                                                                                                        \tag \label{Fxt49}$$
    then the principle of virtual work (the first order necessary condition of equilibrium) states that the inequality
        $$\zy(\sj^k(\bc,q_0),v) \geqs 0
                                                                                                        \tag \label{Fxt50}$$
    holds for each $k$-jet with volume $(\sj^k(\bc,q_0),v)$ belonging to an admissible process $(\bc,q_0)$ initiating at a
configuration of equilibrium.

        \sect{Potential systems and the Legendre transformation.}

    A {\it potential} order 1 of a {\it potential static system} is a differentiable function $U$ on the configuration
space $Q$.  A potential system is unconstrained.  For each process $(\bc,q_0) \in \bC = \pmbb\sP Q$ with terminal point
$q_1$ the function $w_{(\bc,q_0)}$ is defined by
        $$w_{(\bc,q_0)}(q) = U(q) - U(q_0)
                                                                                                        \tag \label{Fxt51}$$
    for $q \in \bc \setminus \{q_1\}$ and the work $W(\bc,q_0) = w_{(\bc,q_0)}(q_1)$ is the limit
        $$W(\bc,q_0) = \lim_{q \rightarrow q_1} w_{(\bc,q_0)}(q).
                                                                                                        \tag \label{Fxt52}$$
    The set $V$ of a potential system is the tangent bundle $\oT Q$ and the function $\zy$ is the function
        $$\zy \colon \oT Q \rightarrow \R \colon v \mapsto \langle \xd u, v\rangle
                                                                                                        \tag \label{Fxt53}$$
    constructed from the differential $\xd U$ of the potential.  The principle of virtual work for a potential system
states that the equality
        $$\langle \xd U, v\rangle = 0
                                                                                                        \tag \label{Fxt54}$$
    holds for every virtual displacement from a configuration of equilibrium.  The appearance of the equality in place of
the inequality is due to the {\it reversibility} of virtual displacements: if $v$ is a admissible virtual displacement from
a configuration of equilibrium, then so is $-v$ and $\langle \xd U, -v\rangle = -\langle \xd U, v\rangle$

    The {\it Legendre transformation} associates with a static system represented by a set $V \subset \oT Q$ and a function
$\zy \colon V \rightarrow \R$ the {\it constitutive set} $S$ of the system defined as
        $$S = \left\{f \in \sT^\*Q ;\; \zp_Q(f) \in C^0, \zy(v) - \langle f, v\rangle \geqs 0 \;\text{ for each }\; v \in V
\;\text{ such that }\; \zt_Q(v) = \zp_Q(f) \right\}
                                                                                                        \tag \label{Fxt55}$$
    The set $C^0$ is obtained by applying the tangent projection $\zt_Q$ to virtual displacements.  If constraints are
holonomic, then $V = \oT C^0$.  A virtual displacement is in $\oT C^0$ if there is a curve $\zx \colon \R \rightarrow Q$
and a number $\ze > 0$ such that $\zx([0,\ze]) \subset C^0$ and $v = \st\zx(0) \neq 0$.  The inequality in the definition
of $S$ is replaced by an equality in the case of reversibility.

    The physics of the Legendre transformation is that of control of a static system by means of potential external
devices.  The covector $f = -\xd U(\zp_Q(f))$ is the {\it external force} applied to the controlled system by the potential
device with potential $U$.  The constitutive set $S$ characterizes the response of a static system to control by potential
devices only.  Under certain conditions (convexity) this characterization is complete since the set $V$ and the function
$\zy$ can by reconstructed from the constitutive set by the {\it inverse Legendre transformation}.

        \sect{Examples of static systems and Legendre transformations.}

    Configuration spaces of most systems considered here are affine spaces.  If $Q$ is an affine space modeled on a vector
space $W$, then the tangent bundle $\sT Q$ is identified with the product $Q \times W$, the cotangent bundle is identified
with $Q \times W^\*$ and the canonical pairing is the mapping
        $$\langle \,\;,\;\rangle \colon (Q \times W^\*) \times (Q \times W) \colon ((q,f),(q,w)) \mapsto \langle f,
w\rangle.
                                                                                                        \tag \label{Fxt56}$$
    We denote by $q_1 - q_0$ the vector associated with the points $q_0$ and $q_1$.

    For the sake of simplicity we will not exclude zero vectors in the set $V$ of virtual displacements.

        \claim \c{e}{Example}{}{\rm                                                                              \label{Cxt4}
    A material point with configuration $q$ in the affine Euclidean space $Q$ of Newtonian mechanics is tied to a fixed
point $q_0$ with a spring of spring constant $k$.  This is an unconstrained potential system.  The internal energy is the
function
        $$U \colon Q \rightarrow \R \colon q \mapsto \frac{k}{2}\|q - q_0\|^2.
                                                                                                        \tag \label{Fxt57}$$
    The set
        $$S = \left\{(q,f) \in Q \times W^\* ;\; f = kg(q - q_0) \right\}
                                                                                                        \tag \label{Fxt58}$$
    is the constitutive set.
        }{\hfill    $\blacktriangle$}\endclaim

        \claim \c{e}{Example}{}{\rm                                                                              \label{Cxt5}
    Let $q_0$ be a point in $Q$ and let $\zw \colon W \times W \rightarrow \R$ be a bilinear mapping.  The function
        $$\zy \colon Q \times W \rightarrow \R \colon (q,w) \mapsto \zw(q - q_0,w)
                                                                                                        \tag \label{Fxt59}$$
    used in the principle of virtual work produces the constitutive set
        $$\align
            S &= \left\{(q,f) \in \sT^\*Q ;\; \zw(q - q_0,v) - \langle f, v\rangle = 0 \;\text{ for each }\; v \in W
\right\} \\
                &= \left\{(q,f) \in \sT^\*Q ;\; \zw(q - q_0,\cdot) - f = 0 \right\}.
                                                                                                        \tag \label{Fxt60}\endalign$$
    The function $\zy$ is obtained from the differential of the quadratic function
        $$U \colon Q \rightarrow \R \colon q \mapsto \frac{1}{2}\zw(q - q_0,q - q_0)
                                                                                                        \tag \label{Fxt61}$$
    only if $\zw$ is symmetric.
        }{\hfill    $\blacktriangle$}\endclaim

        \claim \c{e}{Example}{}{\rm                                                                              \label{Cxt6}
    Let $Q$ be the configuration manifold of a static mechanical system and let $\zr \colon \sT Q \rightarrow \sT^\*Q$ be a
mapping with the properties of a Euclidean metric.  This means that $\zr$ defines an isomorphism of vector bundles
    \vskip1mm
        $$\vcenter{\xymatrix@R+3mm @C+8mm{{\sT Q} \ar[d]_*{\zt_Q} \ar[r]^*{\zr} &
            \sT^\*Q \ar[d]_*{\zp_Q} \\
            Q \ar@{=}[r] & Q}},
                                                                                                        \tag \label{Fxt62}$$
    \vskip2mm
    \noindent the symmetry relation
        $$\langle \zr(u), v\rangle = \langle \zr(v), u\rangle
                                                                                                        \tag \label{Fxt63}$$
    holds for any pair $(u,v) \in \sT Q \fpr{(\zt_Q,\zt_Q)} \sT Q$, the inequality
        $$\langle \zr(v), v\rangle > 0
                                                                                                        \tag \label{Fxt64}$$
    holds for each $v \neq 0$.  The function
        $$\zy \colon \sT Q \rightarrow \R \colon v \mapsto \sqrt{\langle  \zr(v), v\rangle}
                                                                                                        \tag \label{Fxt65}$$
    is positive homogeneous on fibres of the tangent fibration.  It represents the virtual work of a virtual displacement
$v$ due to friction.  The principle of virtual work is the inequality
        $$\sqrt{\langle  \zr(v), v\rangle} - \langle f, v\rangle \geqs 0.
                                                                                                        \tag \label{Fxt66}$$
    It is satisfied if $f$ is in the constitutive set $S$ and $(f,v) \in \sT^\*Q \fpr{(\zp_Q,\zt_Q)} \sT Q$.

    Let $f \in \sT^\*Q$ be in the constitutive set.  By using $v = \zr^{-1}(f)$ in the principle of virtual work we arrive
at the inequality
        $$\langle f, \zr^{-1}(f)\rangle \leqs \sqrt{\langle f, \zr^{-1}(f)\rangle}.
                                                                                                        \tag \label{Fxt67}$$
    Hence,
        $$\langle f, \zr^{-1}(f)\rangle \leqs 1.
                                                                                                        \tag \label{Fxt68}$$

    The inequality
        $$\langle f, v\rangle \leqs \sqrt{\langle f, \zr^{-1}(f)\rangle} \sqrt{\langle  \zr(v), v\rangle}.
                                                                                                        \tag \label{Fxt69}$$
    is the result of the Schwarz inequality
        $$\langle \zr(u), v\rangle \leqs \sqrt{\langle  \zr(u), u\rangle}\sqrt{\langle  \zr(v), v\rangle}
                                                                                                        \tag \label{Fxt70}$$
    applied to the pair of vectors $u = \zr^{-1}(f)$ and $v$.  If $\langle f, \zr^{-1}(f)\rangle \leqs 1$, then
        $$\langle f, v\rangle \leqs \sqrt{\langle  \zr(v), v\rangle}.
                                                                                                        \tag \label{Fxt71}$$
    Hence, $f \in S$.

    The constitutive set of the system is the set
        $$S = \left\{f \in \sT^\*Q ;\; \langle f, \zr^{-1}(f)\rangle \leqs 1 \right\}.
                                                                                                        \tag \label{Fxt72}$$
        }{\hfill    $\blacktriangle$}\endclaim

        \claim \c{e}{Example}{}{\rm                                                                              \label{Cxt7}
    Let a material point with configuration $q$ in the Euclidean affine space $Q$ be tied with a rigid rod of length $a$ to
a point with configuration $q_0$.  The configuration $q$ is constrained to the sphere
        $$C^0 = \left\{q \in Q ;\; \|q - q_0\| = a \right\}.
                                                                                                        \tag \label{Fxt73}$$
    This is a system with holonomic bilateral constraints.  The set
        $$V = \left\{(q,\zd q) \in Q \times W ;\; \|q - q_0\| = a,\; \langle g(q - q_0), \zd q\rangle = 0 \right\}
                                                                                                        \tag \label{Fxt74}$$
    of admissible virtual displacements is the tangent set $\sT C^0$ of the holonomic constraint $C^0$.  With the work form
$\zy = 0$ the constitutive set is the set
        $$S = \left\{(q,f) \in Q \times W^\* ;\; \|q - q_0\| = a,\; f = a^{-2}\langle f, q - q_0\rangle g(q - q_0) \right\}.
                                                                                                        \tag \label{Fxt75}$$
        }{\hfill    $\blacktriangle$}\endclaim

        \claim \c{e}{Example}{}{\rm                                                                              \label{Cxt8}
    Let $q_0$ be a point in the Euclidean affine space $Q$ of dimension 3 and let $u_1$, $u_2$ and $u_3$ be mutually
orthogonal unit vectors in the model space $W$.  Let the configuration $q \in Q$ of a material point be constrained to the
set
        $$C^0 = \left\{q \in Q ;\; \langle g(u_1), q - q_0\rangle \geqs 0,\; \langle g(u_2), q - q_0\rangle \geqs 0
\right\}.
                                                                                                        \tag \label{Fxt76}$$
    This is a system with holonomic one-sided constraints with the tangent set
        $$\align
        V &= \left\{(q,\zd q) \in Q \times W ;\; \langle g(u_1), q - q_0\rangle \geqs 0,\; \langle g(u_2), q - q_0\rangle
\geqs 0, \right. \\
        &\hskip5mm\left. \langle g(u_1), \zd q\rangle \geqs 0 \;\text{ if }\; \langle g(u_1), q - q_0\rangle = 0,\;\langle
g(u_2), \zd q\rangle \geqs 0 \;\text{ if }\; \langle g(u_2), q - q_0\rangle = 0 \right\}.
                                                                                                        \tag \label{Fxt77}\endalign$$
    of $C^0$ as the set of admissible virtual displacements.  With $\zy = 0$ the constitutive set is the set
        $$\align
        S &= \left\{(q,f) \in Q \times W^\* ;\; \langle g(u_1), q - q_0\rangle \geqs 0,\; \langle g(u_2), q - q_0\rangle
\geqs 0, \right. \\
        &\hskip20mm f = 0 \;\text{ if }\; \langle g(u_1), q - q_0\rangle > 0 \;\text{ and }\; \langle g(u_2), q -
q_0\rangle > 0, \\
        &\hskip10mm \langle f, u_1\rangle \leqs 0,\; \langle f, u_2\rangle = 0,\;\text{ and }\; \langle f, u_3\rangle = 0 \\
        &\hskip30mm \text{ if }\; \langle g(u_1), q - q_0\rangle = 0 \;\text{ and }\; \langle g(u_2), q - q_0\rangle > 0, \\
        &\hskip10mm \langle f, u_1\rangle = 0,\; \langle f, u_2\rangle \leqs 0,\;\text{ and }\; \langle f, u_3\rangle = 0 \\
        &\hskip30mm \text{ if }\; \langle g(u_1), q - q_0\rangle > 0 \;\text{ and }\; \langle g(u_2), q - q_0\rangle = 0, \\
        &\hskip10mm \langle f, u_1\rangle \leqs 0,\; \langle f, u_2\rangle \leqs 0,\;\text{ and }\; \langle f, u_3\rangle =
0 \\
        &\hskip30mm \left. \text{ if }\; \langle g(u_1), q - q_0\rangle = 0 \;\text{ and }\; \langle g(u_2), q - q_0\rangle
= 0 \right\}.
                                                                                                        \tag \label{Fxt78}\endalign$$
        }{\hfill    $\blacktriangle$}\endclaim

        \claim \c{e}{Example}{}{\rm                                                              \label{Cxt9}
    Let $Q = X \times D$ be the configuration space of a skate.  The space $X$ is an affine plane modeled on a vector space
$W$ and $D$ is the projective space of directions in $X$.  We use a Euclidean metric in $X$ represented by a mapping $g
\colon W \rightarrow W^\*$ to identify the space $D$ with the unit circle
        $$D = \left\{\zf \in W ;\; \langle g(\zf), \zf\rangle = 1 \right\}.
                                                                                                        \tag \label{Fxt79}$$
    This is the only use we make of the metric.  The tangent bundle $\sT Q$ is identified with $X \times W \times \sT D$,
where
        $$\sT D = \left\{(\zf,\zd\zf) \in D \times W ;\; \langle g(\zf), \zd\zf\rangle = 0 \right\}.
                                                                                                        \tag \label{Fxt80}$$
    The skate is a system with non holonomic constraints.  The set $C^0$ is the entire space $Q$.  The constraint consists
in restricting virtual displacements in $X$ to those parallel to the direction specified by an element of $D$.  Thus
        $$V = \left\{(x,\zd x,\zf,\zd\zf) \in X \times W \times \sT D ;\; \exi{k \in \R} \zd x = k\zf
\right\}.
                                                                                                        \tag \label{Fxt81}$$
    The cotangent bundle $\sT^\*Q$ is the product $X \times W^\* \times \sT^\*D$.  For each $\zf \in D$ the fibre $\sT_\zf
D$ is a subset of $W$.  Hence, the cotangent bundle $\sT^\*D$ can be specified as the set of pairs $(\zf,\zt)$, where $\zf
\in D$ and $\zt$ is in the quotient space $W^\*\big/\sT_\zf^{\dsize\circ} D$. The set
        $$\align
            S &= \left\{(x,f,\zf,\zt) \in \sT^\*Q;\; \langle f, \zx\rangle + \langle \zt, \zd\zf\rangle = 0 \;\text{ for
each }\; (x,\zd x,\zf,\zd\zf) \in V \right\} \\
         &= \left\{(x,f,\zf,\zt) \in \sT^\*Q;\; \langle f, \zf\rangle = 0, \zt = 0 \right\}
                                                                                                        \tag \label{Fxt82}\endalign$$
    is the constitutive set of the system.
        }{\hfill    $\blacktriangle$}\endclaim

        \claim \c{e}{Example}{}{\rm                                                                              \label{Cxt10}
    Let $Q$ be the Euclidean affine space of Newtonian mechanics.  The model space for $Q$ is a vector space $W$ of
dimension 3.  The Euclidean structure is represented by a metric tensor $g \colon W \rightarrow W^\*$.

    The example gives a formal description of experiments performed by Coulomb in his study of static friction.  Let a
material point be constrained to the set
        $$C^0 = \left\{q \in Q;\;\langle g(k), q - q_0 \rangle \geqs 0 \right\},
                                                                                                        \tag \label{Fxt83}$$
    where $q_0$ is a point in $Q$ and $k \in W$ is a unit vector.  The boundary
        $$\partial C^0 = \left\{q \in Q;\;\langle g(k), q - q_0 \rangle = 0 \right\}
                                                                                                        \tag \label{Fxt84}$$
    is a plane passing through $q_0$ and orthogonal to $k$.  In its displacements on the boundary the point encounters
friction proportional to the component of the external force pressing the point against the boundary.  The system is
characterized by the virtual work function $\zy = 0$ defined on the non holonomic constraints
        $$\align
        V = &\left\{(q,\zd q) \in \sT Q;\; \langle g(k), q - q_0 \rangle \geqs 0,\; \langle g(k), \zd q) \rangle \geq
\zn\sqrt{\|\zd q\|^2 - \langle g(k), \zd q) \rangle^2} \right.\hskip20mm\\
             &\hskip80mm\left.\vphantom{\sqrt{\|\zd q\|^2 - \langle g(k), \zd q) \rangle^2}}\text{ if }\; \langle g(k), q -
q_0 \rangle = 0 \right\}. \\
                                                                                                        \tag \label{Fxt85}\endalign$$

    The principle of virtual work states that $(q,f)$ is in the constitutive set $S$ if and only if the inequality
        $$\langle f, \zd q\rangle \leqs 0
                                                                                                        \tag \label{Fxt86}$$
    is satisfied for each $(q,\zd q) \in V$.

    If $\langle g(k), q - q_0 \rangle > 0$, then a pair $(q,f) \in \sT^\*Q$ is in the constitutive set $S$ if and only if
$f = 0$.

    We consider pairs $(q,f)$ with $\langle g(k), q - q_0 \rangle = 0$.  If $f = -\|f\| k$, then $(q,f)$ is in the
constitutive set and $\|f\|^2 - \langle f, k\rangle^2 = 0$.  Let $(q,f)$ be in the constitutive set and let $\|f\|^2 -
\langle f, k\rangle^2 \neq 0$.  The virtual displacement $(q,\zd q)$ with
        $$\zd q = g^{-1}(f) - \langle f, k\rangle k + \zn\sqrt{\|f\|^2 - \langle f, k\rangle^2}k
                                                                                                        \tag \label{Fxt87}$$
    is in $V$ since
        $$\langle g(k), \zd q\rangle = \zn\sqrt{\|f\|^2 - \langle f, k\rangle^2}.
                                                                                                        \tag \label{Fxt88}$$
    From the principle of virtual work and
        $$\langle f, \zd q\rangle = \|f\|^2 - \langle f, k\rangle^2 + \sqrt{\|f\|^2 - \langle f, k\rangle^2}\langle f,
k\rangle
                                                                                                        \tag \label{Fxt89}$$
    it follows that
        $$\|f\|^2 - \langle f, k\rangle^2 + \sqrt{\|f\|^2 - \langle f, k\rangle^2}\langle f, k\rangle \leqs 0
                                                                                                        \tag \label{Fxt90}$$
    and
        $$\sqrt{\|f\|^2 - \langle f, k\rangle^2\rangle} + \zn\langle f, k\rangle \leqs 0
                                                                                                        \tag \label{Fxt91}$$
    since $\|f\|^2 - \langle f, k\rangle^2\rangle > 0$.

    The Schwarz inequality
        $$\langle g(u), v\rangle - \langle g(k), u\rangle\langle g(k), v\rangle \leqs \sqrt{\|u\|^2 - \langle g(k),
u\rangle^2}\sqrt{\|v\|^2 - \langle g(k), v\rangle^2}
                                                                                                        \tag \label{Fxt92}$$
    for the bilinear symmetric form
        $$(u,v) \mapsto \langle g(u), v\rangle - \langle g(k), u\rangle\langle g(k), v\rangle
                                                                                                        \tag \label{Fxt93}$$
    applied to the pair $(g^{-1}(f),\zd q)$ leads to the inequality
        $$\langle f, \zd q\rangle - \langle f, \zd q\rangle\langle g(k), \zd q\rangle \leqs \sqrt{\|f\|^2 - \langle f, k
\rangle^2}\sqrt{\|\zd q\|^2 - \langle g(k), \zd q\rangle^2}.
                                                                                                        \tag \label{Fxt94}$$
    If $\langle g(k), q - q_0 \rangle = 0$, $\sqrt{\|f\|^2 - \langle f, k\rangle^2\rangle} + \zn\langle f, k\rangle \leqs
0$, and $\langle g(k), \zd q) \rangle \geq \zn\sqrt{\|\zd q\|^2 - \langle g(k), \zd q) \rangle^2}$, then $\langle f, \zd
q\rangle \leqs 0$.  Hence, $(q,f)$ is in the constitutive set $S$.

    We have shown that the set
        $$\align
            S &= \left\{(q,f) \in \sT^\*Q;\; \langle g(k), q - q_0 \rangle \geqs 0,\; f = 0 \;\text{ if }\;\langle g(k), q
- q_0 \rangle > 0\; \right.\hskip15mm \\
        &\hskip20mm\left.\vphantom{\sqrt{\|\zd q\|^2 - \langle g(k), \zd q) \rangle^2}}\text{ and }\; \sqrt{\|f\|^2 -
\langle f, k\rangle^2\rangle} + \zn\langle f, k\rangle \leqs 0 \;\text{ if }\; \langle g(k), q - q_0 \rangle = 0  \right\}
                                                                                                        \tag \label{Fxt95}\endalign$$
    is the constitutive set of the system.

        }{\hfill    $\blacktriangle$}\endclaim

        \sect{The Legendre transformation for partially controlled systems.}

    Let $\oQ$ be the internal configuration space of a system, let $Q$ be the control space and let $\zh \colon \oQ
\rightarrow Q$ be a differential fibration relating the two spaces.  The work function $\zy \colon V \rightarrow \R$ is
defined on a set $V \subset \sT\oQ$.  The system is partially controlled by potential systems.  If $\wQ = \oQ$, then
        $$\align
        \overline S &= \left\{(\overline q,f) \in \oQ \fpr{(\zh,\zp_Q)} \sT^\*Q ;\; \overline q \in \zt_{\oQ}(V),\; \zy(v)
- \langle f, \sT\zh(v)\rangle \geqs 0 \right.\hskip25mm \\
        &\hskip60mm \left. \vphantom{\oQ \fpr{(\zh,\zp_Q)} \sT^\*Q} \text{ for each }\; v \in V \;\text{ such that }\;
\zt_{\oQ}(v) = \overline q \right\}.
                                                                                                        \tag \label{Fxt96}\endalign$$
    If $\wQ = Q$, then the constitutive set $S \subset \sT^\*Q$ is the image $pr_{\sT^\*Q}(\overline S)$ of $\overline S$
by the canonical projection
        $$pr_{\sT^\*Q} \colon \oQ \fpr{(\zh,\zp_Q)} \sT^\*Q \rightarrow \sT^\*Q.
                                                                                                        \tag \label{Fxt97}$$

    The concept of the critical set adapted to first order equilibrium criteria is useful.  The {\it critical set} is the
set
        $$Cr = \left\{\overline q \in \oQ ;\; \overline q \in \zt_{\oQ}(V),\; \zy(v) \geq 0 \;\text{ for each }\; v \in V
\;\text{ such that }\; \zt_{\oQ}(v) = \overline q \text{ and } \zh(v) = 0\right\}.
                                                                                                        \tag \label{Fxt98}$$

        \sect{Examples of the Legendre transformation for partially controlled systems.}

    We give examples of partially controlled systems in the affine space $Q$ of Newtonian mechanics modeled on a Euclidean
vector space $W$ with a metric tensor $g \colon W \rightarrow W^\*$.  The tangent bundle $\sT Q$ is identified with the
product $Q \times W$ and the cotangent bundle is identified with $Q \times W^\*$.  The tangent and the cotangent bundle of
the product $Q \times Q$ are identified with $Q \times Q \times W \times W$ and $Q \times Q \times W^\* \times W^\*$
respectively.

        \claim \c{e}{Example}{}{\rm                                                                               \label{Cxt11}
    Three material points with configurations $q_0$, $q_1$, and $q_2$ in the affine space $Q$ are interconnected with
springs with spring constants $k_{10}$, $k_{20}$, and $k_{21}$.  The virtual work of the system is derived from the
internal energy
        $$\oU \colon Q \times Q \times Q \rightarrow \R \colon (q_0,q_1,q_2) \mapsto \frac{k_{10}}{2} \|q_1 - q_0\|^2 +
\frac{k_{20}}{2} \|q_2 - q_0\|^2 + \frac{k_{21}}{2} \|q_2 - q_1\|^2
                                                                                                        \tag \label{Fxt99}$$
    of the springs.  The point $q_0$ is fixed and not controlled.  The configuration space is the product $Q \times Q$. The
two points $q_1$ and $q_2$ are not constrained.  The constitutive set is the set
        $$\align
        \oS &= \left\{(q_1,q_2,f_1,f_2) \in \sT^\*(Q \times Q) ;\; f_1 = k_{10}g(q_1 - q_0) - k_{21}g(q_2 - q_1),\right.
\hskip10mm\\
        &\hskip60mm \left. f_2 = k_{20}g(q_2 - q_0) + k_{21}g(q_2 - q_1)  \right\}.
                                                                                                        \tag \label{Fxt100}\endalign$$

    If the configuration of $q_2$ is not controlled but observed, then $Q \times Q$ is the internal configuration space,
$Q$ is the control configuration space and the canonical projection
        $$pr_1 \colon Q \times Q \rightarrow Q \colon (q_1,q_2) \mapsto q_1
                                                                                                        \tag \label{Fxt101}$$
    is the relation between the two spaces.  The critical set
        $$Cr = \left\{(q_1,q_2) \in Q \times Q ;\; k_{20}g(q_2 - q_0) + k_{21}g(q_2 - q_1) = 0 \right\}
                                                                                                        \tag \label{Fxt102}$$
    is the image of the section
        $$\zm \colon Q \rightarrow Q \times Q \colon q \mapsto (q,q_0 + k_{21}(k_{20} + k_{21})^{-1}(q - q_0))
                                                                                                        \tag \label{Fxt103}$$
    of the projection $pr_1$.  The set
        $$\align
        \wS &= \left\{(q_1,q_2,f_1) \in Q \times Q \times W^\* ;\; q_2 = q_0 + \frac{k_{21}}{k_{20} + k_{21}}(q_1 - q_0),
\right. \hskip15mm \\
        &\hskip60mm \left. f_1 = \left(k_{10} + \frac{k_{20}k_{21}}{k_{20} + k_{21}}\right) g(q_1 - q_0) \right\}
                                                                                                        \tag \label{Fxt104}\endalign$$
    is the constitutive set.

    If the configuration of $q_2$ is neither controlled nor observed, then the set
        $$S = \left\{(q_1,f_1) \in Q \times W^\* ;\; f_1 = \left(k_{10} + \frac{k_{20}k_{21}}{k_{20} + k_{21}}\right) g(q_1
- q_0) \right\}
                                                                                                        \tag \label{Fxt105}$$
    is the constitutive set.  Note that the presence of the material point with configuration $q_2$ can be ignored.  This
is due to the fact that the critical set is the image of a section of the projection \Ref{Fxt101}.
        }{\hfill    $\blacktriangle$}\endclaim

        \claim \c{e}{Example}{}{\rm                                                                           \label{Cxt12}
    The present example gives a simplified discrete model of the buckling of a rod.  One end of the rod is a point in the
affine space $Q$ with configuration $q_1$ constrained to the line
        $$L = \left\{q \in Q ;\; q - q_0 = \langle g(u), q - q_0\rangle u \right\}
                                                                                                        \tag \label{Fxt106}$$
    through a point $q_0$ in the direction of a unit vector $u$.  The other end is a point with configuration $q_2$
constrained to the plane
        $$P = \left\{q \in Q ;\;\langle g(u), q - q_0\rangle = 0 \right\}
                                                                                                        \tag \label{Fxt107}$$
    through $q_0$ perpendicular to $u$.  The rod can be compressed or extended in length but not bent.  Its relaxed length
is $a$ and the elastic constant is $k$.  The buckling of the rod is simulated by displacements of its end point in the
plane $P$ tied elastically to the point $q_0$ with a spring of spring constant $k'$.  The configuration space is the
product $Q \times Q$ with holonomic constraints represented by
        $$C^0 = \left\{(q_1,q_2) \in Q \times Q ;\; q_1 - q_0 = \langle g(u), q_1 - q_0\rangle u,\; \langle g(u), q_2 -
q_0\rangle = 0 \right\}.
                                                                                                        \tag \label{Fxt108}$$
    The set
        $$\align
        V &= \left\{(q_1,q_2,\zd q_1,\zd q_2) \in \sT(Q \times Q) ;\; q - q_0 = \langle g(u), q_1 - q_0\rangle u, \right. \\
        &\hskip35mm \langle g(u), q_2 - q_0\rangle = 0,\; \zd q_1 = \langle g(u), \zd q_1\rangle u,\; \langle g(u), \zd
q_2\rangle = 0 \left. \right\}
                                                                                                        \tag \label{Fxt109}\endalign$$
    of admissible virtual displacements is the tangent set of $C^0$.  The internal energy of the system is the function
        $$\oU \colon C^0 \rightarrow \R \colon (q_1,q_2) \mapsto \frac{k}{2} (\|q_1 - q_2\| - a)^2 + \frac{k'}{2}\|q_2 -
q_0\|.
                                                                                                        \tag \label{Fxt110}$$
    This expression for the internal energy excludes certain non realistic configurations.  The set
        $$\align
        \oS &= \left\{(q_1,q_2,f_1,f_2) \in \sT^\*(Q \times Q) \vphantom{\|^{-1}};\; q_1 - q_0 = \langle g(u), q_1 -
q_0\rangle u,\; \langle g(u), q_2 - q_0\rangle = 0, \right. \\
        &\hskip20mm \left. \langle f_1, u\rangle = k\left(1 - a\|q_2 - q_1\|^{-1}\right)\langle g(q_1 - q_0), u\rangle,
\right. \\
        &\hskip30mm \left. f_2 - \langle f_2, u\rangle g(u) = \left(k + k' - ka\|q_2 - q_1\|^{-1}\right)g(q_2 - q_0)
\right\}
                                                                                                        \tag \label{Fxt111}\endalign$$
    is the constitutive set of the system.

    If the configuration $q_2$ is not controlled, then the critical set is the union of sets
        $$Cr_1 = \left\{(q_1,q_2) \in Q \times Q ;\; q_1 - q_0 = \langle g(u), q_1 - q_0\rangle u,\; q_2 = q_0 \right\}
                                                                                                        \tag \label{Fxt112}$$
    and
        $$\align
        Cr_2 &= \left\{(q_1,q_2) \in Q \times Q ;\; q_1 - q_0 = \langle g(u), q_1 - q_0\rangle u, \right. \\
        &\hskip30mm \left. \langle g(u), q_2 - q_0\rangle = 0,\; (k + k')\|q_2 - q_1\| = ka \right\}.
                                                                                                        \tag \label{Fxt113}\endalign$$

    If the configuration $q_2$ is not controlled but observed, then the constitutive set is the union of sets
        $$\align
        \wS_1 &= \left\{(q_1,q_2,f_1) \in Q \times Q \times W^\*;\; q_1 - q_0 = \langle g(u), q_1 - q_0\rangle u,
\vphantom{\|^{-1}} \right. \\
        &\hskip30mm \left. q_2 = q_1,\; \langle f_1, u\rangle = k\left(1 - a\|q_1 - q_0\|^{-1}\right)\langle g(q_1 - q_0),
u\rangle \right\}
                                                                                                        \tag \label{Fxt114}\endalign$$
    and
        $$\align
        \wS_2 &= \left\{(q_1,q_2,f_1) \in Q \times Q \times W^\* ;\; q_1 - q_0 = \langle g(u), q_1 - q_0\rangle u,\;
\langle g(u), q_2 - q_0\rangle = 0, \right. \\
        &\hskip20mm \left. (k + k')\|q_2 - q_1\| = ka,\; \langle f_1, u\rangle = - k'\langle g(q_1 - q_0), u\rangle
\right\}.
                                                                                                        \tag \label{Fxt115}\endalign$$

    If the configuration $q_2$ is neither controlled nor observed, then the constitutive set is the union of sets
        $$\align
        S_1 &= \left\{(q_1,f_1) \in Q \times W^\*;\; q_1 - q_0 = \langle g(u), q_1 - q_0\rangle u, \vphantom{\|^{-1}}
\right. \\
        &\hskip30mm \left. \langle f_1, u\rangle = k\left(1 - a\|q_1 - q_0\|^{-1}\right)\langle g(q_1 - q_0), u\rangle
\right\}
                                                                                                        \tag \label{Fxt116}\endalign$$
    and
        $$\align
        S_2 &= \left\{(q_1,f_1) \in Q \times W^\* ;\; q_1 - q_0 = \langle g(u), q_1 - q_0\rangle u, \right. \\
        &\hskip20mm \left. (k + k')\|q_1 - q_0\| \leqs ka,\; \langle f_1, u\rangle = - k'\langle g(q_1 - q_0), u\rangle
\right\}.
                                                                                                        \tag \label{Fxt117}\endalign$$
    The constitutive set $S = S_1 \cup S_2$ is not a submanifold of $\sT^\*Q$.
        }{\hfill    $\blacktriangle$}\endclaim

        \claim \c{e}{Example}{}{\rm                                                                              \label{Cxt13}
    A material point with configuration $q_2$ in the affine space $Q$ is connected to a fixed point $q_0$ with a rigid rod
of length $a$.  A second material point with configuration $q_1$ is tied elastically to $q_2$ with a spring of spring
constant $k$.  The configuration space is the product $Q \times Q$ with holonomic constraints represented by
        $$C^0 = \left\{(q_1,q_2) \in Q \times Q ;\; \|q_2 - q_0\| = a \right\}.
                                                                                                        \tag \label{Fxt118}$$
    The set
        $$V = \left\{(q_1,q_2,\zd q_1,\zd q_2) \in \sT(Q \times Q) ;\; \|q_2 - q_0\| = a,\; \langle g(q_2 - q_1), \zd
q_2\rangle = 0 \right\}
                                                                                                        \tag \label{Fxt119}$$
    is the tangent set of $C^0$.  The set
        $$\align
        \oS &= \left\{(q_1,q_2,f_1,f_2) \in \sT^\*(Q \times Q) ;\; \|q_2 - q_0\| = a,\; f_1 = kg(q_1 - q_2), \right.
\hskip10mm\\
        &\hskip30mm \left. f_2 - kg(q_2 - q_1) = a^{-2}\langle f_2 - kg(q_2 - q_1), q_2 - q_0\rangle g(q_2 - q_0)  \right\}.
                                                                                                        \tag \label{Fxt120}\endalign$$
    is the constitutive set and
        $$Cr = \left\{(q_1,q_2) \in Q \times Q ;\; \|q_2 - q_0\| = a, \|q_1 - q_0\|(q_2 - q_0) = \pm a(q_1 - q_0) \right\}.
                                                                                                        \tag \label{Fxt121}$$
    is the critical set.

    If the configuration $q_2$ is not controlled but observed, then
        $$\align
        \wS &= \left\{(q_1,q_2,f_1) \in Q \times Q \times W^\* ;\; \|q_2 - q_0\| = a, \right. \\
            &\hskip30mm  \left. \|q_1 - q_0\|(q_2 - q_0) = \pm a(q_1 - q_0),\; f_1 = kg(q_1 - q_2) \right\}.
                                                                                                        \tag \label{Fxt122}\endalign$$
    is the constitutive set.

    If $q_2$ is not observed, then we have the constitutive set
        $$\align
        S &= \left\{(q_1,f_1) \in \sT^\*Q ;\; \|f_1\| = ka \;\text{ if }\; q_1 = q_0, \vphantom{\|^{-1}}\right. \\
            &\hskip20mm  \left. f_1 = k\left(1 \pm a\|q_1 - q_0\|^{-1}\right)g(q_1 - q_0) \;\text{ if }\; q_1 \neq q_0
\right\}.
                                                                                                        \tag \label{Fxt123}\endalign$$
    This set is the image of the injective mapping

        $$\zf \colon K \times \R \rightarrow Q \times W^\* \colon (w,r) \mapsto (q_0 + rw, k(r - a)g(w)),
                                                                                                        \tag \label{Fxt124}$$
    where $K$ is the unit sphere
        $$K = \left\{w \in W ;\; \|w\| = 1 \right\}.
                                                                                                        \tag \label{Fxt125}$$
    The space $\sT(K \times \R)$ is identified with the set
        $$\left\{(w,r,\zd w,\zd r) \in W \times \R \times (W \oplus \R) ;\; \|w\| = 1,\; \langle g(w), \zd w\rangle = 0
\right\}
                                                                                                        \tag \label{Fxt126}$$
    and the space $\sT(Q \times W^\*)$ is identified with $Q \times W^\* \times Q \times W^\*$.  The tangent mapping
        $$\align
    \sT\zf &\colon \sT(K \times \R) \rightarrow \sT(Q \times W^\*) \\
    &\colon (w,r,\zd w,\zd r) \mapsto (q_0 + rw,k(r - a)g(w),\zd rw + r\zd w,k\zd rg(w) + k(r - a)g(\zd w))
                                                                                                        \tag \label{Fxt127}\endalign$$
    maps injectively each tangent space of $K \times \R$ at $(w,r)$ into the tangent space of $Q \times W^\*$ at
$\zf(w,r)$. It follows that $\zf$ is an injective immersion.

    The space $\sT\sT^\*Q \fpr{\sT^\*Q} \sT\sT^\*Q$ is identified with $Q \times W^\* \times (W \oplus W^\*) \times (W
\oplus W^\*)$.  The symplectic form on $Q \times W^\*$ is introduced as the bilinear mapping
        $$\align
    \zw_Q &\colon Q \times W^\* \times (W \oplus W^\*) \times (W \oplus W^\*) \rightarrow \R \\
    &\colon (q,f,\zd_1 q,\zd_1 f,\zd_2 q,\zd_2 f) \mapsto \langle \zd_1 f, \zd_2 q\rangle - \langle \zd_2 f, \zd_1 q\rangle.
                                                                                                        \tag \label{Fxt128}\endalign$$
    The product $\sT(K \times \R) \fpr{K \times \R} \sT(K \times \R)$ is the set
        $$\align
        &\left\{(w,r,\zd_1 w,\zd_1 r,\zd_2 w,\zd_2 r) \in W \times \R \times (W \oplus \R) \times (W \oplus \R) ;\; \|w\| =
1, \right. \\
        &\hskip60mm\left. \langle g(w), \zd_1 w\rangle = 0,\; \langle g(w), \zd_2 w\rangle = 0 \right\}
                                                                                                        \tag \label{Fxt129}\endalign$$
    and the pull back form
        $$\align
    \zf^\*\zw_Q &\colon \sT(K \times \R) \fpr{K \times \R} \sT(K \times \R) \rightarrow \R \\
    &\colon (w,r,\zd_1 w,\zd_1 r,\zd_2 w,\zd_2 r) \mapsto \langle k\zd_1 rg(w) + k(r - a)g(\zd_1 w), \zd_2 rw + r\zd_2
w\rangle \\
        &\hskip45mm - \langle k\zd_2 rg(w) + k(r - a)g(\zd_2 w), \zd_1 rw + r\zd_1 w\rangle
                                                                                                        \tag \label{Fxt130}\endalign$$
    is zero.  We have shown that the image $S$ of $\zf$ is an immersed Lagrangian submanifold of $\sT^\*Q$.  It can be
shown to be embedded.  The mapping
        $$\zp_Q \circ \zf \colon K \times \R \rightarrow Q \colon (w,r) \mapsto q_0 + rw
                                                                                                        \tag \label{Fxt131}$$
    is not injective.  The tangent mapping
        $$\sT(\zp_Q \circ \zf) \colon \sT(K \times \R) \rightarrow \sT Q \colon (w,r,\zd w,\zd r) \mapsto (q_0 + rw,\zd rw +
r\zd w)
                                                                                                        \tag \label{Fxt132}$$
    is of rank 3 if $r \neq 0$ and of rank 1 if $r = 0$.  This indicates the presence of a {\it Lagrangian singularity} of
$S$ above the point $q_0$.

    Note that the critical set is not the image of a section of $\zh$.  For each control configuration $(x,y)$ we have two
different internal equilibrium configurations if $x^2 + y^2 > 0$ and an infinity of internal equilibrium configurations if
$x^2 + y^2 = 0$.  The external force necessary to maintain the control configuration $(x,y)$ depends on the internal
configuration.  Thus even if the internal configuration is not directly observed its presence can not be completely ignored.
        }{\hfill    $\blacktriangle$}\endclaim

        \sect{The Legendre transformation of composed static systems.}

    Let two static systems in a configuration manifold $Q$ be characterized by sets $(C^0{}_1,V_1,\zy_1,S_1)$ and
$(C^0{}_2,V_2,\zy_2S_2)$ respectively.  Let the {\it composition} of the systems be characterized by $(C^0,V,\zy,S)$.  The
constraints $C^0$ and $V$ are the intersections $C^0 = C^0{}_1 \cap C^0{}_2$ and $V = V_1 \cap V_2$.  The form $\zy \colon
V \rightarrow \R$ is the sum $\zy = \zy_1|V + \zy_2|V$.  Sets  $S_1{}_q = S_1 \cap \sT^\*_q Q$, $S_2{}_q = S_2 \cap
\sT^\*_q Q$, and $S_q = S \cap \sT^\*_q Q$ are defined by
        $$S_1{}_q = \left\{f \in \sT_q Q;\; f|V_1{}_q = \zy_1{}_q \right\},
                                                                                                        \tag \label{Fxt133}$$
        $$S_2{}_q = \left\{f \in \sT_q Q;\; f|V_2{}_q = \zy_2{}_q \right\},
                                                                                                        \tag \label{Fxt134}$$
    and
        $$S_q = \left\{f \in \sT_q Q;\; f|V_q = \zy_q \right\}.
                                                                                                        \tag \label{Fxt135}$$
    If $f_1 \in S_1{}_q$ and $f_2 \in S_2{}_q$, then $f_1 + f_2 \in S_q$.  This establishes the inclusion $S_1{}_q +
S_2{}_q \subset S_q$.  Assuming that sets $V_1{}_q = V_1 \cap \sT_q Q$, $V_2{}_q = V_2 \cap \sT_q Q$, and $V_q = V \cap
\sT_q Q$ are subspaces of $\sT_q Q$ we will show that the inclusion $S_q \subset S_1{}_q + S_2{}_q$ holds as well.  We
choose complements $V'_1{}_q$ of $V_q$ in $V_1{}_q$, $V'_2{}_q$ of $V_q$ in $V_2{}_q$, and $V_0{}_q$ of $V_1{}_q + V_2{}_q$
in $\sT_q Q$.  Relations
        $$\sT_q Q = V_1{}_q + V'_2{}_q + V_0{}_q = V_2{}_q + V'_1{}_q + V_0{}_q
                                                                                                        \tag \label{Fxt136}$$
        $$V_1{}_q \cap V'_2{}_q = \{0\},\;\;V'_2{}_q \cap V_0{}_q = \{0\},\;\;V_0{}_q \cap V_1{}_q = \{0\},
                                                                                                        \tag \label{Fxt137}$$
    and
        $$V_2{}_q \cap V'_1{}_q = \{0\},\;\;V'_1{}_q \cap V_0{}_q = \{0\},\;\;V_0{}_q \cap V_2{}_q = \{0\}
                                                                                                        \tag \label{Fxt138}$$
    hold.  If $f \in S_q$, then $f$ is the sum of covectors $f_1 \in S_1{}_q Q$ and $f_2 \in S_2{}_q Q$ characterized by
        $$f_1|V_1{}_q = \zy_1{}_q,\;\;f_1|V'_2{}_q = f|V'_2{}_q - \zy_2{}_q|V'_2{}_q,\;\;f_1|V_0{}_q =
\frac{1}{2}f|V_0{}_q,
                                                                                                        \tag \label{Fxt139}$$
    and
        $$f_2|V_2{}_q = \zy_2{}_q,\;\;f_2|V'_1{}_q = f|V'_2{}_q - \zy_1{}_q|V'_1{}_q,\;\;f_2|V_0{}_q = \frac{1}{2}f|V_0{}_q.
                                                                                                        \tag \label{Fxt140}$$

    We have established rules of composition of static systems is sufficiently simple cases.  These rules are expressed by

        $$C^0 = C^0{}_1 \cap C^0{}_2,
                                                                                                        \tag \label{Fxt141}$$
        $$V = V_1 \cap V_2,
                                                                                                        \tag \label{Fxt142}$$
        $$\zy = \zy_1|V + \zy_2|V,
                                                                                                        \tag \label{Fxt143}$$
    and
        $$S_q = S_1{}_q + S_2{}_q
                                                                                                        \tag \label{Fxt144}$$
    for each $q \in C^0$.

    Two submanifolds $C_1$ and $C_2$ of a manifold $Q$ are said to have {\it clean intersection} if the intersection $C_1
\cap C_2$ is a submanifold and $\sT C_1 \cap \sT C_2 = \sT(C_1 \cap C_2)$.  Examples show that the simple rules listed
above do not apply to the composition of systems with holonomoic constraints if the intersection of constraints is not
clean.

        \claim \c{e}{Example}{}{\rm                                                                              \label{Cxt14}
    We consider the composition of two holonomic systems of the type described in Example \Ref{Cxt7}.  The constraints and
the constitutive set for the first system are represented by the sets
        $$C^0{}_1 = \left\{q \in Q ;\; \|q - q_1\| = a \right\},
                                                                                                        \tag \label{Fxt145}$$
        $$V_1 = \left\{(q,\zd q) \in Q \times W ;\; \|q - q_1\| = a,\; \langle g(q - q_1), \zd q\rangle = 0 \right\},
                                                                                                        \tag \label{Fxt146}$$
    and
        $$S_1 = \left\{(q,f) \in Q \times W^\* ;\; \|q - q_2\| = a,\; f = a^{-2}\langle f, q - q_1\rangle g(q - q_1)
\right\}.
                                                                                                        \tag \label{Fxt147}$$
    For the second system we have
        $$C^0{}_2 = \left\{q \in Q ;\; \|q - q_2\| = a \right\},
                                                                                                        \tag \label{Fxt148}$$
        $$V_2 = \left\{(q,\zd q) \in Q \times W ;\; \|q - q_2\| = a,\; \langle g(q - q_2), \zd q\rangle = 0 \right\},
                                                                                                        \tag \label{Fxt149}$$
    and
        $$S_2 = \left\{(q,f) \in Q \times W^\* ;\; \|q - q_2\| = a,\; f = a^{-2}\langle f, q - q_2\rangle g(q - q_2)
\right\}.
                                                                                                        \tag \label{Fxt150}$$

    If the distance $\|q_2 - q_1\|$ between the centres of the spheres $C^0{}_1$ and $C^0{}_2$ is less than $2a$, then the
composed system is a system with holonomic constraints.  The intersection
        $$C^0 = C^0{}_1 \cap C^0{}_2 = \left\{q \in Q ;\; \|q - q_1\| = a,\; \|q - q_2\| = a \right\}
                                                                                                        \tag \label{Fxt151}$$
    is clean.  The intersection
        $$\align
        V = V_1 \cap V_2 &= \left\{(q,\zd q) \in Q \times W ;\; \|q - q_1\| = a,\; \|q - q_2\| = a, \right. \\
    &\hskip40mm \left. \langle g(q - q_1), \zd q\rangle = 0,\; \langle g(q - q_2), \zd q\rangle = 0 \right\}
                                                                                                        \tag \label{Fxt152}\endalign$$
    is the tangent set $\sT C^0$.  The constitutive set
        $$\align
    S &= \left\{(q,f) \in Q \times W^\* ;\; \|q - q_1\| = a,\; \|q - q_2\| = a,\; \langle f, \zd q\rangle = 0\; \text{ for
each} \right. \\
        &\hskip20mm \left. \zd q \in W \;\text{ such that }\; \langle g(q - q_1), \zd q\rangle = 0\;
\text{ and }\; \langle g(q - q_2), \zd q\rangle = 0 \right\}.
                                                                                                        \tag \label{Fxt153}\endalign$$
    is obtained from the principle of virtual work.  At each $q \in C^0$ the set
        $$S_q = \left\{f \in W^\* ;\; (q,f) \in S \right\}
                                                                                                        \tag \label{Fxt154}$$
    is the sum
        $$\left\{f \in W^\* ;\; (q,f) \in S_1 \right\} + \left\{f \in W^\* ;\; (q,f) \in S_2 \right\}.
                                                                                                        \tag \label{Fxt155}$$
    The simple rules are followed in this case.

    If $\|q_2 - q_1\| = 2a$, then the set
        $$C^0 = C^0{}_1 \cap C^0{}_2 = \left\{q \in Q ;\; \|q - q_1\| = a,\; \|q - q_2\| = a \right\}
                                                                                                        \tag \label{Fxt156}$$
    has only one element $q = q_1 + \frac{1}{2}(q_2 - q_1)$.  The intersection $V_1 \cap V_2$ is the set
        $$\left\{(q,\zd q) \in Q \times W ;\; q = q_1 + \frac{1}{2}(q_2 - q_1),\; \langle g(q_2 - q_1), \zd q\rangle =
0 \right\}.
                                                                                                        \tag \label{Fxt157}$$
    This is not the tangent set of $C^0$.  The intersection of $C^0{}_1$ with $C^0{}_2$ is not clean.  With
        $$V = \sT C^0 = \left\{(q,f) \in Q \times W^\* ;\; q = q_1 + \frac{1}{2}(q_2 - q_1),\; \zd q = 0 \right\}
                                                                                                        \tag \label{Fxt158}$$
    the principle of virtual work produces the constitutive set
        $$S = \left\{(q,f) \in Q \times W^\* ;\; q = q_1 + \frac{1}{2}(q_2 - q_1) \right\}.
                                                                                                        \tag \label{Fxt159}$$
    This is not the correct constitutive set for the composed system.  The reason of this failure is that the approximative
assumption of perfect rigidity of the separate constraints is no longer realistic in the case of composition which is not
clean.  To obtain a complete description of the composed system the precise elastic properties of the constraints must be
known.  A partial characterization of the system is provided by the constitutive set
        $$S = \left\{(q,f) \in Q \times W^\* ;\; q = q_1 + \frac{1}{2}(q_2 - q_1),\; f = a^{-2}\langle f, q - q_1\rangle
g(q - q_1) \right\}
                                                                                                        \tag \label{Fxt160}$$
    generated by the non holonomic constraints $V = V_1 \cap V_2$.  Note that these constraints are not integrable since
the inclusion $V \subset \sT C^0$ does not hold.  The simple composition rules can not be followed.
        }{\hfill    $\blacktriangle$}\endclaim

        \sect{Configuration spaces for the dynamics of autonomous machanical systems.}

    We consider the space $\sQ(M|\R;[t_0,t_1])$ of differentiable mappings $\zg \colon [t_0,t_1] \rightarrow \R$ of an
interval $[t_0,t_1] \subset \R$ in a manifold $M$. This space serves as the configuration space for dynamics in the time
interval $[t_0,t_1]$.  It is not a differential manifold.  Basic constructions of differential geometry are nevertheless
possible.  Tangent vectors and covectors can be defined if a class of functions and a class of curves are chosen to act as
differentiable mappings.

    From a function
        $$\zf \colon \sT M \rightarrow \R
                                                                                                        \tag \label{Fxt161}$$
    we derive the function
        $$F \colon \sQ(M|\R;[t_0,t_1]) \rightarrow \R \colon \zg \mapsto \int_{\textstyle t_0}^{\textstyle t_1}\zf \circ
\st\zg.
                                                                                                        \tag \label{Fxt162}$$
    Functions on $\sQ(M|\R;[t_0,t_1])$ obtained by this construction are declared differentiable.

    Starting with any differentiable mapping
        $$\zq \colon \R \times [t_0,t_1] \rightarrow M
                                                                                                        \tag \label{Fxt163}$$
    we introduce the curve
        $$\zG \colon \R \rightarrow \sQ(M|\R;[t_0,t_1]) \colon s \mapsto \zq(s,\,\cdot\,)
                                                                                                        \tag \label{Fxt164}$$
    in $\sQ(M|\R;[t_0,t_1])$ considered differentiable.

    The restriction of a differentiable curve to an interval $[0,a]$ is a parametrization of a process.  Virtual
displacements are tangent vectors defined as equivalence classes of differentiable curves.  Curves $\zG$ and $\zG'$ are
equivalent if
        $$(F \circ \zG')(0) = (F \circ \zG)(0)
                                                                                                        \tag \label{Fxt165}$$
    and
        $$\xD(F \circ \zG')(0) = \xD(F \circ \zG)(0)
                                                                                                        \tag \label{Fxt166}$$
    for each differentiable function $F$.  Let $\zG$ and $\zG'$ be derived from mappings $\zq \colon \R \times [t_0,t_1]
\rightarrow M$ and $\zq' \colon \R \times [t_0,t_1] \rightarrow M$ and let $F$ be constructed from a mapping $\zf \colon
\sT M \rightarrow \R$.  Equivalence conditions involve the function
        $$F \circ \zG \colon \R \rightarrow \R \colon s \mapsto \int_{\textstyle t_0}^{\textstyle t_1}\zf \circ
\st\zq(s,\,\cdot\,)
                                                                                                        \tag \label{Fxt167}$$
    and a similar function with $\zq'$.  Equivalence conditions will be satisfied if $\zq$ and $\zq'$ produce the same
mapping
        $$w \colon [t_0,t_1] \rightarrow \sT M \colon t \mapsto \st\zq(\,\cdot\,,t)(0).
                                                                                                        \tag \label{Fxt168}$$
    This mapping is a natural choice of a representative of the equivalence class of $\zG$.

    A covector is an equivalence class of pairs $(q,F)$ of a configuration $q \in \sQ(M|\R;[t_0,t_1])$ and a differentiable
function $F$ on $\sQ(M|\R;[t_0,t_1])$.  Pairs $(q,F)$ and $(q',F')$ are equivalent if $q' = q$ and
        $$\xD(F' \circ \zG)(0) = \xD(F \circ \zG)(0)
                                                                                                        \tag \label{Fxt169}$$
    for each differentiable curve $\zG$.

    Let $F$, $F'$ and $\zG$ be derived from mappings $\zf \colon \sT M \rightarrow \R$, $\zf \colon \sT M \rightarrow \R$
and $\zq \colon \R \times [t_0,t_1] \rightarrow M$ respectively.  It follows from the formula
        $$\xD(F \circ \zG)(0) = \int_{\textstyle t_0}^{\textstyle t_1}\langle \xd\zf, \st w\rangle
                                                                                                        \tag \label{Fxt170}$$
    that the mapping $\xd\zf \circ \st\zq(0,\,\cdot\,)$ could be used to represent the equivalence class of $(q,F)$.  This
is not a convenient representation since $\st w$ is not a generic mapping from $[t_0,t_1]$ to $\sT\sT M$.  A better
representation is derived from the equality
        $$\xD(F \circ \zG)(0) = -\int_{\textstyle t_0}^{\textstyle t_1}\langle \cE\zf \circ \st^2\zq(0,\,\cdot\,), w\rangle
+ \langle(\cP\zf \circ \st\zq(0,\,\cdot\,)), w(t_1) \rangle - \langle(\cP\zf \circ \st\zq(0,\,\cdot\,)), w(t_0) \rangle,
                                                                                                        \tag \label{Fxt171}$$
    where
        $$\cE\zf \colon \sT^2 M \rightarrow \sT^\*M
                                                                                                        \tag \label{Fxt172}$$
    and
        $$\cP\zf \colon \sT M \rightarrow \sT^\*M
                                                                                                        \tag \label{Fxt173}$$
    are mappings defined in [DA].  The equality suggests that a covector should be represented by a mapping
        $$f \colon [t_0,t_1] \rightarrow \sT^*M
                                                                                                        \tag \label{Fxt174}$$
    and two covectors $p_0$ and $p_1$ in $\sT^*M$.  The pairing between a covector represented by $(f,p_0,p_1)$ and a
vector represented by $w$ is expressed by
        $$\langle (f,p_0,p_1)\rangle = -\int_{\textstyle t_0}^{\textstyle t_1}\langle f, w\rangle + \langle p_1, w(t_1)
\rangle - \langle p_0, w(t_0) \rangle.
                                                                                                        \tag \label{Fxt175}$$

    In dynamics the term {\it action} replaces the term {\it work} in statics.  For simple autonomous systems usually
considered in analytical mechanics all processes are admissible and the action of a process is the difference of values at
the end points of the process of a function
        $$L \colon \sQ(M|\R;[t_0,t_1]) \rightarrow \R \colon \zg \mapsto \int_{\textstyle t_0}^{\textstyle t_1}\zl \circ
\st\zg
                                                                                                        \tag \label{Fxt176}$$
    derived from a {\it Lagrangian}
        $$\zl \colon \sT M \rightarrow \R.
                                                                                                        \tag \label{Fxt177}$$
    The variational principle
        $$\int_{\textstyle t_0}^{\textstyle t_1}\langle \xd\zl, \st w\rangle = -\int_{\textstyle t_0}^{\textstyle
t_1}\langle f, w\rangle + \langle p_1, w(t_1) \rangle - \langle p_0, w(t_0) \rangle
                                                                                                        \tag \label{Fxt178}$$
    produces the {\it Euler-Lagrange equations}
        $$\cE\zl \circ \st^2\zg = f
                                                                                                        \tag \label{Fxt179}$$
    in $[t_0,t_1]$ and the {\it momentum-velocity relations}
        $$(\cP\zl \circ \st\zg)(t_0) = p_0
                                                                                                        \tag \label{Fxt180}$$
    and
        $$(\cP\zl \circ \dot\zg)(t_1) = p_1.
                                                                                                        \tag \label{Fxt181}$$

    The formulation of analytical mechanics in a finite time interval has its infinitesimal limit with the interval
shrinking to a point.  The passage to the infinitesimal limit is greatly simplified by observing that in a slightly modified
version of the formulation the interval $[t_0,t_1]$ is used exclusively as a domain of integration and can be treated as a
current or distribution.  Replacing the interval by the Dirac delta $\zd_t$ at $t \in \R$ produces the infinitesimal limit.
We will describe the modified version of the formulation using currents in $\R$ in place of intrvals.

    We introduce an equivalence relation in the set of pairs $(\bc,\zg)$, where $\bc$ is a current in $\R$ and $\zg$ is a
curve in $M$.  Pairs $(\bc,\zg)$ and $(\bc',\zg')$ are equivalent if $\bc' = \bc$ and
        $$\int_{\displaystyle\bc} \zf \circ \st\zg' = \int_{\displaystyle\bc} \zf \circ \st\zg
                                                                                                        \tag \label{Fxt182}$$
    for each differentiable function $\zf \colon \sT M \rightarrow \R$.  The space $\sQ(M|\R;\bc)$ of equivalence classes is
the {\it configuration space} corresponding to the current $\bc$.  The equivalence class of a pair $(\bc,\zg)$ will be
denoted by $\sq(\bc,\zg)$.

    Each differentiable function $\zf \colon \sT M \rightarrow \R$ defines a function
        $$F \colon \sQ(M|\R;\bc) \rightarrow \R \colon \sq(\bc,\zg) \mapsto \int_{\displaystyle\bc} \zf \circ \st\zg
                                                                                                        \tag \label{Fxt183}$$
    considered differentiable.

    From a mapping
        $$\zq \colon \R \times \R \rightarrow M
                                                                                                        \tag \label{Fxt184}$$
    we construct the curve
        $$\zG \colon \R \rightarrow \sQ(M|\R;\bc) \colon s \mapsto \sq(\bc,\zq(s,\,\cdot\,))
                                                                                                        \tag \label{Fxt185}$$
    in $\sQ(M|\R;\bc)$ differentiable by declaration.

    Tangent vectors are defined as equivalence classes of differentiable curves.  Equalities \Ref{Fxt165} and \Ref{Fxt166}
define equivalence of curves.  The space of tangent vectors is denoted by $\sT\sQ(M|\R;\bc)$.  A natural representative of a
tangent vector is an equivalence class of pairs $(\bc,w)$, where $\bc$ is a current in $\R$ and $w$ is a curve in $\sT M$.
Pairs $(\bc,w)$ and $(\bc',w')$ are equivalent if $\bc' = \bc$ and
        $$\int_{\displaystyle\bc}\langle \xd\zf, \st w'\rangle = \int_{\displaystyle\bc}\langle \xd\zf, \st w\rangle.
                                                                                                        \tag \label{Fxt186}$$
    The space of equivalence classes will be denoted by $\sQ(\sT M|\R;\bc)$.

    It is convenient to consider the construction of spaces $\sQ(M|\R;\bc)$ and $\sQ(\sT M|\R;\bc)$ as the results of
applications of a covariant functor $\sQ(\,\cdot\,|\R;\bc)$ to manifolds $M$ and $\sT M$.  This functor applied to the
morphism $\zt_M \colon \sT M \rightarrow M$ results in the morphism
        $$\sQ(\zt_M|\R;\bc) \colon \sQ(\sT M|\R;\bc) \rightarrow \sQ(M|\R;\bc) \colon \sq(\bc,w) \mapsto \sq(\bc,\zt_M \circ w)
                                                                                                        \tag \label{Fxt187}$$
    which can be treated as a vector fibration.  The morphism
        $$\zt_{\sQ(M|\R;\bc)} \colon \sT\sQ(M|\R;\bc) \rightarrow \sQ(M|\R;\bc) \colon \st\zG(0) \mapsto \zG(0)
                                                                                                        \tag \label{Fxt188}$$
    can be turned into a vector fibration by requiring that the diagram
    \vskip2mm
    \vskip-1mm$$\xy!<10mm>
        $$\xymatrix@R+8mm @C+22mm{{\sT\sQ(M|\R;\bc)} \ar[d]_*{\zt_{\sQ(M|\R;\bc)}} \ar[r]^*{\zk_{\sQ(M|\R;\bc)}} &
            \sQ(\sT M|\R;\bc) \ar[d]_*{\sQ(\zt_M|\R;\bc)} \\
            \sQ(M|\R;\bc) \ar@{=}[r] & \sQ(M|\R;\bc)}$$\endxy
                                                                                                        \tag \label{Fxt189}$$
    \vskip2mm
    \noindent be a vector fibration morphism with the morphism $\zk_{\sQ(M|\R;\bc)}$ defined by
        $$\zk_{\sQ(M|\R;\bc)} \colon \sT\sQ(M|\R;\bc) \rightarrow \sQ(\sT M|\R;\bc) \colon \st\zG(0) \mapsto \sq(\bc,w)
                                                                                                        \tag \label{Fxt190}$$
    in terms of a curve
        $$\zG \colon \R \rightarrow \sQ(M|\R;[t_0,t_1]) \colon s \mapsto \zq(s,\,\cdot\,)
                                                                                                        \tag \label{Fxt191}$$
    and and a curve
        $$w \colon \R \rightarrow \sT M \colon t \mapsto \st\zq(\,\cdot\,,t)(0).
                                                                                                        \tag \label{Fxt192}$$
    both derived from the same mapping $\zq \colon \R \times \R \rightarrow M$.  Note that $\zk_{\sQ(\,\cdot\,|\R;\bc)}$ is a
natural transformation from the composition $\sT\sQ(\,\cdot\,|\R;\bc)$ to the composition $\sQ(\sT\,\cdot\,|\R;\bc)$ of
functors $\sT$ and $\sQ(\,\cdot\,|\R;\bc)$ composed in reverse order.

    Covectors are equivalence classes of pairs $(q,F)$ of a configuration $q \in \sQ(M|\R;\bc)$ and a differentiable function
$F$ on $\sQ(M|\R;\bc)$.  Pairs $(q,F)$ and $(q',F')$ are equivalent if $q' = q$ and
        $$\xD(F' \circ \zG)(0) = \xD(F \circ \zG)(0)
                                                                                                        \tag \label{Fxt193}$$
    for each differentiable curve $\zG$.  The space of cotangent vectors will be denoted by $\sT^\*\sQ(M|\R;\bc)$.  The
equivalence class of $(q,F)$ is denoted by $\xd F(q)$.  The mapping $\zp_{\sQ(M|\R;\bc)} \colon \sT^\*\sQ(M|\R;\bc)
\rightarrow \sQ(M|\R;\bc)$ is a vector fibration.  There is an alternative reprsesentation of covectors.  A covector can be
represented as an equivalence class of triples $(\bc,p,f)$, where $\bc$ is a current in $\R$ and $p$ and $f$ are curves $p
\colon \R \rightarrow \sT^\*M$ and $f \colon \R \rightarrow \sT^\*M$ such that $\zp_M \circ f = \zp_M \circ p$.  Triples
$(\bc,p,f)$ and $(\bc',p',f')$ are equivalent if $\bc' = \bc$ and
        $$-\int_{\displaystyle\bc} \langle f', w\rangle + \int_{\displaystyle\partial\bc} \langle p', w\rangle =
-\int_{\displaystyle\bc} \langle f, w\rangle + \int_{\displaystyle\partial\bc} \langle p, w\rangle
                                                                                                        \tag \label{Fxt194}$$
    for each mapping $w \colon \R \rightarrow \sT M$ such that $\zt_M \circ w = \zp_M \circ f$.  The space of equivalence
classes will be denoted by $\sP(\sT^\*M|\R;\bc)$ and $\sp(\bc,p,f)$ will denote the class of $(\bc,p,f)$.  The mapping
        $$\sP(\zp_M|\R;\bc) \colon \sP(\sT^\*M|\R;\bc) \rightarrow \sQ(M|\R;\bc) \colon \sp(\bc,p,f) \mapsto \sq(\bc,\zp_M
\circ f)
                                                                                                        \tag \label{Fxt195}$$
    is a vector fibration dual to the fibration $\sQ(\zt_M|\R;\bc)$ with the pairing
        $$\langle\,\;,\; \rangle \colon \sP(\sT^\*M|\R;\bc) \times_{\sQ(M|\R;\bc))} \sQ(\sT M|\R;\bc)
\rightarrow \R \colon (\sp(\bc,p,f),\sq(\bc,w)) \mapsto -\int_{\displaystyle\bc} \langle f, w\rangle +
\int_{\displaystyle\partial\bc} \langle p, w\rangle.
                                                                                                        \tag \label{Fxt196}$$
    The diagram
    \vskip2mm
    \vskip-1mm$$\xy!<10mm>
        $$\xymatrix@R+8mm @C+22mm{{\sP(\sT^\*M|\R;\bc)} \ar[d]_*{\sP(\zp_M|\R;\bc)} \ar[r]^*{\za_{\sQ(M|\R;\bc)}} &
            \sT^\*\sQ(M|\R;\bc) \ar[d]_*{\zp_{\sQ(M|\R;\bc)}} \\
            \sQ(M|\R;\bc) \ar@{=}[r] & \sQ(M|\R;\bc)}$$\endxy
                                                                                                        \tag \label{Fxt197}$$
    \vskip2mm
    \noindent is a vector fibration morphism dual to \Ref{Fxt189}.  The mapping $\za_{\sQ(M|\R;\bc)}$ is defined by
        $$\za_{\sQ(M|\R;\bc)}(\sp(\bc,p,f)) = \xd F(\sq(\bc,\zg)),
                                                                                                        \tag \label{Fxt198}$$
    where the function $F$ and the curves $p$ and $f$ are obtained from the constructions
        $$F \colon \sQ(M|\R;\bc) \rightarrow \R \colon \sq(\bc,\zg) \mapsto \int_{\displaystyle\bc} \zf \circ \st\zg,
                                                                                                        \tag \label{Fxt199}$$
        $$p = \cP\zf \circ \st^2\zg,
                                                                                                        \tag \label{Fxt200}$$
        $$f = \cE\zf \circ \st\zg
                                                                                                        \tag \label{Fxt201}$$
    in terms of a mapping $\zf \colon \sT M \rightarrow \R$ and a curve $\zg \colon \R \rightarrow M$.

    For a Lagrangian mechanical system the dynamics is contained in the variational principle
        $$\int_{\displaystyle\bc} \langle \xd\zl, w\rangle = -\int_{\displaystyle\bc} \langle f, w\rangle +
\int_{\displaystyle\partial\bc} \langle p, w\rangle.
                                                                                                        \tag \label{Fxt202}$$

    There are two important types of currents in $\R$: an interval $\bc = [t_0,t_1]$ and a Dirac current $\bc = \zd_t$.  If
$\bc = [t_0,t_1]$, then the configuration space $\sQ(M|\R;[t_0,t_1])$ is the space introduced earlier.  If $\bc = \zd_t$
then $\sQ(M|\R;\bc) = \sT M$.  The equivalence class of $(\zd_t,\zg)$ is the velocity $\st\zg(t)$.  Since the space
$\sQ(M|\R;\zd_t)$ is a differential manifold differentiable functions and curves are well defined.  Spaces
$\sT\sQ(M|\R;\bc)$ and $\sQ(\sT M|\R;\bc)$ are both equal to $\sT\sT M$.  Fibrations $\zt_{\sQ(M|\R;\bc)}$ and
$\sQ(\zt_M\R;\bc)$ assume the form $\zt_{\sT M} \colon \sT\sT M \rightarrow \sT M$ and $\sT\zt_M \colon \sT\sT M
\rightarrow \sT M$ respectively.  The diagram \Ref{Fxt189} is in this case the familiar isomorphism
    \vskip2mm
    \vskip-1mm$$\xy!<10mm>
        $$\xymatrix@R+8mm @C+22mm{{\sT\sT M} \ar[d]_*{\zt_{\sT M}} \ar[r]^*{\zk_M} &
            \sT\sT M \ar[d]_*{\sT\zt_M} \\
            \sT M \ar@{=}[r] & \sT M}$$\endxy
                                                                                                        \tag \label{Fxt203}$$
    \vskip2mm
    \noindent of vector fibrations [DA].  The space is the cotangent bundle $\sT^\*\sT M$.  The space $\sT^\*\sQ(M|\R;\bc)$
is the fibre product
        $$\sT\sT^\*M \times_M \sT^\*M = \left\{(\dot p,f) \in \sT\sT^\*M \times \sT^\*M ;\; \zp_M(f) =
\zp_M(\zt_{\sT^\*M}(p)) \right\}.
                                                                                                        \tag \label{Fxt204}$$
    The equivalence class of $(\zd_t,p,f)$ is the pair $(\st p(t),f(t)$.  The pairig \Ref{Fxt196} assumes the form
        $$\langle\,\;,\; \rangle \colon (\sT\sT^\*M \times_M \sT^\*M) \times_{\sT M} \sT\sT M \rightarrow \R \colon ((\dot
p,f),w) \mapsto \langle \dot p, w\rangle^\ssT - \langle f, \zt_{\sT M}(w)\rangle.
                                                                                                        \tag \label{Fxt205}$$
    The domain of this pairing is the set
        $$\left\{((\dot p,f),w) \in (\sT\sT^\*M \times_M \sT^\*M) \times \sT\sT M ;\; \sT\zp_M(\dot p) = \sT\zt_M(w)\right\}
                                                                                                        \tag \label{Fxt206}$$
    and the pairing $\langle \,\;,\;\rangle^\ssT$ is the tangent pairing defined in [DA].  The diagram
    \vskip1mm
        $$\xymatrix@R+8mm @C+12mm{{\sT\sT^\*M \times_M \sT^\*M} \ar[d]_*{\sT\zp_M} \ar[r]^*{\za_M} &
            \sT^\*\sT M \ar[d]_*{\zp_{\sT M}} \\
            \sT M \ar@{=}[r] & \sT M}
                                                                                                        \tag \label{Fxt207}$$
    \vskip2mm
    \noindent is a morphism dual to the vector fibration isomorphism \Ref{Fxt203}.  The better known version
    \vskip1mm
        $$\hskip5mm\xymatrix@R+8mm @C+20mm{{\sT\sT^\*M} \ar[d]_*{\sT\zp_M} \ar[r]^*{\za_M} &
            \sT^\*\sT M \ar[d]_*{\zp_{\sT M}} \\
            \sT M \ar@{=}[r] & \sT M}
                                                                                                        \tag \label{Fxt208}$$
    \vskip2mm
    \noindent of this morphism is obtained by setting $f = 0$ and using the tangent pairing
        $$\langle\,\;,\; \rangle \colon \sT\sT^\*M \times_{\sT M} \sT\sT M \rightarrow \R \colon (\dot
p,w) \mapsto \langle \dot p, w\rangle^\ssT
                                                                                                        \tag \label{Fxt209}$$
    in place of \Ref{Fxt205}.

    The infinitesimal limit of dynamics is formulated in [DA].

        \sect{Configuration spaces for the statics of continuous media.}
\vskip3mm
        \ssca{The configuration of a body, strain.}

    Let $M$ be the {\it material space} and $N$ the {\it physical space} of a medium.  The physical space is a Euclidean
affine space of dimension 3.  By treating this space as a Riemannian differential manifold we simplify the presentation of
the geometrical framework for statics.  The material space is a differential manifold.  The set of local diffeomorphisms
$\zg \colon M \rightarrow N$ each defined on an open subset $\Dm(\zg) \subset M$ will be denoted by $\cQ\NM$.  Elements of
this set are called {\it placement maps}.  The 1-jet prolongation $\sj^1\zg \colon M \rightarrow \sJ^1\NM$ of a placement
map is an important ingredient in the definition of $strain$.  This jet prolongation is equivalentlently represented by the
tangent map $\sT\zg$.  A {\it body} is represented by an odd current $\bc$ of dimension $m = \dim(M)$ and compact support
$\Sup(\bc)$.  The vector space of such currents will be denoted by $\bC M$.  We consider commutative diagrams
    \vskip2mm
    \vskip-1mm$$\xy!<10mm>
        $$\xymatrix@R+6mm @C+10mm{{\sJ^1\NM} \ar[d]_*{\zs_1{}_{\NM}} \ar[r]^*{\zk} &
            \tW^m\sT^\* M \ar[d]_*{\tilde\zp_m{}_M} \\
            M \ar@{=}[r] & M}$$\endxy
                                                                                                        \tag \label{Fxt210}$$
    \vskip2mm
    \noindent where $\tzp_m{}_M \colon \tW^m\sT^\* M \rightarrow M$ is the fibration of odd $m$-covectors in $M$ and
$\zs_1{}_{\NM} \colon \sJ^1\NM \rightarrow M$ is the jet-source projection.  The space of such diagrams will be denoted by
$\cK\NM$.  Each diagram is well represented by the mapping $\zk \colon \sJ^1\NM \rightarrow \tW^m\sT^\* M$.

    Let $\bC M \ux \cQ\NM$ denote the set
        $$\left\{(\bc,\zg) \in \bC M \times \cQ\NM ;\; \Sup(\bc) \subset \Dm(\zg) \right\}.
                                                                                                        \tag \label{Fxt211}$$
    In this {\it restricted product} set we introduce an equivalence relation.  Pairs $(\bc,\zg)$ and
$(\bc',\zg')$ are equivalent if $\bc' = \bc$ and
        $$\int_{\displaystyle\bc} \zk \circ \sj^1\zg' = \int_{\displaystyle\bc} \zk \circ \sj^1\zg
                                                                                                        \tag \label{Fxt212}$$
    for each $\zk \in \cK\NM$.  The composition $\zk \circ \sj^1\zg$ can be integrated since as a section of the fibration
$\tzp_m{}_M \colon \tW^m\sT^\* M \rightarrow M$ it is an odd $m$-form on $M$ also called a {\it scalar density}.  The
equivalence class of $(\bc,\zg)$ denoted by $\bq(\bc,\zg)$ is a {\it configuration} of the body $\bc$.  The set of all
equivalence classes is denoted by $\bQ\NM$. There is the natural projection $\bzg_{\NM} \colon \bQ\NM \rightarrow \bC M
\colon \bq(\bc,\zg) \mapsto \bc$. A fibre $\bQ\NMc = \bzg_{\NM}^{-1}(\bc)$ is the {\it configuration space} of the body
$\bc$ and the set $\bQ\NM$ is the {\it configuration bundle}.

\vskip3mm
        \sscx{Vectors and covectors.}

    Each element $\zk \in \cK\NM$ induces a function
        $$\bzl \colon \bQ\NM \rightarrow \R \colon \bq(\bc,\zg) \mapsto \int_{\displaystyle\bc} \zk \circ
\sj^1\zg.
                                                                                                        \tag \label{Fxt213}$$
    Functions constructed in this way are considered {\it differentiable}.  The space of differentiable functions on
$\bQ\NM$ will be denoted by $\bK\NM$.

    The tangent bundle of the set $\bC M$ can be identified with the product $\bC M \times \bC M$ since this set is a
vector space.

    We consider the set $\cG\NM$ of differentiable mappings $\zq \colon \R \times M \rightarrow N$ such that for each $s
\in \R$ the mapping $\zq(s,\cdot)$ is a local diffeomorphism defined on an open subset $\Dm(\zq(s,\cdot)) \subset M$.  With
each element $(\bc,\zd\bc,\zq)$ of the {\it restricted product}
        $$\left\{(\bc,\zd\bc,\zq) \in \bC M \times \bC M \times \cQ\NM ;\; \all{s \in \R} \Sup(\bc + s\zd\bc) \subset
\Dm(\zq(s,\cdot)) \right\}
                                                                                                        \tag \label{Fxt214}$$
    we associate the mapping
        $$\bg(\bc,\zd\bc,\zq) \colon \R \rightarrow \bQ\NM \colon s \mapsto \bq(\bc + s\zd\bc,\zq(s,\cdot)).
                                                                                                        \tag \label{Fxt215}$$
    Mappings constructed in this way will be considered {\it differentiable curves}.  The set of differentiable curves will
be denoted by $\bG\NM$.

    {\it Tangent vectors} in $\bQ\NM$ are defined as equivalence classes of differentiable curves.  Curves $\lg$ and $\lg'$
are equivalent if
        $$\lg'(0) = \lg(0)
                                                                                                        \tag \label{Fxt216}$$
    and
        $$\xD(\bzl \circ \lg')(0) = \xD(\bzl \circ \lg)(0)
                                                                                                        \tag \label{Fxt217}$$
    for each differentiable function $\bzl$.  The equivalence class of $\lg$ will be denoted by $\bt\lg(0)$.  The space of
tangent vectors will be denoted by $\bT\bQ\NM$.  It will be shown that the fibres of the projection
        $$\bzt_{\NM} \colon \bT\bQ\NM \rightarrow \bQ\NM \colon \bt\lg(0) \mapsto \lg(0)
                                                                                                        \tag \label{Fxt218}$$
    are vector spaces.

    {\it Covectors} in $\bQ\NM$ are equivalence classes of pairs $(\lq,\bzl) \in \bQ\NM \times \bK\NM$.  Two pairs
$(\lq,\bzl)$ and $(\lq',\bzl')$ are equivalent if $\lq' = \lq$ and
        $$\xD(\bzl' \circ \bg(\bc,\zd\bc,\zq))(0) = \xD(\bzl \circ \bg(\bc,\zd\bc,\zq))(0)
                                                                                                        \tag \label{Fxt219}$$
    for each differentiable curve $\lg \colon \R \rightarrow \bQ\NM$ such that $\lg(0) = \lq$.  The equivalence class of
$(\lq,\bzl)$ will be denoted by $\xd\bzl(\lq)$.  The bundle of covectors will be denoted by $\bT^\*\bQ\NM$.  Fibres of the
fibration
        $$\bzp_\NM \colon \bT^\*\bQ\NM \rightarrow \bQ\NM \colon \xd\bzl(\lq) \mapsto \lq
                                                                                                        \tag \label{Fxt220}$$
    are vector spaces.

\vskip3mm
        \sscx{Vertical vectors and covectors.}

    A {\it vertical differentiable curve} is a mapping
        $$\bg(\bc,\zq) \colon \R \rightarrow \bQ\NM \colon s \mapsto \bq(\bc,\zq(s,\cdot))
                                                                                                        \tag \label{Fxt221}$$
    associated with an element $(\bc,\zq)$ of the {\it restricted product}
        $$\left\{(\bc,\zq) \in \bC M \times \cQ\NM ;\; \all{s \in \R} \Sup(\bc) \subset \Dm(\zq(s,\cdot)) \right\}.
                                                                                                        \tag \label{Fxt222}$$
    Vertical curves $\lg$ and $\lg'$ are equivalent if
        $$\lg'(0) = \lg(0)
                                                                                                        \tag \label{Fxt223}$$
    and
        $$\xD(\bzl \circ \lg')(0) = \xD(\bzl \circ \lg)(0)
                                                                                                        \tag \label{Fxt224}$$
    for each differentiable function $\bzl$.  Equivalence classes are {\it vertical vectors}.  The equivalence class of a
vertical curve $\lg$ will be denoted by $\bt\lg(0)$ and the space of vertical vectors will be denoted by $\bV\bQ\NM$.
Fibres of the fibration
        $$\bzt^\sv_{\NM} \colon \bV\bQ\NM \rightarrow \bQ\NM \colon \bt\lg(0) \mapsto \lg(0)
                                                                                                        \tag \label{Fxt225}$$
    are vector spaces.

    The dual bundle is the set $\bV^\*\bQ\NM$ is the set of equivalence classes of pairs $(\lq,\bzl) \in \bQ\NM \times
\bK\NM$.  Two pairs $(\lq,\bzl)$ and $(\lq',\bzl')$ are equivalent if $\lq' = \lq$ and
        $$\xD(\bzl' \circ \bg(\bc,\zd\bc,\zq))(0) = \xD(\bzl \circ \bg(\bc,\zd\bc,\zq))(0)
                                                                                                        \tag \label{Fxt226}$$
    for each vertical differentiable curve $\lg \colon \R \rightarrow \bQ\NM$ such that $\lg(0) = \lq$.  This set is
isomorphic to the quotient of the bundle $\bT^\*\bQ\NM$ of covectors by the polar $\bV^\polar\bQ\NM$ of the subbundle of
vertical vectors.  We have the fibration
        $$\bzp^\sv_{\NM} \colon \bV^\*\bQ\NM \rightarrow \bQ\NM \colon \bt\lg(0) \mapsto \lg(0)
                                                                                                        \tag \label{Fxt227}$$

\vskip3mm
        \sscx{Virtual displacements, force and stress.}

    Let $\zg \colon M \rightarrow N$ be a placement map.  A commutative diagram
    \vskip-1mm$$\xy!<9mm>
        $$\xymatrix@R+6mm @C+10mm{{M} \ar[rd]_*{\zg} \ar[r]^*{\zd\zg} &
            \sT N \ar[d]_*{\zt_N} \\ & M}$$\endxy
                                                                                                        \tag \label{Fxt228}$$
    \vskip2mm
    \noindent is called a {\it deformation} of $\zg$.  A deformation is well represented by the mapping $\zd\zg$.  Let
$\cD\NM$ be the space of deformations and let $\bC M \ux \cD\NM$ denote the {\it restricted product} set
        $$\left\{(\bc,\zd\zg) \in \bC M \times \cQ\NM ;\; \Sup(\bc) \subset \Dm(\zd\zg) \right\}.
                                                                                                        \tag \label{Fxt229}$$

    We consider commutative diagrams
    \vskip-1mm$$\xy!<9mm>
        $$\xymatrix@R+6mm @C+15mm{{\sT N} \ar[d]_*{\zt_N} \ar[r]^*{\zf} &
            \tW^m\sT^\* M \ar[d]_*{\tzp_m{}_M} \\
            N \ar[r]^*{\zg^{-1}} & M}$$\endxy
                                                                                                        \tag \label{Fxt230}$$
    \vskip2mm
    \noindent and
    \vskip-1mm$$\xy!<10mm>
        $$\xymatrix@R+6mm @C+14mm{{\sT N} \ar[d]_*{\zt_N} \ar[r]^*{\zt} &
            \tW^{m-1}\sT^\* M \ar[d]_*{\tzp_{m-1}{}_M} \\
            N \ar[r]^*{\zg^{-1}} & M}$$\endxy
                                                                                                        \tag \label{Fxt231}$$
    \vskip2mm
    \noindent where $\zg$ is a placement map and $\tzp_{m-1}{}_M \colon \tW^{m-1}\sT^\* M \rightarrow M$ is the fibration
of odd $(m-1)$-covectors in $M$.  The diagram \Ref{Fxt230} represented by the mapping $\zf$ is called a {\it force field}
associated with the placement map $\zg$.  The diagram \Ref{Fxt231} is represented by the mapping $\zt$ and is called a
{\it stress field} associated with the placement map $\zg$.  The space of force fields will be denoted by $\cF\NM$ and the
space of stress fields will be denoted by $\cT\NM$. We will use the symbol $\bC M \ux (\cF\NM \ux \cT\NM)$ to denote the
{\it restricted product} composed of triples $(\bc,\zf,\zt) \in \bC M \times \cF\NM \times \cT\NM$.  The force field $\zf$
and the stress field $\zt$ in a triple $(\bc,\zf,\zt) \in \bC M \ux (\cF\NM \ux \cT\NM)$ are associated with the same
placement map $\zg$, and $\Sup(\bc) \subset \Dm(\zg)$.

    In the set $\bC M \ux \cD\NM$ we introduce an equivalence relation.  Two pairs $(\bc,\zd\zg)$ and
$(\bc',\zd\zg')$ are equivalent if $\bc' = \bc$ and
        $$\int_{\displaystyle\bc} \zf \circ \zd\zg' + \int_{\displaystyle\partial\bc} \zt \circ \zd\zg' =
\int_{\displaystyle\bc} \zf \circ \zd\zg + \int_{\displaystyle\partial\bc} \zt \circ \zd\zg
                                                                                                        \tag \label{Fxt232}$$
    for each $(\bc,\zf,\zt) \in \bC M \times \cF\NM \times \cT\NM$.  The equivalence class of $(\bc,\zd\zg)$ will be called
a {\it displacement} of the body $\bc$ and will be denoted by $\bd(\bc,\zd\zg)$. The space of all displacements will be
denoted by $\bD\NM$.  There is a dual equivalence relation in the set $\bC M \ux (\cF\NM \ux \cT\NM)$.  Triples
$(\bc,\zf,\zt)$ and $(\bc',\zf',\zt')$ are equivalent if $\bc' = \bc$ and
        $$\int_{\displaystyle\bc} \zf' \circ \zd\zg + \int_{\displaystyle\partial\bc} \zt' \circ \zd\zg =
\int_{\displaystyle\bc} \zf \circ \zd\zg + \int_{\displaystyle\partial\bc} \zt \circ \zd\zg
                                                                                                        \tag \label{Fxt233}$$
    for each $(\bc,\zd\zg) \in \bC M \ux \cQ\NM$.  The equivalence class of $(\bc,\zf,\zt)$ will be called a {\it load}
applied to the body $\bc$ and will be denoted by $\bl(\bc,\zf,\zt)$.  The space of loads will be denoted by $\bL\NM$.

        \sscx{The Euler-Lagrange mapping.}

    Fibrations
    \vskip-1mm$$\xy!<10mm>
        $$\xymatrix@R+7mm{*+<2mm>!<4mm,0mm>{\sJ^1\NM \fpr{(\zt_1{}_\NM,\zt_N)} \sT N}
\ar[d]_*{\zt_1{}_\NM{}^\*(\zt_N)} \\ \sJ^1\NM}$$\endxy
                                                                                                        \tag \label{Fxt234}$$
    \vskip2mm
    \noindent and
    \vskip-1mm$$\xy!<11mm>
        $$\xymatrix@R+7mm{*+<2mm>!<4mm,0mm>{\sJ^2\NM \fpr{(\zt_2{}_\NM,\zt_N)} \sT N}
\ar[d]_*{\zt_2{}_\NM{}^\*(\zt_N)} \\ \sJ^2\NM}$$\endxy
                                                                                                        \tag \label{Fxt235}$$
    \vskip2mm
    \noindent are the inverse images of
    \vskip-1mm$$\xy!<10mm>
        $$\hskip3mm\xymatrix@R+3mm{{\sT N} \ar[d]_*{\zt_N} \\ N}$$\endxy
                                                                                                        \tag \label{Fxt236}$$
    \vskip2mm
    \noindent under the jet-target projections $\zt_1{}_\NM \colon \sJ^1\NM \rightarrow N$ and $\zt_2{}_\NM \colon \sJ^2\NM
\rightarrow N$ respectively.  The projections $\zt_1{}_\NM{}^\*(\zt_N)$ and $\zt_2{}_\NM{}^\*(\zt_N)$ are obtained by
restricting the projections
        $$pr_{\sJ^1\NM} \colon \sJ^1\NM \times \sT N \rightarrow \sJ^1\NM
                                                                                                        \tag \label{Fxt237}$$
     and
        $$pr_{\sJ^2\NM} \colon \sJ^2\NM \times \sT N \rightarrow \sJ^2\NM
                                                                                                        \tag \label{Fxt238}$$
     to $\sJ^1\NM \fpr{(\zt_1{}_\NM,\zt_N)} \sT N$ and $\sJ^2\NM \fpr{(\zt_2{}_\NM,\zt_N)} \sT N$ respectively.  Both
inverse images are vector fibrations.  Fibres $(\zt_1{}_\NM{}^\*(\zt_N))^{-1}(\sj^1\zg(x))$ and
$(\zt_2{}_\NM{}^\*(\zt_N))^{-1}(\sj^2\zg(x))$ are both isomorphic to the fibre $\zt_N{}^{-1}(\zg(x))$.

    The {\it Euler-Lagrange mapping} is a morphism
    \vskip-1mm$$\xy!<11mm>
        \hskip-5mm$$\xymatrix@R+7mm @C+20mm{*+<3mm>!<4.3mm,1.7mm>{\sJ^2\NM \fpr{(\zt_2{}_\NM,\zt_N)} \sT N}
\ar[d]_*{\zt_2{}_\NM{}^\*(\zt_N)}\ar[r]^*{\cE\xd_V u} & \tW^m\sT^\* M \ar[d]_*{\tzp_m{}_M} \\ \sJ^2\NM
\ar[r]^*{\zs_2{}_\NM} & M}$$\endxy
                                                                                                        \tag \label{Fxt239}$$
    \vskip2mm
    \noindent of vector fibrations.  The {\it stress-strain mapping} is a morphism
    \vskip-1mm$$\xy!<11mm>
        \hskip-5mm$$\xymatrix@R+7mm @C+20mm{*+<3mm>!<4.3mm,1.7mm>{\sJ^1\NM \fpr{(\zt_1{}_\NM,\zt_N)} \sT N}
\ar[d]_*{\zt_1{}_\NM{}^\*(\zt_N)}\ar[r]^*{\cP\xd_V u} & \tW^{m-1}\sT^\*M \ar[d]_*{\tzp_{m-1}{}_M} \\ \sJ^1\NM
\ar[r]^*{\zs_1{}_\NM} & M}$$\endxy
                                                                                                        \tag \label{Fxt240}$$
    \vskip2mm
    \noindent of vector fibrations.

    A {\it vertical curve} in $\sJ^1\NM$ is a mapping $\zg \colon \R \rightarrow \sJ^1\NM$ such that the composition
$\zs_1{}_\NM \circ \zg$ is a constant mapping.  The image of the composition $\zk \circ \zg$ of a vertical curve $\zg$ with
a mapping $\zk$ in the commutative diagram \Ref{Fxt210} is included in the vector space
$(\tzp_m{}_M)^{-1}(\zs_1{}_\NM(\zg(0)) \subset R^m M$.  The derivative $\xD(\zk \circ \zg)(0)$ is well defined.  The {\it
vertical differential} $\xd_V\zk \colon \sV\sJ^1\NM \rightarrow R^m M$ of the mapping $\zk$ is defined in terms of the
directional derivative $\xd_V\zk(\st\zg(0)) = \xD(\zk \circ \zg)(0)$.  The diagram
    \vskip-1mm$$\xy!<10mm>
        $$\xymatrix@R+6mm @C+17mm{{\sV\sJ^1\NM} \ar[d]_*{\zt^\sv_{\sJ^1\NM}} \ar[r]^*{\xd_V\zk} &
            \tW^m\sT^\* M \ar[d]_*{\tzp_m{}_M} \\
            \sJ^1\NM \ar[r]^*{\zs_1{}_{\NM}} & M}$$\endxy
                                                                                                        \tag \label{Fxt241}$$
    \vskip2mm
    \noindent is a morphism of vector fibrations.

    With a function $a \colon M \rightarrow \R$ we associate the diagram
    \vskip-1mm$$\xy!<10mm>
        $$\xymatrix@R+6mm @C+15mm{{\sV\sJ^k\NM} \ar[d]_*{\zt^\sv_{\sJ^k\NM}}
\ar[r]^*{F(a)} &
            \sV\sJ^k\NM \ar[d]_*{\zt^\sv_{\sJ^k\NM}} \\
            \sJ^k\NM \ar@{=}[r] & \sJ^k\NM}$$\endxy
                                                                                                        \tag \label{Fxt242}$$
    \vskip2mm
    \noindent The first row is the mapping
        $$F(a) \colon \sV\sJ^k\NM \rightarrow \sV\sJ^k\NM \colon \sj^{1,k}\zq(0,x_0) \mapsto \sj^{1,k}\zq^a(0,x_0)
                                                                                                        \tag \label{Fxt243}$$
     with the mapping $\zq^a \colon \R \times M \rightarrow N$ defined by $\zq^a(s,x) = \zq(s(a(x) - a(x_0)),x)$.

    The diagram \Ref{Fxt242} acts on a diagram
    \vskip-1mm$$\xy!<10mm>
        $$\xymatrix@R+6mm @C+17mm{{\sV\sJ^k\NM} \ar[d]_*{\zt^\sv_{\sJ^k\NM}} \ar[r]^*{\zm} &
            \tW^m\sT^\* M \ar[d]_*{\tzp_m{}_M} \\
            \sJ^1\NM \ar[r]^*{\zs_1{}_{\NM}} & M}$$\endxy
                                                                                                        \tag \label{Fxt244}$$
    \vskip2mm
    \noindent producing the diagram
    \vskip-1mm$$\xy!<10mm>
        $$\xymatrix@R+6mm @C+17mm{{\sV\sJ^k\NM} \ar[d]_*{\zt^\sv_{\sJ^k\NM}}
\ar[r]^*{\xi_{F(a)}\zm} &
            \tW^m\sT^\* M \ar[d]_*{\tzp_m{}_M} \\
            \sJ^1\NM \ar[r]^*{\zs_1{}_{\NM}} & M}$$\endxy
                                                                                                        \tag \label{Fxt245}$$
    \vskip2mm
    \noindent where the mapping $\xi_{F(a)}\zm$ is defined by
        $$\xi_{F(a)}\zm(w) = \zm(F(a)(w)).
                                                                                                        \tag \label{Fxt246}$$

    The inverse image
    \vskip-1mm$$\xy!<10mm>
        $$\xymatrix@R+7mm{*+<2mm>!<4mm,0mm>{\sJ^1\NM \fpr{(\zs_1{}_\NM,\zp_M)} \sT^\* M}
\ar[d]_*{\zs_1{}_\NM{}^\*(\zp_M)} \\ \sJ^1\NM}$$\endxy
                                                                                                        \tag \label{Fxt247}$$
    \vskip2mm
    \noindent diagram is a vector fibration and the mapping
        $$\xi_{F(\cdot)}\zm \colon \bigl(\sJ^1\NM \fpr{(\zs_1{}_\NM,\zp_M)} \sT^\* M\bigr) \times_{\sJ^1\NM} \sV\sJ^1\NM
\rightarrow \tW^m\sT^\* M \colon (\xd a(x),w) \mapsto \xi_{F(a)}\zm(w)
                                                                                                        \tag \label{Fxt248}$$
    is bilinear.  It induces a morphism
    \vskip-1mm$$\xy!<10mm>
        \hskip3mm$$\xymatrix@R+6mm @C+20mm{{\sV\sJ^1\NM} \ar[d]_*{\zt^\sv_{\sJ^1\NM}} \ar[r]^*{\widetilde{\xi_{F}\zm}} &
            \sT M \otimes \tW^m\sT^\* M \ar[d]_*{\zp} \\
            \sJ^1\NM \ar[r]^*{\zs_1{}_\NM} & M}$$\endxy
                                                                                                        \tag \label{Fxt249}$$
    \vskip2mm
    \noindent of vector fibrations.  The mapping $\widetilde{\xi_{F}\zm}$ in the upper row is characterized by
        $$\langle \xd a(x) \otimes \tilde c, \widetilde{\xi_{F}\zm}(w)\rangle = \langle \tilde c,
\xi_{F(a)}\zm(w)\rangle
                                                                                                        \tag \label{Fxt250}$$
    with $a \colon M \rightarrow \R$, $x \in M$, $w \in \sT\sJ^1(N|M,x) \subset \sV\sJ^1\NM$, and $\tilde c \in \tW^m\sT_x
M$.  We are using in \Ref{Fxt248} the symbol $\times_{\sJ^1\NM}$ for the fibre product of two vector fibrations.  The
symbol $\zp$ in \Ref{Fxt249} has an obvious meaning.  The diagram
    \vskip-1mm$$\xy!<10mm>
        $$\xymatrix@R+6mm @C+17mm{{\sV\sJ^1\NM} \ar[d]_*{\zt^\sv_{\sJ^1\NM}}
\ar[r]^*{\xi_{F}\zm} &
            \tW^{m-1}\sT^\* M \ar[d]_*{\zp} \\
            \sJ^1\NM \ar[r]^*{\zs_1{}_\NM} & M}$$\endxy
                                                                                                        \tag \label{Fxt251}$$
    \vskip2mm
    \noindent is obtained from the diagram \Ref{Fxt249} with the use of the isomorphism of $\sT M \otimes \tW^m\sT^\* M$
with $\tW^{m-1}\sT^\* M$.

    The mapping $\xi_{F}\zm$ is vertical with respect to the projection $\zt_1{}_\NM$ in the sense that if
$\sT\zt_1{}_\NM(w) = 0$, then $\xi_{F}\zm(w) = 0$.  This property makes it possible to construct a mapping
        $$\cP\zm \colon \sJ^1\NM \fpr{(\zt_1{}_\NM,\zt_N)} \sT N \rightarrow \tW^{m-1}\sT^\*M
                                                                                                        \tag \label{Fxt252}$$
     characterized by $\cP\zm(j,v) = \xi_{F}\zm(w)$, where $w$ is a vertical vector such that $\zt^\sv_{\sJ^1\NM}(w) = j$
and $\sT\zt_1{}_\NM(w) = v$.  The diagram
    \vskip-1mm$$\xy!<11mm>
        \hskip-5mm$$\xymatrix@R+7mm @C+20mm{*+<3mm>!<4.3mm,1.7mm>{\sJ^1\NM \fpr{(\zt_1{}_\NM,\zt_N)} \sT N}
\ar[d]_*{\zt_1{}_\NM{}^\*(\zt_N)}\ar[r]^*{\cP\zm} & \tW^{m-1}\sT^\*M \ar[d]_*{\tzp_{m-1}{}_M} \\ \sJ^1\NM
\ar[r]^*{\zs_1{}_\NM} & M}$$\endxy
                                                                                                        \tag \label{Fxt253}$$
    \vskip2mm
    \noindent is a morphism of vector fibrations.

    Let $\zd\zg \colon M \rightarrow \sT N$ be a deformation of a placement map $\zg \colon M \rightarrow N$. The jet
prolongations $\sj^1\zd\zg \colon M \rightarrow \sJ^1(\sT N|M)$ and $\sj^2\zd\zg \colon M \rightarrow \sJ^2(\sT N|M)$ are
transformed in deformations $\zd\sj^1\zg = \zk^{1,1}{}_\NM \circ \sj^1\zd\zg$ and $\zd\sj^2\zg = \zk^{2,1}{}_\NM \circ
\sj^2\zd\zg$ of the jet prolongations $\sj^1\zg$ and $\sj^2\zg$ respectively.  Mappings $\zg$, $\zd\zg$, $\zd\sj^1\zg$, and
$\zd\sj^2\zg$ can be constructed from a mapping $\zq \colon \R \times M \colon N$ by the following definitions
        $$\zg \colon M \rightarrow N \colon x \mapsto \zq(0,x),
                                                                                                        \tag \label{Fxt254}$$
        $$\zd\zg \colon M \rightarrow \sT N \colon x \mapsto \st\zq(\cdot,x)(0),
                                                                                                        \tag \label{Fxt255}$$
        $$\zd\sj^1\zg \colon M \rightarrow \sV\sJ^1\NM \colon x \mapsto \sj^{1,1}\zq(0,x),
                                                                                                        \tag \label{Fxt256}$$
    and
        $$\zd\sj^2\zg \colon M \rightarrow \sV\sJ^2\NM \colon x \mapsto \sj^{1,2}\zq(0,x),
                                                                                                        \tag \label{Fxt257}$$

    The {\it total differential} is a differential operator applied to linear morphisms from $\sV\sJ^1\NM$ to
$\tW^{m-1}\sT^\*M$.  The total differential of a morphism $\zn \colon \sV\sJ^1\NM \rightarrow \tW^{m-1}\sT^\*M$ is a linear
morphism $\xd_T\zn \colon \sV\sJ^1\NM \rightarrow \tW^{m-1}\sT^\*M$ characterized by $\xd_T\zn \circ \zd\sj^2\zg = \xd(\zn
\circ \zd\sj^1\zg)$ for each deformation $\zd\zg$ of a placement $\zg$.

\vskip3mm
        \sscx{The internal energy.}

    The {\it internal energy density} of a material is represented by a mapping $u \colon \sJ^1\NM \rightarrow \tW^m\sT^\*
M$.  The {\it internal energy} of a configuration $\bq(\bc,\zg)$ of a body $\bc$ is the integral
        $$U(\bq(\bc,\zg)) = \int_{\displaystyle\bc} u \circ \sj^1\zg.
                                                                                                        \tag \label{Fxt258}$$
    The internal energy is a function on $\bQ\NM$.

    Variational principles of statics require the definition of the internal energy of a configuration.  The form of the
internal energy varies with the type of the medium.  The study of the variational principles for different types of media
is underway.

    \centerline{A REFERENCE}

[DA] Giusepe Marmo, W\l odzimierz Tulczyjew, Pawe\l \ Urba\'nski, Dynamics of autonomous systems with external forces, Acta
Physica Polonica B, {\bf 33} (2002), 1181--1240

        \end